\documentclass[12pt,a4paper]{article}
\usepackage[british]{babel}
\usepackage[latin1]{inputenc}
\usepackage{color}
\usepackage{ulem}

\usepackage[intlimits]{amsmath}
\usepackage{amssymb}
\usepackage{mathrsfs} 

\usepackage{url}
\usepackage{graphicx}

\usepackage[bf,small]{caption}

\usepackage{cite}
\usepackage{booktabs}
\usepackage{slashed}
\usepackage{hyperref}

\hypersetup{
 bookmarksnumbered=true,
 pdftitle = {Antiproton Limits on Decaying Gravitino Dark Matter}, 
 pdfauthor = {Timur Delahaye, Michael Grefe}, 
 pdfsubject = {antiproton limits on decaying gravitino dark matter}, 
 pdfkeywords = {gravitino, dark matter, R-parity violation, cosmic rays, antiproton},
 colorlinks=true,
 linkcolor=blue,
 citecolor=blue,
 urlcolor=blue,
}

\numberwithin{equation}{section}
\allowdisplaybreaks[1]

\providecommand{\abs}[1]{\ensuremath{\left\lvert #1\right\rvert }}
\providecommand{\MP}{\ensuremath{M_{\text{Pl}}}}
\DeclareMathOperator{\BR}{BR}
\DeclareMathOperator{\RE}{Re}

\begin{document}

\date{\mbox{ }}

\title{
{\normalsize
FTUAM-13-10\hfill\mbox{}\\
IFT-UAM/CSIC-13-055\hfill\mbox{}\\
LAPTH-017/13\hfill\mbox{}\\
December 2013\hfill\mbox{}\\}
\vspace{1.5cm}
\bf Antiproton Limits on \\ Decaying Gravitino Dark Matter\\[8mm]}

\author{Timur Delahaye$^{a,b,}$\thanks{Electronic address: timur.delahaye@fysik.su.se}\ ~and Michael Grefe$^{c,}$\thanks{Electronic address: michael.grefe@uam.es}\\[2mm]
{\normalsize\it \hspace*{-1cm}$^a$LAPTh, Universit\'e de Savoie, CNRS, BP 110, F-74941 Annecy-le-Vieux Cedex, France\hspace*{-1cm}} \\
{\normalsize\it \hspace*{-1cm}$^b$Institut d'Astrophysique de Paris, UMR 7095 -- CNRS, Universit\'e Pierre \& Marie Curie,\hspace*{-1cm}} \\[-2mm]
{\normalsize\it 98 bis boulevard Arago, 75014, Paris, France} \\
{\normalsize\it $^c$Instituto de F\'isica Te\'orica UAM/CSIC and Departamento de F\'isica Te\'orica,} \\[-2mm]
{\normalsize\it Universidad Aut\'onoma de Madrid, Cantoblanco, E-28049 Madrid, Spain} 
}
\maketitle

\thispagestyle{empty}

\begin{abstract}

We derive 95\,\% CL lower limits on the lifetime of decaying dark matter in the channels $Z\nu$, $W\ell$ and $h\nu$ using measurements of the cosmic-ray antiproton flux by the PAMELA experiment. Performing a scan over the allowed range of cosmic-ray propagation parameters we find lifetime limits in the range of $8\times10^{28}$\,s to $5\times10^{25}$\,s for dark matter masses from roughly 100\,GeV to 10\,TeV. We apply these limits to the well-motivated case of gravitino dark matter in scenarios with bilinear violation of $R$-parity and find a similar range of lifetime limits for the same range of gravitino masses. Converting the lifetime limits to constraints on the size of the $R$-parity violating coupling we find upper limits in the range of $10^{-8}$ to $8\times10^{-13}$.

\end{abstract}

\newpage

\section{Introduction}

Despite a lot of theoretical and experimental efforts, the nature of the dark matter (DM) in the Universe still remains one of the biggest unresolved problems in cosmology and particle physics. At the same time the explanation of the DM by a yet-unknown particle is one of the best motivations for the existence of particle physics beyond the Standard Model. A multitude of candidate particles has been proposed in the literature, the most thoroughly studied candidates being weakly interacting massive particles (WIMPs) and in particular the lightest neutralino in the framework of the Minimal Supersymmetric Standard Model (for recent reviews see~\cite{Bertone:2004pz,Feng:2010gw,Bergstrom:2012fi}).\smallskip

In the last years several possible hints of DM have been observed: To name a few, the DAMA and CoGeNT underground experiments observed an annual modulation signal that is consistent with light WIMPs from the Galactic halo scattering off detector nuclei~\cite{Bernabei:2010mq,Aalseth:2011wp}. Also the CRESST-II and CDMS II experiments report a few signal events that could be interpreted as a light WIMP signal~\cite{Angloher:2011uu,Agnese:2013rvf}. In addition, the PAMELA satellite observed a rise in the cosmic-ray positron fraction above 10\,GeV~\cite{Adriani:2008zr},\footnote{This rise was actually expected and announced back in 1989 due to the positron production by nearby pulsars~\cite{Boulares:1989}.} and it has been claimed that data of the Fermi-LAT experiment feature a gamma-ray signal in the Galactic centre region that could originate from the annihilation of light DM particles~\cite{Hooper:2010mq}, as well as a gamma-ray line at around 130\,GeV that could be a signal of DM annihilation into photons~\cite{Bringmann:2012vr,Weniger:2012tx}. 

So far, all of these hints have been inconclusive. In particular, some of them could be explained by other processes (the rise in the positron fraction could, for instance, originate from nearby astrophysical sources~\cite{Delahaye:2010ji,Ahlers:2009ae}), while others are in strong tension with further experimental results (the XENON experiment excludes the parameter space compatible with light WIMP signals claimed by other underground experiments~\cite{Aprile:2012nq}), or could be due to systematics or just be a statistical fluctuation (the origin of the Fermi gamma-ray line around 130\,GeV still remains unknown~\cite{Fermi-LAT:2013uma}). For these reasons we are not focussing on those DM hints in the present work.\smallskip

In this paper we are studying the gravitino as a candidate for the DM in the Universe~\cite{Pagels:1981ke}.\footnote{The gravitino is the spin-3/2 superpartner of the graviton in supergravity theories.} In particular, we are considering gravitino DM in a supersymmetric model with bilinear violation of $R$-parity~\cite{Takayama:2000uz,Buchmuller:2007ui}. The main motivation for this scenario is that a gravitino as the lightest supersymmetric particle (LSP) together with a small violation of $R$-parity leads to a cosmology that is consistent with thermal leptogenesis~\cite{Fukugita:1986hr} and avoids cosmological gravitino problems~\cite{Buchmuller:2007ui}. Due to the double suppression of the gravitino interactions by the Planck scale and by the small $R$-parity violating couplings, the gravitino remains a viable DM candidate with a lifetime well exceeding the age of the Universe~\cite{Takayama:2000uz}.

An interesting feature of this scenario is that the next-to-lightest supersymmetric particle (NLSP) is expected to be metastable on collider time scales. This is due to the fact that the decay into the gravitino LSP is suppressed by the Planck scale such that the NLSP decay is dominated by $R$-parity violating channels into Standard Model particles. The main signatures are displaced vertices with distinctive decay signatures or tracks of charged massive particles leaving the detector, depending on the nature of the NLSP and the smallness of $R$-parity violation. The collider phenomenology of scenarios with small bilinear $R$-parity violation has been studied in detail, ranging from simple estimates~\cite{Buchmuller:2007ui,Bobrovskyi:2010ps} to more sophisticated studies for different NLSP candidates including a generic detector simulation~\cite{Bobrovskyi:2011vx,Bobrovskyi:2012dc}. 

Another signature of gravitino DM could come from indirect DM searches. Although the gravitino lifetime is extremely long, its decays might be observed in the spectra of cosmic rays. Signatures of decaying gravitino DM in several cosmic-ray channels have been discussed in the literature: gamma rays~\cite{Takayama:2000uz,Buchmuller:2007ui,Bertone:2007aw,Ibarra:2007wg,Choi:2009ng,Vertongen:2011mu,Restrepo:2011rj,Huang:2011xr,Ishiwata:2008cu,Buchmuller:2009xv,Bomark:2009zm,Grefe:2011kh}, charged cosmic rays~\cite{Ibarra:2008qg,Ishiwata:2008cu,Buchmuller:2009xv,Bomark:2009zm,Grefe:2011kh} and neutrinos~\cite{Bomark:2009zm,Covi:2008jy,Grefe:2011kh}. In particular, in most of these works gravitino DM was discussed in the context of the EGRET gamma-ray excess above a few GeV~\cite{Strong:2004ry} and later also in the context of the PAMELA and Fermi-LAT excesses in the cosmic-ray electron and positron spectra~\cite{Adriani:2008zr,Ackermann:2010ij}. Since the time of these studies the EGRET excess has been falsified by Fermi-LAT observations~\cite{Abdo:2010nz}, while the rise in the positron fraction has been confirmed by Fermi LAT~\cite{FermiLAT:2011ab} and very recently with great precision by AMS-02~\cite{Aguilar:2013qda}.\footnote{The AMS-02 result was studied recently in the context of gravitino DM with $R$-parity violation~\cite{Ibe:2013nka}.} Moreover, new data on antiprotons have been published by the PAMELA collaboration~\cite{Adriani:2010rc,Adriani:2012paa}.\footnote{One can also refer to~\cite{Maurin:2013lwa} to get access to all the existing data on charged cosmic rays.}

Let us add one remark about the recent activity related to the hint for a gamma-ray line around 130\,GeV in the Fermi-LAT data. Although the expected gamma-ray spectrum from gravitino DM with a Wino NLSP could be compatible with the line observation, the angular distribution of the signal is found to be generically in tension with the one expected for the case of gravitino decays~\cite{Buchmuller:2012rc}. We therefore do not further pursue this possibility in this work.

The aim of the present study is rather to provide new lower limits on the gravitino lifetime derived from antiproton observations and to convert these limits into constraints on the amount of $R$-parity violation in gravitino DM scenarios. Besides new data, we use an updated calculation of the gravitino two-body decay widths that was obtained in~\cite{Grefe:2011dp}. In addition, we update the generation of the decay spectra with the help of \textsc{Pythia}~\cite{Sjostrand:2006za} by increasing statistics and employing a new simulation that includes the effect of QED final state radiation from leptons in the two-body final state. The latter effect is not of particular importance for the present study of antiproton signals, but will allow to use the same simulation also for the study of other cosmic-ray channels in the future.

One of the main novelties of the current study is that we derive antiproton limits by performing a scan over a large set of allowed cosmic-ray propagation parameters instead of working with a predefined set. This method allows us to estimate more reliably the uncertainty range introduced by the propagation of charged cosmic rays in the Milky Way.\smallskip

The outline of the paper is as follows: In Section~\ref{gravitino} we shortly review the relevant gravitino cosmology, introduce the scenario of bilinear $R$-parity violation, and discuss the decay channels and branching ratios of the unstable gravitino LSP. In Section~\ref{spectra} we describe the simulation of the spectra of final state particles created in two-body decays of gravitinos. In Section~\ref{limits} we derive limits from antiproton observations on the lifetime of the individual decay channels that appear in gravitino decays, while we apply these limits to the particular case of decaying gravitino DM in Section~\ref{lifetime}. Before concluding we also transform these limits into constraints on the amount of $R$-parity violation in this scenario. In Appendices~\ref{App:RPV}--\ref{App:CR} we give additional information on bilinear $R$-parity violation, the gravitino decay widths, the generation of decay spectra in \textsc{Pythia}, and cosmic-ray propagation in the Milky Way, respectively.

\section{Gravitino Dark Matter}
\label{gravitino}

\subsection{Gravitino Cosmology}
\label{cosmo}

According to current standard cosmological scenarios, an inflationary phase in the early Universe dilutes any primordial abundance of gravitinos and in most cases gravitinos do not reach thermal equilibrium with the hot plasma during reheating after inflation~\cite{Ellis:1982yb}. Anyway, gravitinos can still be produced abundantly in scattering processes in the thermal bath~\cite{Nanopoulos:1983up}. The gravitino relic density from this thermal production is given by~\cite{Pradler:2006qh}:\footnote{The prefactor of this formula has an $\mathcal{O}(1)$ uncertainty from unknown higher order contributions and non-perturbative effects~\cite{Bolz:2000fu}. Resummation of thermal masses leads to an increase of the gravitino relic density by about a factor of two~\cite{Rychkov:2007uq}. Keeping these caveats in mind, for the qualitative discussion in this section it will be sufficient to work with the results of Eq.~(\ref{eq:relicdensity}).}
\begin{equation}
  \Omega_{3/2}^{\text{TP}}h^2\simeq\sum_{i=1}^3\omega_i\,g_i^2\left(1+\frac{M_i^2}{3\,m_{3/2}^2}\right)\ln\left(\frac{k_i}{g_i}\right)\left(\frac{m_{3/2}}{100\,\text{GeV}}\right)\left(\frac{T_R}{10^{10}\,\text{GeV}}\right),
\label{eq:relicdensity}
\end{equation}
where the sum runs over the Standard Model gauge groups. The gauge couplings $g_i$ and the gaugino masses $M_i$ are understood to be evaluated at an energy corresponding to the reheating temperature $T_R$. Assuming a reheating temperature larger than a common SUSY mass scale $m_\text{SUSY}$, the one-loop renormalization group equations for these parameters are given by
\begin{equation}
\begin{split}
  g_i(T_R) &=\left[g_i(m_Z)^{-2}-\frac{\beta_i^\text{SM}}{8\,\pi^2}\ln\left(\frac{m_\text{SUSY}}{m_Z}\right)-\frac{\beta_i^\text{SUSY}}{8\,\pi^2}\ln\left(\frac{T_R}{m_\text{SUSY}}\right)\right]^{-1/2}, \\
  M_i(T_R) &=\left(\frac{g_i(T_R)}{g_i(m_Z)}\right)^2M_i(m_Z)\,.
\end{split}
\end{equation}
For the calculations in this section we assume $m_\text{SUSY}=1\,$TeV. The other parameters for the evaluation of these expressions are found in Table~\ref{tab:relicdensity}.
\begin{table}[t]
 \centering
 \begin{tabular}{cccccccc}
  \toprule
  gauge group & $i$ & $\omega_i$ & $k_i$ & $\beta_i^\text{SM}$ & $\beta_i^\text{SUSY}$ & $g_i$ & $g_i(m_Z)$ \\
  \midrule
  $U(1)_Y$ & 1 & 0.018 & 1.266 & 41/6 &  11 & $g'$ & 0.36 \\
  $SU(2)_L$ & 2 & 0.044 & 1.312 & -19/6 &  1 & $g$ & 0.65 \\
  $SU(3)_C$ & 3 & 0.117 & 1.271 & -7 &  -3 & $g_s$ & 1.22 \\
  \bottomrule
 \end{tabular}
 \caption{Collection of parameters for the calculation of the gravitino abundance from thermal production according to Eq.~(\ref{eq:relicdensity}). The values for $\omega_i$ and $k_i$ are taken from~\cite{Pradler:2006qh}.}
 \label{tab:relicdensity}
\end{table}
Decays of the NLSP into the gravitino and Standard Model particles do not contribute to the gravitino relic density in scenarios with broken $R$-parity. This is because decay processes involving a gravitino in the final state are suppressed compared to $R$-parity violating decays unless the amount of $R$-parity violation is extremely small.

\begin{figure}[t]
\centering
  \includegraphics[width=0.8\linewidth]{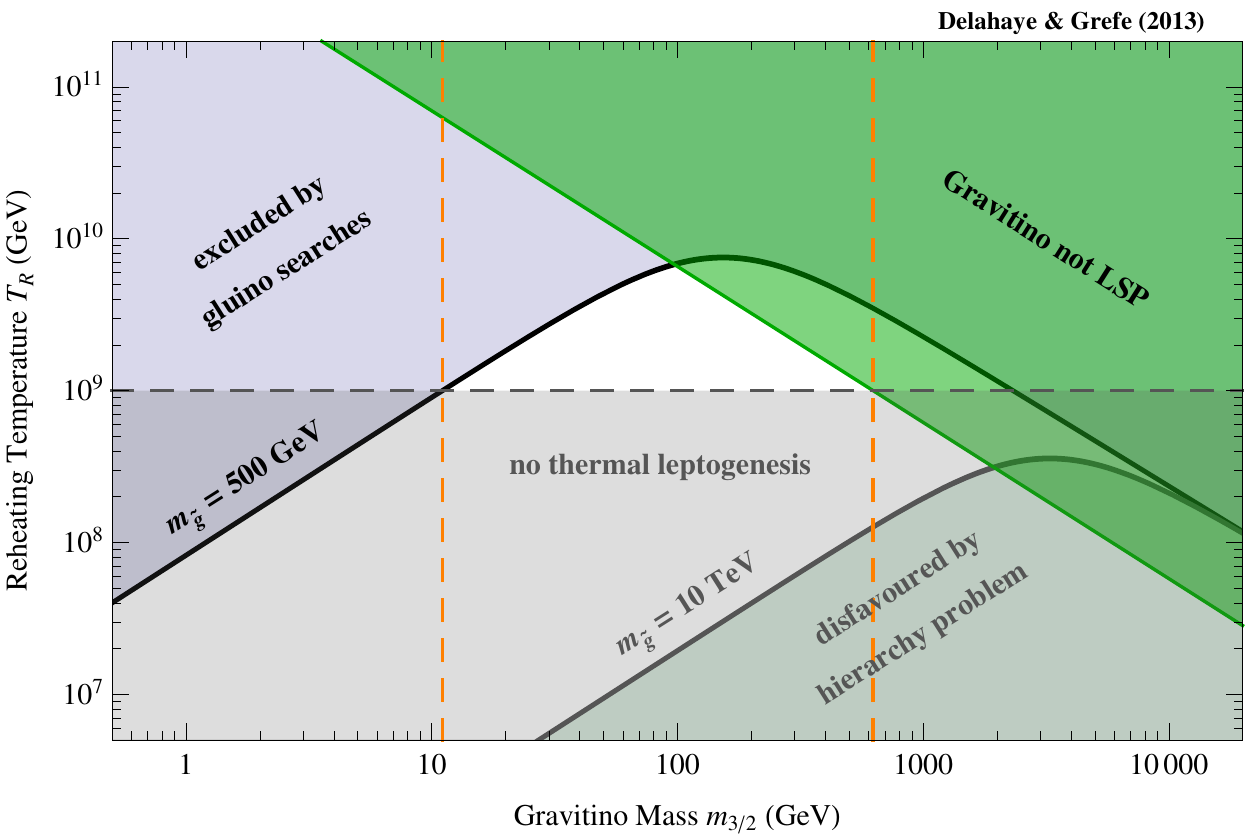}
  \caption{Gravitino mass--reheating temperature plane. The solid black lines show the contours where the gravitino relic density matches the observed DM density for a given value of the gluino mass according to Eq.~\eqref{eq:relicdensity}. We present here the case of universal gaugino masses at the GUT scale. Low values of the gluino mass are excluded by collider searches (\textit{violet region}) and large values are disfavoured by the hierarchy problem (\textit{grey region}). In the upper right corner (\textit{green region}) the gravitino is heavier than the Bino and thus cannot be a DM candidate. Standard thermal leptogenesis requires $T_R\gtrsim10^9\,$GeV, constraining the allowed gravitino parameter space to the white area between the orange dashed lines, \textit{i.e.} $10\,\text{GeV}\lesssim m_{3/2}\lesssim500$\,GeV. See text for details.}
  \label{fig:relicdensity}
\end{figure}
In order to illustrate the parameter space of thermally produced gravitino DM, in Fig.~\ref{fig:relicdensity} we present the plane spanned by the gravitino mass and the reheating temperature. Here we fix the gravitino relic abundance to the density of cold DM in the Universe as determined by a combination of most recent CMB and BAO data, \textit{i.e.} $\Omega_{\text{DM}}h^2\nobreak=\nobreak0.1187\nobreak\pm\nobreak0.0017$~\cite{Ade:2013lta}. According to Eq.~\eqref{eq:relicdensity} in this case every choice for the gaugino masses corresponds to a line in this plane. If one considers cases where the gravitino is not the only DM particle but makes up only a fraction of the measured energy density, also reheating temperatures below the indicated line can be allowed. By contrast, higher values of the reheating temperature are excluded since they would lead to overproduction of the gravitino relic abundance.

Standard thermal leptogenesis requires $T_R\gtrsim10^9\,$GeV in order to thermally produce the heavy Majorana neutrinos~\cite{Davidson:2002qv}. We observe from Fig.~\ref{fig:relicdensity} that this lower bound on the reheating temperature constrains the allowed range of gravitino masses. Under the assumption of universal gaugino masses at the GUT scale (evaluated at $m_\text{SUSY}=1\,$TeV) we find an allowed range between $m_{3/2}=\mathcal{O}(10)$\,GeV and $m_{3/2}=\mathcal{O}(500)$\,GeV for a lower limit on the gluino mass of $500$\,GeV.\footnote{The ATLAS and CMS collaborations at the LHC have performed several searches for signals of supersymmetry, so far without success. Depending on the model assumptions they find lower limits on the gluino mass ranging from several hundred GeV to above a TeV~\cite{ATLAS_SUSY,CMS_SUSY}. Since no studies exist for the particular case of supersymmetric models with gravitino DM and $R$-parity violation, for definiteness we use in Fig.~\ref{fig:relicdensity} a conservative lower limit of 500\,GeV for the gluino mass.} The upper limit in this case is given when the gravitino mass equals the Bino mass parameter as the gravitino has to be the LSP for being a DM candidate. This result is qualitatively equivalent to the result of a more sophisticated numerical analysis discussing also the possibility of other candidates for the NLSP~\cite{Buchmuller:2008vw}. A particular scenario that motivates a reheating temperature of the order of $10^{9\text{--}10}$\,GeV is spontaneous breaking of $B$--$L$ symmetry at the GUT scale followed by hybrid inflation and tachyonic preheating~\cite{Buchmuller:2012wn}. In this case a gravitino in the mass range $10\,\text{GeV}\lesssim m_{3/2}\lesssim700$\,GeV is found to be a viable DM candidate for a gluino mass of 1\,TeV.

Allowing for other models of leptogenesis (see for instance~\cite{Fong:2013wr}) the lower bound on the reheating temperature could be relaxed and also gravitino masses below $\mathcal{O}(10)$\,GeV could become viable. Likewise also the upper limit on the gravitino mass could be relaxed a bit. However, for a gravitino LSP with a mass above $\mathcal{O}(1)$\,TeV one would have to push the spectrum of supersymmetric particles to very large masses. In that case supersymmetry would not give a natural solution to the hierarchy problem. 

Apart from producing the correct relic density one could also worry about other cosmological gravitino problems~\cite{Khlopov:1984pf}. In fact, a gravitino that is not the LSP would be unstable with a decay width suppressed by the Planck scale and could therefore decay around the time of big bang nucleosynthesis (BBN). If the gravitino is sufficiently abundant, such late decays can easily influence the relic densities of the light elements produced through BBN, leading to conflicts with observations~\cite{Ellis:1984eq}. Therefore, in those cases the reheating temperature is constrained to be roughly lower than $10^5$--$10^6$\,GeV if the gravitino is not very heavy, \textit{i.e.} $m_{3/2}\lesssim10$\,TeV~\cite{Kawasaki:2008qe}. 

The choice of a stable gravitino LSP leads to a metastable NLSP, since it can only decay into the gravitino and Standard Model particles via interactions suppressed by the Planck scale. In this case there are also constraints from BBN, however not on the reheating temperature but on the gravitino mass. Since the NLSP lifetime roughly depends on the square of the gravitino mass, $m_{3/2}$ is constrained to be rather small, depending on the nature and mass of the NLSP~\cite{Kawasaki:2008qe}.

Once $R$-parity is broken, the NLSP dominantly decays into Standard Model particles via $R$-violating interactions. In this case the BBN constraints are easily evaded. Even a very small amount of $R$-parity violation leads to a decay of the NLSP before the time of BBN~\cite{Buchmuller:2007ui}. In Section~\ref{RPVconstraints} we will give a quantitative estimate of the corresponding constraints on the amount of $R$-parity violation.

\subsection{Gravitino Decay Through Bilinear \texorpdfstring{\boldmath$R$}{R}-Parity Violation}
\label{gravitinoRPV}

In models with bilinear $R$-parity breaking the distinction between down-type Higgs and lepton supermultiplets is lost. Since lepton number is not a conserved quantity anymore, the left-handed neutrinos mix with the neutralinos to form new mass eigenstates. Similarly, the charged leptons mix with the charginos, and the sneutrinos and sleptons mix with the Higgs bosons (see Appendix~\ref{App:RPV} for more details). These mixings, together with the gravitino interaction Lagrangian~\cite{Bolz:2000fu},
\begin{equation}
 \begin{split}
  \mathscr{L}_{\text{int}}= &-\frac{i}{\sqrt{2}\,\MP}\left[ \left( D_{\mu}^*\phi^*\right) \bar{\psi}_{\nu}\gamma^{\mu}\gamma^{\nu}P_L\chi-\left( D_{\mu}\phi\right) \bar{\chi}P_R\gamma^{\nu}\gamma^{\mu}\psi_{\nu}\right] \\
  &\qquad-\frac{i}{8\MP}\,\bar{\psi}_{\mu}\left[ \gamma^{\nu},\,\gamma^{\rho}\right] \gamma^{\mu}\lambda^aF_{\nu\rho}^a+\mathcal{O}(\MP^{-2}),
 \label{interaction}
 \end{split}
\end{equation}
lead to decays of the gravitino LSP and are therefore a crucial ingredient for the study of gravitino DM decay in this work.

\subsubsection*{Decay Channels and Decay Widths}

Based on the interaction Lagrangian given above, gravitinos with a mass around the electroweak scale can decay into up to four different two-body final states~\cite{Ishiwata:2008cu,Grefe:2008zz}:
\begin{equation}
  \psi_{3/2}\rightarrow\gamma\nu_i\,,\qquad \psi_{3/2}\rightarrow Z\nu_i\,,\qquad \psi_{3/2}\rightarrow W\ell_i\,,\qquad \psi_{3/2}\rightarrow h\nu_i\,, \nonumber
\end{equation}
where the index $i$ indicates the lepton flavour. For masses above the TeV scale also three-body decays into several massive gauge bosons~\cite{Grefe:2011dp} as well as electroweak bremsstrahlung~\cite{Ciafaloni:2010ti} could play a role but we will neglect these contributions in this work. For light gravitinos in general the two-body decay into a photon and a neutrino is the dominant decay channel, but three-body decays with virtual intermediate gauge bosons can also give important contributions~\cite{Choi:2010xn,Choi:2010jt,Diaz:2011pc,Grefe:2011dp}. For the purpose of the present work, however, we will concentrate on the two-body decays listed above and therefore restrict the analysis to gravitino masses larger than the mass of the $W$ boson.

The decay widths for the different two-body decay channels (including the corresponding conjugate final states) have been calculated in previous works and read\footnote{For more details on the decay widths see Appendix~\ref{App:decaywidths} and references therein.}
\begin{align}
  \Gamma_{\psi_{3/2}\rightarrow\gamma\nu_i} &\simeq\frac{\xi_i^2\,m_{3/2}^3}{32\,\pi\,\MP^2}\,m_Z^2\,s^2_{\theta_W}c^2_{\theta_W}\left(\frac{M_2-M_1}{M_1\,M_2}\right)^2, \label{gammanu}\\
  \Gamma_{\psi_{3/2}\rightarrow Z\nu_i} &\simeq\frac{\xi_i^2\,m_{3/2}^3\,\beta_Z^2}{32\,\pi\,\MP^2}\,\Bigg\{m_Z^2\left( \frac{s^2_{\theta_W}}{M_1}+\frac{c^2_{\theta_W}}{M_2}\right) ^2\!\!\!f_Z+\frac{1}{6}\abs{1+m_Z^2\,s_{2\beta}\,\frac{M_1\,c^2_{\theta_W}+M_2\,s^2_{\theta_W}}{M_1\,M_2\,\mu}}^2\!\!h_Z \nonumber\\
    &\quad\qquad-\frac{8}{3}\frac{m_Z^2}{m_{3/2}}\left( \frac{s^2_{\theta_W}}{M_1}+\frac{c^2_{\theta_W}}{M_2}\right) \left( 1+m_Z^2\,s_{2\beta}\,\frac{M_1\,c^2_{\theta_W}+M_2\,s^2_{\theta_W}}{M_1\,M_2\,\mu}\right) j_Z \Bigg\}\,, \\
  \Gamma_{\psi_{3/2}\rightarrow W\ell_i} &\simeq\frac{\xi_i^2\,m_{3/2}^3\,\beta_W^2}{16\,\pi\,\MP^2}\,\Bigg\{\frac{m_W^2}{M_2^2}\,f_W+\frac{1}{6}\abs{1+s_{2\beta}\frac{m_W^2}{M_2\,\mu}}^2\!\!h_W\nonumber\\
    &\quad\qquad-\frac{8}{3}\frac{m_W^2}{m_{3/2}M_2}\left( 1+s_{2\beta}\frac{m_W^2}{M_2\,\mu}\right) j_W\Bigg\}\,, \\
  \Gamma_{\psi_{3/2}\rightarrow h\nu_i} &\simeq\frac{\xi_i^2\,m_{3/2}^3\,\beta_h^4}{192\,\pi\,\MP^2}\abs{\frac{m_{\tilde{\nu}_i}^2+\frac{1}{2}\,m_Z^2\,c_{2\beta}}{m_h^2-m_{\tilde{\nu}_i}^2}}^2, \label{hnu}
\end{align}
where $s_{2\beta}\equiv\sin2\beta$, $c_{2\beta}\equiv\cos2\beta$, $s^2_{\theta_W}\equiv\sin^2\theta_W$, and $c^2_{\theta_W}\equiv\cos^2\theta_W$ for a compact notation. In these expressions $\MP=2.4\times10^{18}$\,GeV is the reduced Planck mass and $\xi_i$ parametrizes the strength of $R$-parity violation. The kinematic functions $\beta_X, f_X, j_X$, and $h_X$ are given by
\begin{alignat}{2}
  \beta_X &=1-\frac{m_X^2}{m_{3/2}^2}\,, &\qquad\qquad f_X &=1+\frac{2}{3}\,\frac{m_X^2}{m_{3/2}^2}+\frac{1}{3}\,\frac{m_X^4}{m_{3/2}^4}\,, \nonumber\\
  j_X &=1+\frac{1}{2}\,\frac{m_X^2}{m_{3/2}^2}\,, &\qquad\qquad h_X &=1+10\,\frac{m_X^2}{m_{3/2}^2}+\frac{m_X^4}{m_{3/2}^4}\,, \label{kinematicfunctions}
\end{alignat}
where $X=Z,\,W,\,h$ indicates the nature of the massive boson in the final state. Since the lepton flavour of the decay channels has practically no impact on the resulting antiproton spectra, for the purpose of the present work there is no need to make any assumption about the flavour structure of $R$-parity violation. Thus in the rest of the paper we will always work with the sum over the different flavours:
\begin{equation}
  \Gamma_{\psi_{3/2}\rightarrow\gamma\nu} = \sum_i\Gamma_{\psi_{3/2}\rightarrow\gamma\nu_i} \qquad\text{and}\qquad \xi^2 = \sum_i\xi_i^2\,.
\end{equation}

\begin{figure}[t]
 \centering
 \includegraphics[width=0.8\linewidth]{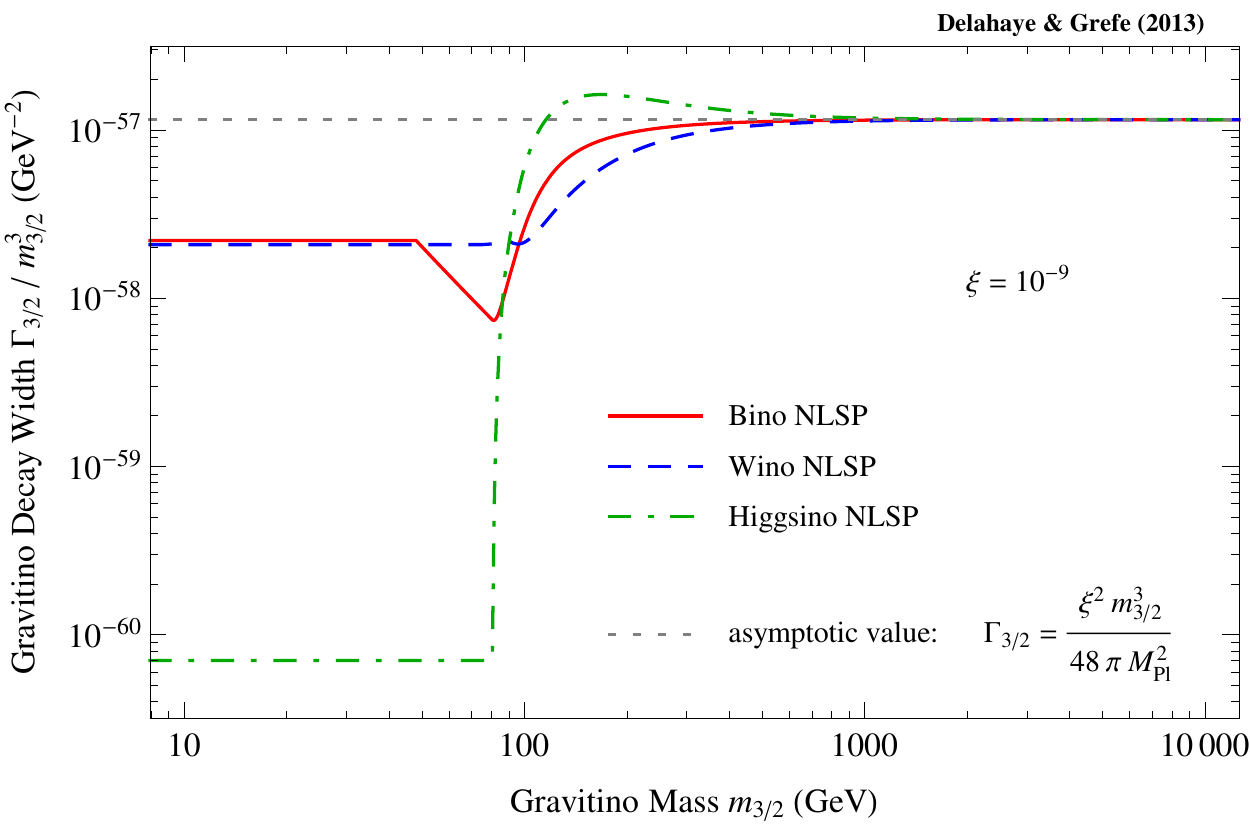} 
 \caption{Total decay width of the gravitino as a function of the gravitino mass for three choices of supersymmetry parameters: Bino NLSP, Wino NLSP and Higgsino NLSP. The decay width is rescaled by a factor $m_{3/2}^{-3}$ in order to make the asymptotic behaviour and features around the electroweak scale more obvious. See text for details.}
 \label{fig:width}
\end{figure}
To give an impression of the behaviour of the gravitino decay width as a function of the gravitino mass, we present in Fig.~\ref{fig:width} a calculation of the total decay width
\begin{equation}
  \Gamma_{3/2}=\Gamma_{\psi_{3/2}\rightarrow\gamma\nu}+\Gamma_{\psi_{3/2}\rightarrow Z\nu}+\Gamma_{\psi_{3/2}\rightarrow W\ell}+\Gamma_{\psi_{3/2}\rightarrow h\nu}
\end{equation}
for an arbitrarily fixed value of the $R$-parity breaking parameter $\xi=10^{-9}$. As can be seen from Eq.~(\ref{gammanu})--(\ref{hnu}) the gravitino decay width generally scales with the third power of the gravitino mass. For a better visibility of the asymptotic behaviour and features around the electroweak scale, we thus scale the decay width by a factor $m_{3/2}^{-3}$. Motivated by the discussion in~\cite{Buchmuller:2012rc} we consider three example cases for the relevant supersymmetry parameters: a case with a Bino-like NLSP, a case with a Wino-like NLSP, and a case with a Higgsino-like NLSP. See Table~\ref{parameters} for the corresponding choices of parameters.\footnote{For low gravitino masses we have to adjust the mass parameters to guarantee that the Wino mass parameter and the $\mu$-parameter obey the LEP limits on the chargino and charged Higgs masses, \textit{i.e.} $M_2, \mu\gtrsim100\,$GeV~\cite{Heister:2003zk,Abbiendi:2013hk}. In these cases we set either $M_2$ or $\mu$ to 100\,GeV and calculate the remaining masses accordingly, \textit{i.e.} keeping the ratios between the different masses as in Table~\ref{parameters}, but decoupling the gravitino mass.} At low gravitino masses the gaugino mass hierarchy has a strong impact on the total width as can be seen from the decay width of the photon + neutrino channel in Eq.~(\ref{gammanu}). In this region the curves are flat as the parameters $M_1$, $M_2$ and $\mu$ are fixed to their experimental lower limits and do not vary with the gravitino mass. For the case of Bino NLSP, around 50\,GeV $M_1$ and $M_2$ start to increase with the gravitino mass, thus suppressing the decay width for the $\gamma\nu$ channel. Above the $W$ mass, additional decay channels become kinematically available and the total decay width increases rapidly. In this region the scaling of the supersymmetric parameters causes a suppression of the decay width for all parameter sets but this suppression is overcompensated by the increase in the kinematic factors of Eq.~(\ref{kinematicfunctions}). At around 1\,TeV the gravitino decay width is completely dominated by Feynman diagrams that do not include mixing matrices and it approaches an asymptotic behaviour independent of the choice of the supersymmetry parameters~\cite{Grefe:2011dp} (\textit{cf.} Eq.~(\ref{gammanu})--(\ref{hnu})):
\begin{equation}
  \Gamma_{3/2}\rightarrow\frac{\xi^2\,m_{3/2}^3}{48\,\pi\,\MP^2}\,.
\end{equation}
\begin{table}[t]
 \centering
 \begin{tabular}{lccccc}
  \toprule
  Setup & $M_1$ & $M_2$ & $\mu$ & $m_{\tilde{\nu}_i}$ & $\tan\beta$ \\
  \midrule
  Bino NLSP & $1.1\,m_{3/2}$ & $1.9\,M_1$ & $10\,m_{3/2}$ & $2\,m_{3/2}$ & 10 \\
  Wino NLSP & $10\,m_{3/2}$ & $1.1\,m_{3/2}$ & $10\,m_{3/2}$ & $2\,m_{3/2}$ & 10 \\
  Higgsino NLSP & $10\,m_{3/2}$ & $1.9\,M_1$ & $1.1\,m_{3/2}$ & $2\,m_{3/2}$ & 10 \\
  \bottomrule
 \end{tabular}
 \caption{Choices of the relevant supersymmetry parameters for the three example cases of gravitino DM with a Bino-like NLSP, a Wino-like NLSP, or a Higgsino-like NLSP. The ratio of 1.9 between $M_2$ and $M_1$ for the cases of Bino and Higgsino NLSP is motivated by the assumption of universal gaugino masses at the GUT scale. We fix the ratio at a scale of 1\,TeV and do not implement any mass running as these numbers are only meant to represent three examples for different hierarchies of $M_1$, $M_2$ and $\mu$.}
 \label{parameters}
\end{table}

\subsubsection*{Branching Ratios}

In contrast to the absolute values of the decay widths, the branching ratios for the different decay channels do not depend on the strength of $R$-parity violation since all the decay widths are proportional to $\xi^2$. In this case the main dependence is on the gravitino mass and the supersymmetric mass spectrum. 

\begin{figure}[t]
 \centering
 \includegraphics[width=0.8\linewidth]{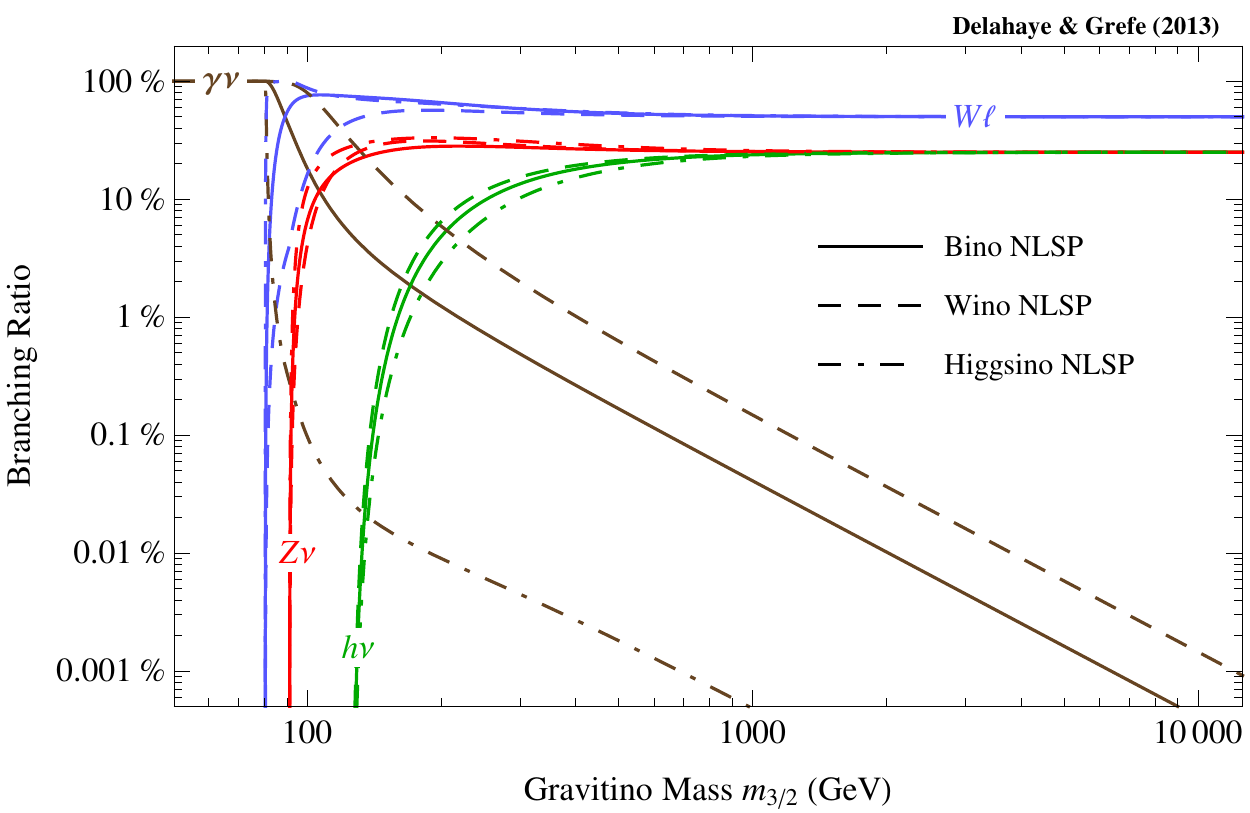} 
 \caption{Branching ratios of the different gravitino two-body decay channels as a function of the gravitino mass for three choices of supersymmetry parameters: Bino NLSP, Wino NLSP and Higgsino NLSP. In particular the branching ratio for the channel $\gamma\nu$ strongly depends on the supersymmetric mass spectrum. See text for details.}
 \label{BRplot}
\end{figure}
\begin{table}[t]
 \centering
\begin{footnotesize}
 \setlength{\tabcolsep}{0.84mm}
 \begin{tabular}{@{}c@{\hskip 3mm}cccc@{\hskip 3mm}cccc@{\hskip 3mm}cccc@{}}
  \toprule
  & \multicolumn{4}{c}{Bino NLSP} & \multicolumn{4}{c}{Wino NLSP} & \multicolumn{4}{c}{Higgsino NLSP} \\
   \cmidrule(lr){2-5}
   \cmidrule(lr){6-9}
   \cmidrule(l){10-13}
  $m_{3/2}$ & $\gamma\nu$ & $W\ell$ & $Z\nu$ & $h\nu$ & $\gamma\nu$ & $W\ell$ & $Z\nu$ & $h\nu$ & $\gamma\nu$ & $W\ell$ & $Z\nu$ & $h\nu$ \\
  \midrule
  85\,GeV & 77\,\% & 23\,\% & --- & --- & 99\,\% & 1.2\,\% & --- & --- & 1.1\,\% & 99\,\% & --- & --- \\
  100\,GeV & 18\,\% & 75\,\% & 6.6\,\% & --- & 80\,\% & 16\,\% & 4.1\,\% & --- & 0.097\,\% & 88\,\% & 12\,\% & --- \\
  150\,GeV & 2.7\,\% & 71\,\% & 25\,\% & 0.43\,\% & 15\,\% & 55\,\% & 30\,\% & 0.68\,\% & 0.016\,\% & 68\,\% & 32\,\% & 0.21\,\% \\
  200\,GeV & 1.2\,\% & 66\,\% & 28\,\% & 4.9\,\% & 5.9\,\% & 57\,\% & 31\,\% & 6.5\,\% & 0.0090\,\% & 64\,\% & 33\,\% & 2.9\,\% \\
  300\,GeV & 0.49\,\% & 59\,\% & 28\,\% & 13\,\% & 2.0\,\% & 54\,\% & 29\,\% & 15\,\% & 0.0045\,\% & 59\,\% & 31\,\% & 10\,\% \\
  500\,GeV & 0.17\,\% & 53\,\% & 26\,\% & 20\,\% & 0.63\,\% & 52\,\% & 26\,\% & 21\,\% & 0.0018\,\% & 54\,\% & 28\,\% & 18\,\% \\
  1\,TeV & 0.041\,\% & 51\,\% & 25\,\% & 24\,\% & 0.15\,\% & 50\,\% & 25\,\% & 24\,\% & 0.00048\,\% & 51\,\% & 26\,\% & 23\,\% \\
  2\,TeV & 0.010\,\% & 50\,\% & 25\,\% & 25\,\% & 0.036\,\% & 50\,\% & 25\,\% & 25\,\% & 0.00012\,\% & 50\,\% & 25\,\% & 25\,\% \\
  3\,TeV & 0.0045\,\% & 50\,\% & 25\,\% & 25\,\% & 0.016\,\% & 50\,\% & 25\,\% & 25\,\% & 0.000054\,\% & 50\,\% & 25\,\% & 25\,\% \\
  5\,TeV & 0.0016\,\% & 50\,\% & 25\,\% & 25\,\% & 0.0058\,\% & 50\,\% & 25\,\% & 25\,\% & 0.000020\,\% & 50\,\% & 25\,\% & 25\,\% \\
  10\,TeV & 0.00041\,\% & 50\,\% & 25\,\% & 25\,\% & 0.0015\,\% & 50\,\% & 25\,\% & 25\,\% & 0.0000049\,\% & 50\,\% & 25\,\% & 25\,\% \\
  \bottomrule
 \end{tabular}
\end{footnotesize}
 \caption{Branching ratios of the different gravitino two-body decay channels for several gravitino masses for the three example cases of Bino NLSP, Wino NLSP and Higgsino NLSP.}
 \label{tab:BRtable}
\end{table}
The resulting branching ratios are shown in Fig.~\ref{BRplot} and given in tabulated form in Table~\ref{tab:BRtable}. The most apparent difference of the three considered cases is the large separation of the respective branching ratios for the photon + neutrino channel above the kinematic threshold for the other decay channels. This is due to the different hierarchies for the Bino and Wino mass parameters that have a strong influence on the decay width for the photon + neutrino channel (see Eq.~(\ref{gammanu})). The other decay channels only have a strong dependence on the mass spectrum in the vicinity of their kinematic thresholds. As mentioned before, for gravitino masses above several hundreds of GeV the decay widths are dominated by Feynman diagrams that do not include mixing matrices and therefore the result becomes practically independent of the supersymmetric mass spectrum.

\section{Decay Spectra}
\label{spectra}

\begin{figure}[ht]
 \centering
 \includegraphics[width=0.325\textwidth]{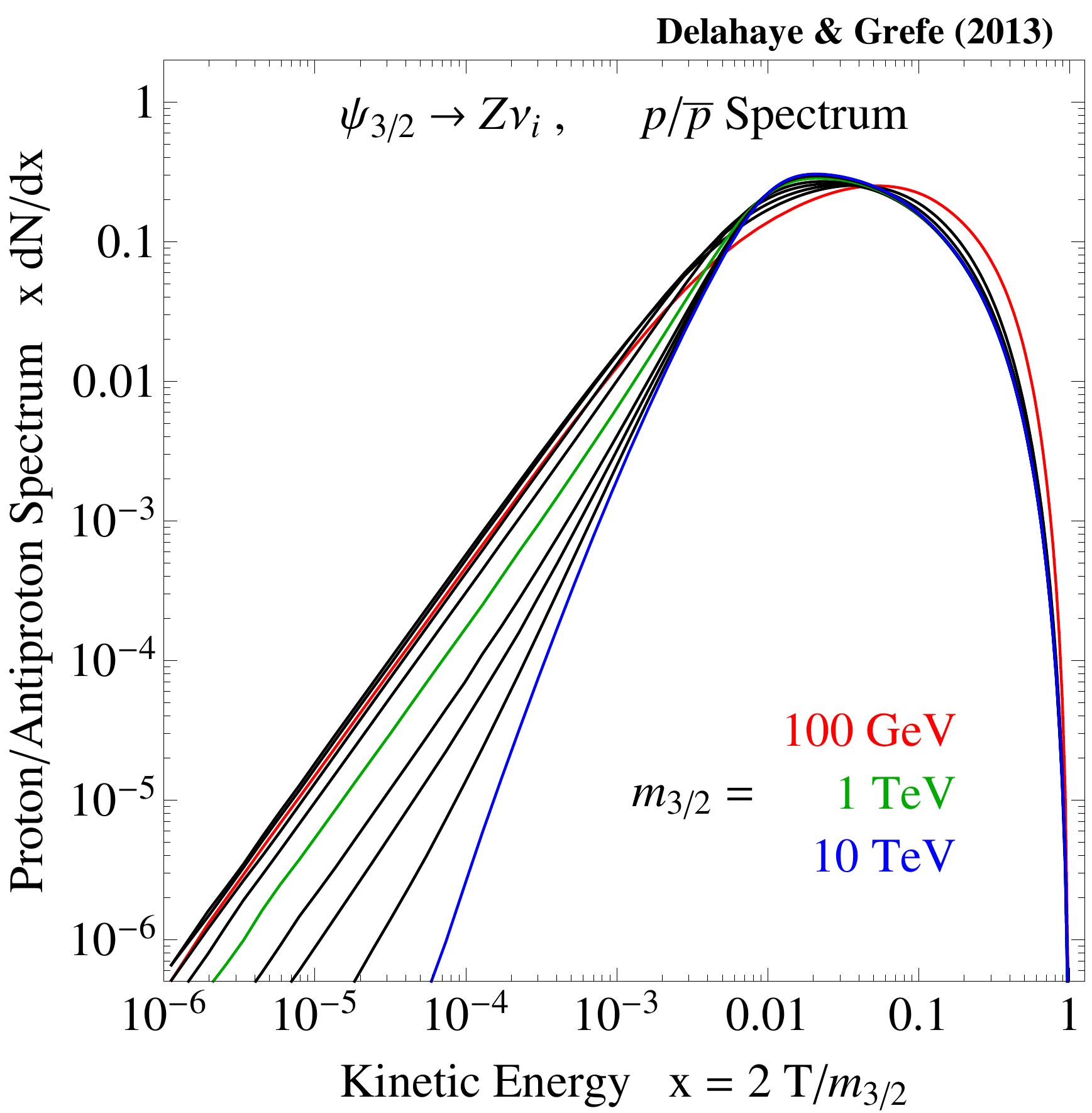} 
 \hfill
 \includegraphics[width=0.325\textwidth]{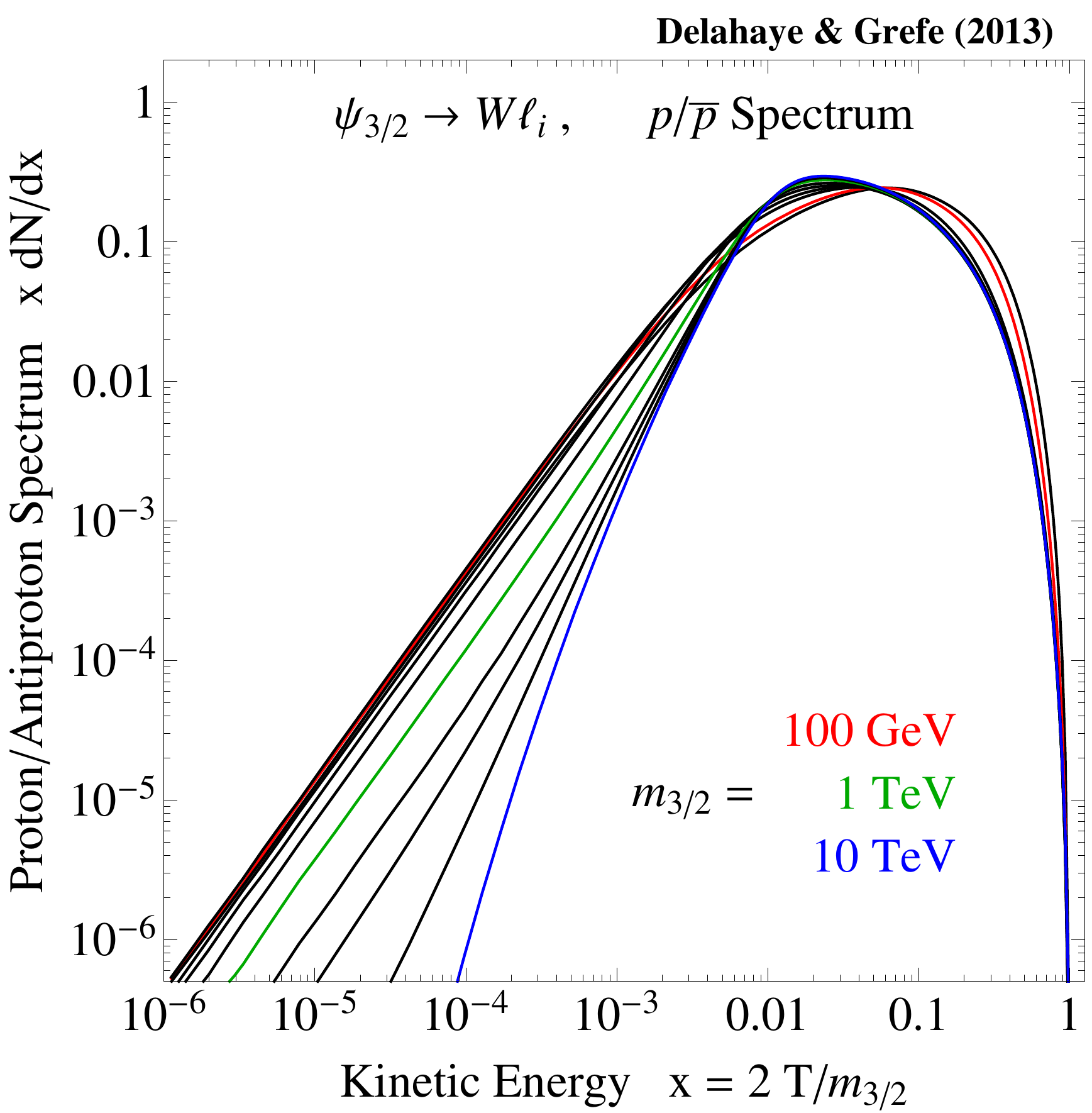} 
 \hfill
 \includegraphics[width=0.325\textwidth]{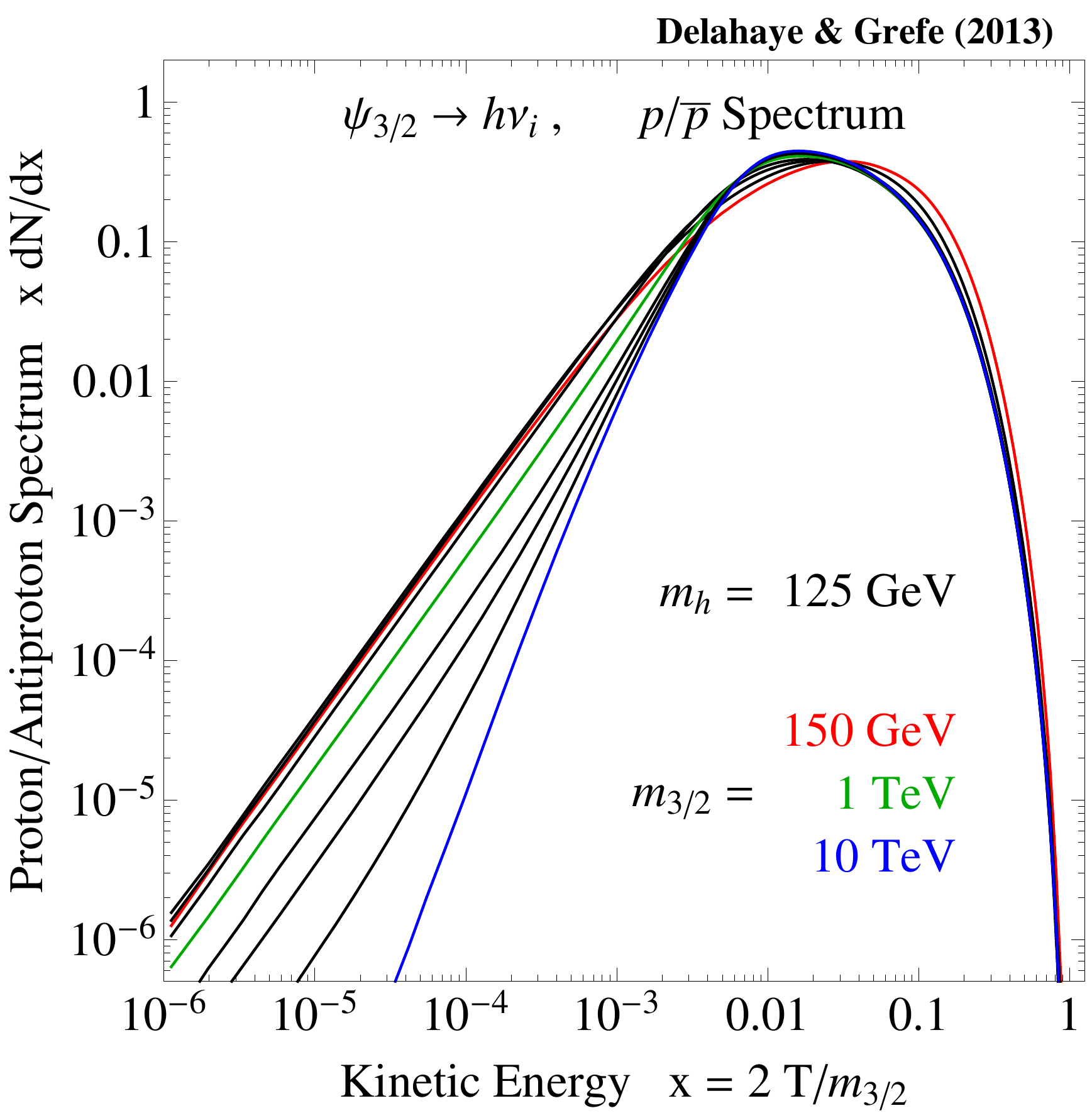} 
 \caption{Proton/antiproton spectra from the decay of a DM particle into $Z\nu_i$ (\textit{left}), $W\ell_i$ (\textit{centre}) and $h\nu_i$ (\textit{right}). The spectra are shown for DM masses of 100\,GeV (\textit{red}), 150\,GeV, 200\,GeV, 300\,GeV, 500\,GeV, 1\,TeV (\textit{green}), 2\,TeV, 3\,TeV, 5\,TeV, and 10\,TeV (\textit{blue}). For $W\ell_i$ we also show the spectrum for a DM mass of 85\,GeV. For the channel $h\nu_i$ we assumed a Standard Model-like lightest Higgs boson with a mass of 125\,GeV~\cite{:2012gk,:2012gu}. In this case the lightest DM mass considered is 150\,GeV (\textit{red}). The spectra are universal for all lepton flavours in the decay as no protons/antiprotons are produced in lepton decays and the influence of the different lepton masses is negligible. All the spectra are normalized to the respective DM mass.}
 \label{Znuspectra}
\end{figure}

For the generation of the spectra of protons/antiprotons from gravitino decay we simulated $5\times10^7$ events with the event generator \textsc{Pythia} 6.4~\cite{Sjostrand:2006za} for each of the decay channels and for a set of gravitino masses of roughly equal distance on a logarithmic scale: $m_{3/2}=100\,$GeV, 150\,GeV, 200\,GeV, 300\,GeV, 500\,GeV, 1\,TeV, 2\,TeV, 3\,TeV, 5\,TeV, and 10\,TeV. For the decay channel $W\ell$ we also considered a DM mass of 85\,GeV, while for the Higgs decay channel the smallest DM mass considered is 150\,GeV. The reason for the choice of these masses is as follows: For each decay channel, the lowest gravitino mass is chosen such that the corresponding two-body decay is kinematically allowed. The highest masses chosen for the analysis go beyond the range motivated by cosmology, but it is anyway interesting to include them for a phenomenological analysis. The resulting spectra are presented in Fig.~\ref{Znuspectra}, while Table~\ref{decaymultiplicities} summarizes the total number of protons/antiprotons produced in the respective decay channels.\footnote{The decay spectra are available in tabulated form at \url{http://www.desy.de/\~mgrefe/files.html}.}$^,$\footnote{The results for the Higgs boson + neutrino channel differ from those reported in~\cite{Grefe:2011dp}.  In a cross-check of our results with those of~\cite{Grefe:2011dp} we found that the method to simulate the $h\nu$ channel in that work (described in~\cite{Grefe:2008zz}) lead to erroneous results.} The multiplicities do not change with the DM mass as all protons and antiprotons are produced in massive boson decays that are independent of the boost factor.\footnote{This conclusion is expected to change once electroweak final state radiation is taken into account. Following the results of~\cite{Ciafaloni:2010ti} we estimate that the multiplicity of protons and antiprotons could increase by a factor of about two for DM masses in the TeV range. Most of these additional protons and antiprotons would appear, however, at energies below the peak in the spectrum, while at larger energies the spectrum is not expected to change significantly. This could have an impact on our results, but it is not expected to change the limits by orders of magnitude. In any case our limits are conservative, \textit{i.e.} with additional antiprotons they would only become stronger.} For details on the \textsc{Pythia} simulation we refer to Appendix~\ref{App:spectra}.
\begin{table}
 \centering
 \begin{tabular}{cccc}
  \toprule
  Particle type & $Z\nu$ & $W\ell$ & $h\nu$ \\
  \midrule
  $p+\bar{p}$ & 1.67 & 1.60 & 2.35 \\
  \bottomrule
 \end{tabular}
 \caption{Multiplicities of protons and antiprotons from gravitino decays after the fragmentation of the different on-shell intermediate particles as simulated with \textsc{Pythia}.}
 \label{decaymultiplicities}
\end{table}

In order to simulate the gravitino decay into the different channels we started the \textsc{Pythia} simulation with a resonance decay into two particles, $Z$ boson and neutrino, $W$ boson and charged lepton, and Higgs boson and neutrino, respectively. In this way the $Z$, $W$ and Higgs bosons were treated as decaying isotropically in their rest frames. This is only correct for spin-0 particles like the Higgs boson but not for the $Z$ and $W$ cases. However, it is a good assumption for generic decay spectra from fermionic DM as the exact decay spectra can only be obtained using the full matrix element for the decay of a particular DM candidate. From a comparison of the fermion spectra from $Z$, $W$ or $h$ decay generated in this way to the corresponding spectra obtained from the gravitino three-body decay formul\ae\ in~\cite{Grefe:2011dp} we found that the behaviour of the spectra in the very vicinity of the endpoint can differ by an $\mathcal{O}(1)$ factor. In contrast to that, we expect the soft part of the decay spectra to be practically independent of the exact treatment.

In particular, the protons and antiprotons considered in this work are generated in decays along the hadronization process of the massive bosons (including also late neutron decays). Therefore, their spectra are soft and featureless, and practically do not suffer from the uncertainty pointed out above. As mentioned before, in our current \textsc{Pythia} treatment no electroweak bremsstrahlung is included. Since all the decay channels we consider already include massive gauge or Higgs bosons, these electroweak corrections do not qualitatively alter the decay spectra as, for instance, in the case of leptophilic DM~\cite{Kachelriess:2009zy}. Therefore we do not include this effect in the current work, although, however, this contribution might dominate the low-energy part of the spectra and thus become relevant in indirect searches for multi-TeV DM~\cite{Ciafaloni:2010ti}.

We also cross-checked our spectra against other results in the literature. To compare our proton/antiproton decay spectra for the channels $Z\nu, W\ell$ and $h\nu$ with annihilation spectra in the channels $ZZ, WW$ and $hh$ we exploited the fact that far above the kinematic threshold, \textit{i.e.} for $m_\text{DM}\gg m_{Z,\,W,\,h}$, the following relation holds:
\begin{equation}
  2\left(\frac{dN_{\bar{p}}}{dT}\right)_{\text{decay}}\simeq\left(\frac{dN_{\bar{p}}}{dT}\right)_{\text{annihilation}},
\end{equation}
taking $m_\text{DM}^{\text{decay}}=2\,m_\text{DM}^{\text{annihilation}}$ to account for the different kinematics of DM annihilations and decays. We find good agreement with the results presented in~\cite{Cirelli:2010xx} (excluding electroweak bremsstrahlung).

\section{Limits from Cosmic-Ray Observations}
\label{limits}

As first suggested by~\cite{Silk:1984zy}, if Galactic DM annihilates, its annihilation products could have a large energy and contribute significantly to the Galactic cosmic-ray budget. The same reasoning holds for DM decays and among the various possible decay products the search for antiprotons is one of the most promising because the astrophysical background is low and well understood~\cite{Donato:2001ms}. However, since charged particles are subject to a complex propagation in the interstellar medium, the estimation of their flux at the Earth has to be calculated with caution (see Appendix~\ref{App:CR}).

\subsection{Propagation of Charged Particles in the Milky Way}

The current understanding of Galactic cosmic-ray propagation translates into a diffusion equation that takes into account the scattering of cosmic rays off magnetic field inhomogeneities, convection by stellar winds, interaction with the interstellar gas (elastic and inelastic scattering, production of tertiary antiprotons) and of course the source of the cosmic rays. The various parameters of this equation and of its boundary conditions are not fixed by theory but rather constrained by observational data (see for instance~\cite{Maurin:2001sj,Maurin:2002hw,Maurin:2010zp,Putze:2010zn}). This leads to some uncertainty in the expected fluxes which can be quite important, especially when dealing with DM signals. One major aspect of this work is to fully size the uncertainty due to the variation of propagation parameters within the range constrained by measurements of the boron-to-carbon ratio\footnote{Indeed, boron, being very scarce in stars, is known to be a secondary cosmic ray, \textit{i.e.} it is produced by the interaction of heavier nuclei (mainly carbon) with the interstellar gas. The boron-to-carbon ratio is independent of the prescription for the cosmic-ray sources and depends only on the propagation parameters.} and to derive the corresponding uncertainty on the limits on the DM model.

The computation of the secondary flux, namely the flux of antiprotons created by the spallation process of cosmic protons and $\alpha$ particles off interstellar hydrogen and helium, has been detailed in various papers (see for instance the appendices of~\cite{Donato:2001ms} for technical aspects). As the propagation equation for antiprotons is extremely similar to the one of boron (only the interaction cross-sections differ), the constraints from the boron-to-carbon ratio fully play their part and the resulting uncertainty is very small, comparable with the observational error bars of PAMELA~\cite{Adriani:2012paa}. This is a major achievement of cosmic-ray physics and proves that the model in use, though far from being perfect and complete, is at least self-consistent to describe various cosmic-ray populations at once.

Concerning the DM signal, the propagation equation and its treatment are just the same as in the case of astrophysical secondaries. Note that in both cases we also took into account the production of tertiary antiprotons (\textit{i.e.} inelastic production of cosmic-ray antiprotons by the interaction of higher-energy antiprotons with the interstellar gas). However, of course, the source term is very different from the case of secondaries. In the following we fix the parameters concerning the description of the DM halo profile. In the case of DM decays this choice only marginally affects the fluxes of charged cosmic rays (see for instance~\cite{Ibarra:2008qg}). For definiteness we use a Navarro--Frenk--White (NFW) profile with a scale radius of $r_s= 20$\,kpc:
\begin{equation}
   \rho_{\text{halo}}(r)=\rho_{\text{loc}}\,\frac{\left( R_{\odot}/r_s\right) \left( 1+R_{\odot}/r_s\right) ^2}{\left( r/r_s\right) \left( 1+r/r_s\right) ^2}\,.
\end{equation}
We consider a local DM density of $\rho_\text{loc}=0.4\,$GeV/cm$^3$~\cite{Nesti:2012zp} and a distance $R_\odot=8.3\,$kpc from the Sun to the centre of the Milky Way~\cite{Ghez:2008ms,Gillessen:2008qv}. Changing the local DM density would have a straightforward impact on the results as the signal scales linearly with the density. Changing the distance of the Sun to the Galactic centre $R_\odot$ needs to be done consistently with a change of the local DM density. Such a change would have little impact on the final results but is not a mere linear scaling.\footnote{Comparisons with constraints derived in other works could be influenced by the fact that there is no standard value for the distance $R_\odot$. Indeed, while preparing the Dark Matter Les Houches Agreement (DLHA)~\cite{Brooijmans:2012yi}, it has been decided not to set a recommended value for this parameter because it is still under important revision by astronomers (see for instance~\cite{Majaess:2010zu,Malkin:2012nu,deGrijs:2012dq,Malkin:2013ac}). Even though many authors claim a precision around 5\% for this quantity, the spread among different measurements is of the order of 10\%. The same also is true for the local DM density (see~\cite{deBoer:2010eh,Bovy:2012tw,Garbari:2012ff,Inoue:2013mga,Nesti:2013uwa}). It is important to note that all estimates of the local dark matter density heavily depend on $R_\odot$, so if consistently taken into account, this can affect the cosmic-ray flux estimate by a non-negligible quantity (a few 10\%).}
All the other parameters, namely the DM particle mass, its decay channel, its lifetime, and of course the propagation parameters are varied. Since the source spatial distribution is very different in the case of DM (the whole DM halo or at least the part of it that lies in the diffusion region) than in the case of astrophysical secondary cosmic rays (only the Galactic disk), the boron-to-carbon ratio constraints lead to much larger uncertainties when applied to antiprotons generated in DM decays. Depending on the energy, the effect can be larger than one order of magnitude. For this reason it is of utmost importance to consider the propagation uncertainty when deriving constraints on DM scenarios. Moreover, it is important to stress that, though they give a fair approximation of the spread of the signal, using only the MIN, MED and MAX propagation parameter sets of~\cite{Donato:2003xg} is not sufficient to size the uncertainties in all energy bins, especially at high energies. Nevertheless, since these sets of propagation parameters have been widely used in the literature, in our results we will indicate them separately in addition to the full range derived from a scan over the allowed propagation parameter sets.

Finally, after having travelled throughout the Galaxy, cosmic rays enter the Sun's magnetosphere and their flux is hence affected by the so-called solar modulation. Major theoretical work is ongoing to describe correctly this phenomenon (see for instance~\cite{Roberts:2012eq,Potgieter:2013cwj,Maccione:2012cu}). This issue being still unsettled we have adopted two methods: in the first case we use the standard Fisk potential model that we know does not rely on firm theoretical grounds but gives reasonable results for antiprotons; in the second case we dismiss all data below 10\,GeV to keep only those that are believed not to be too much affected by solar modulation. 

The strength of indirect detection constraints derived from the antiproton channel has been demonstrated many times already (see for instance~\cite{Cirelli:2013hv,Cerdeno:2011tf}). Thanks to the use of semi-analytical methods to solve the propagation equations, computing a flux requires little time on a usual machine and that allows us to perform scans of the full parameter space in a couple of hours. A distinctive feature of this work is to perform such a scan over propagation parameter sets instead of limiting ourselves to the MIN/MED/MAX cases.

\subsection{Antiproton Limits}

To determine limits on the DM lifetime we use a chi-square goodness-of-fit test. In this kind of test the chi-square statistic is defined as
\begin{equation}
  \chi^2=\sum_i^k\frac{(\mathcal{O}_i-\mathcal{E}_i)^2}{\sigma_i^2}\,,
\end{equation}
where the sum runs over the set of data bins used for the test. The $\mathcal{O}_i$ are the observed values, the $\mathcal{E}_i$ are the expected values from the background and signal models, and $\sigma_i$ are the errors of the observed data.

For each individual set of cosmic-ray propagation parameters in agreement with the observed boron-over-carbon ratio we proceed as following:

First, we perform a goodness-of-fit test for the background model against the data. For our analysis we use the most recent antiproton data from the PAMELA experiment. These data are based on 3.5 years of observation from July 2006 to December 2009 and were published in early 2013~\cite{Adriani:2012paa}. 

When we are following the method including the effects of solar modulation, we vary the Fisk potential $\phi_F$ from 200\,MV to 2\,GV in steps of 50\,MV and select the value that gives the lowest $\chi^2$ statistics. Usually we find the best fit for values in the range $\phi_F\sim 350$--700\,MV, leading to values for the $\chi^2$ statistics in the range $\chi^2\sim8$--11, thus indicating a very good fit for a $\chi^2$ distribution with $\nu=22$ degrees of freedom (the PAMELA data consist of 23 bins and the Fisk potential is our fit parameter). Therefore, we conclude that the secondary antiproton background fits well the PAMELA measurements and that there is no indication for an additional contribution from a primary antiproton source like DM.

When following the method without solar modulation and considering only the data bins above 10\,GeV we find $\chi^2\sim22$--38 for $\nu=8$. The fit is not quite good because solar modulation still has some impact at energies around 10\,GeV. However, going to even higher energies would leave us with too few data points for the analysis. Let us stress that the small number of energy bins together with the fact that without solar modulation the secondary flux does not lead to a satisfactory fit of the data, can have an effect on the derived limits on the DM lifetime. The reason to show the obtained results anyway is to be able to compare to constraints derived without any assumption on the model of solar modulation.

Then, we add the DM contribution to the fit and compute the value of the DM lifetime $\tau_{\text{best fit}}$ for which $\chi^2$ is minimal. While in the case without solar modulation a DM contribution almost never leads to an improved fit (\textit{i.e.} usually we find $\tau_{\text{best fit}}=\infty$), in the case including solar modulation in many cases we find a nonvanishing DM contribution in the best fit, in particular for heavier DM masses. In these cases a small DM contribution can lead to a slightly better fit to the data than secondaries alone, since the expected flux of secondary antiprotons lies marginally below the central values of the PAMELA data in the highest energy bins. However, since for the case including solar modulation the $\chi^2$ is very good already for the secondaries alone, we conclude that this improvement has no statistical significance.

We then proceed to calculate limits on the DM lifetime at 95\,\% CL by requiring that the limiting $\chi^2(\tau_{95\,\%\text{ CL}})$ deviates from the minimal or best-fit $\chi^2(\tau_{\text{best fit}})$ by an amount $\Delta\chi^2$ corresponding to a $2\sigma$ exclusion:
\begin{equation}
  \chi^2(\tau_{95\,\%\text{ CL}})=\chi^2(\tau_{\text{best fit}})+\Delta\chi^2\,.
\end{equation}
For the case without solar modulation we have one free parameter in the fit, namely the DM lifetime. Therefore, we use a value of $\Delta\chi^2=4$ to derive the limits. For the analysis including the effects of solar modulation we have the same limiting value. Since the value of the Fisk potential has already been determined along with $\tau_{\text{best fit}}$, again the only fit parameter is the DM lifetime. We hence finally obtain a value for $\tau_{95\,\%\text{ CL}}$ and proceed to the next parameter-space point.

\begin{figure}[t!]
\centering  
  \includegraphics[width=0.49\linewidth]{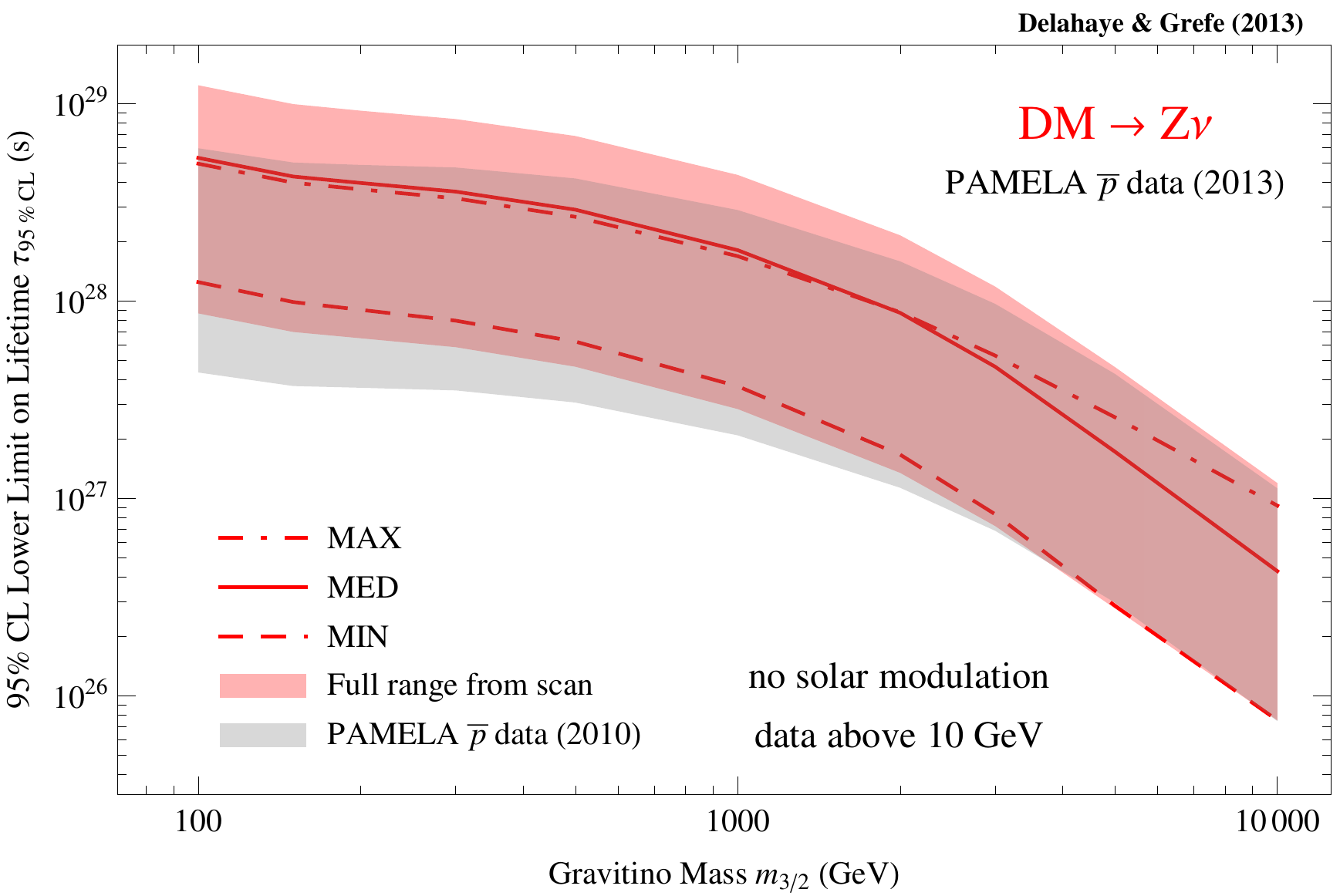}
  \includegraphics[width=0.49\linewidth]{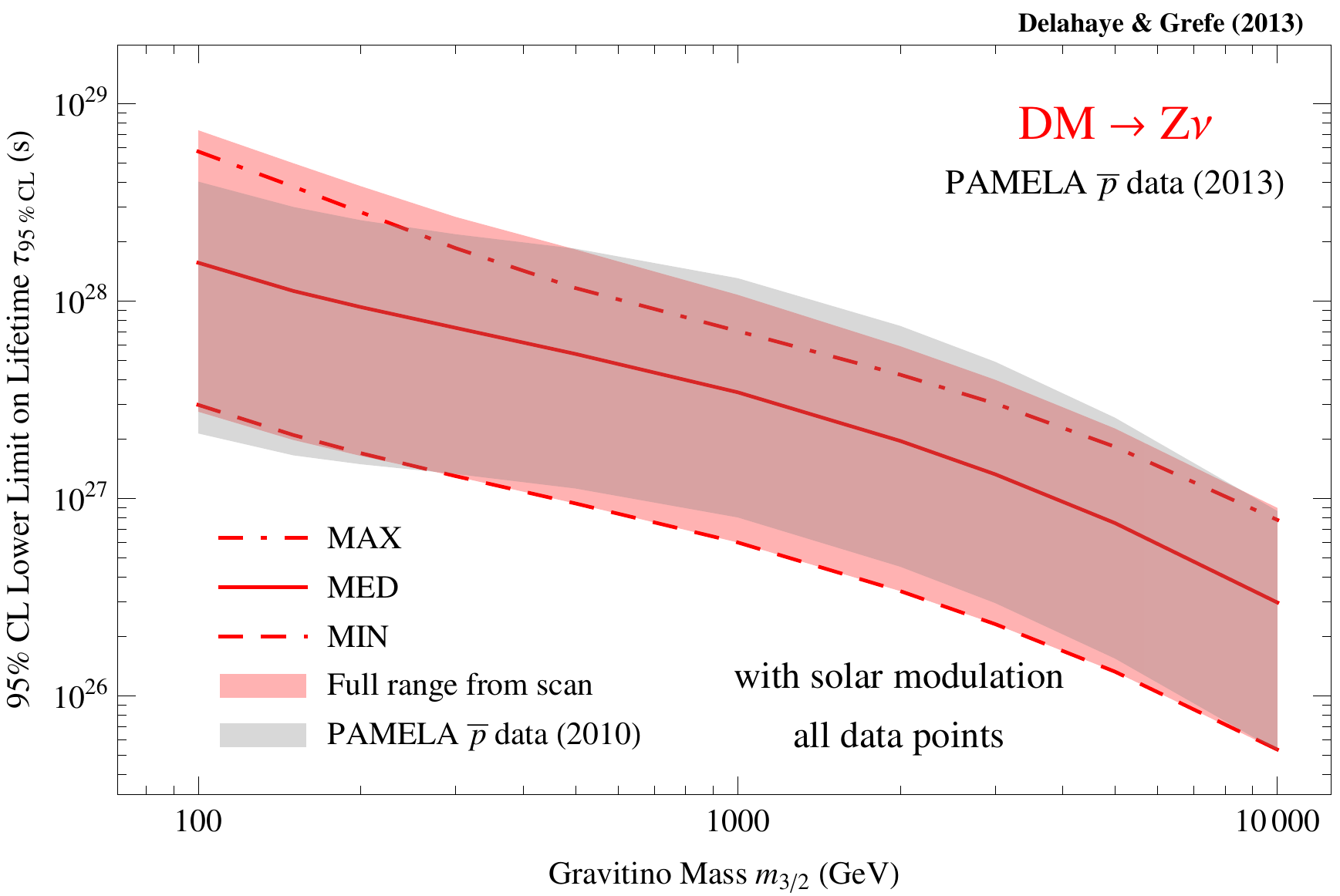}\\
  \includegraphics[width=0.49\linewidth]{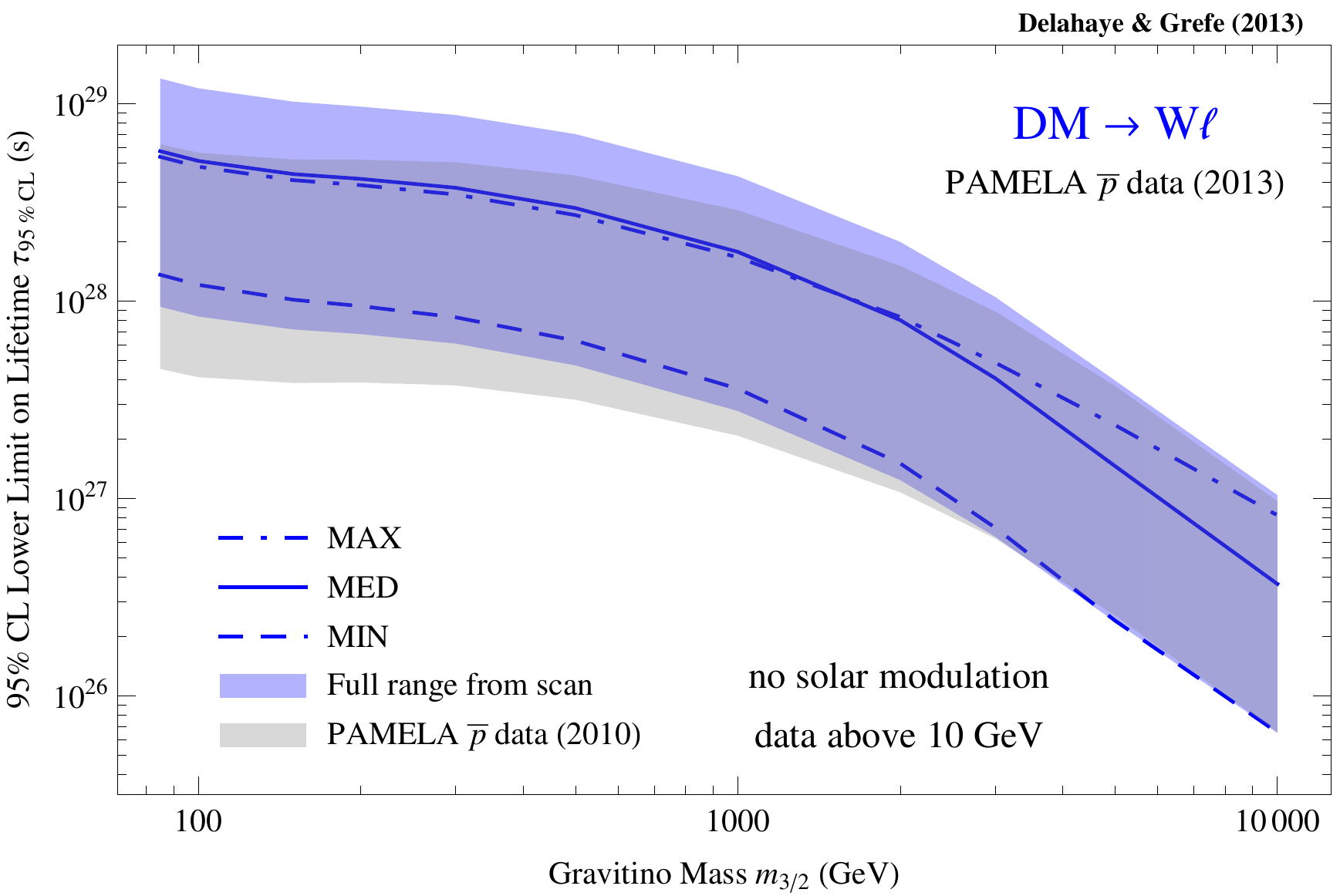}
  \includegraphics[width=0.49\linewidth]{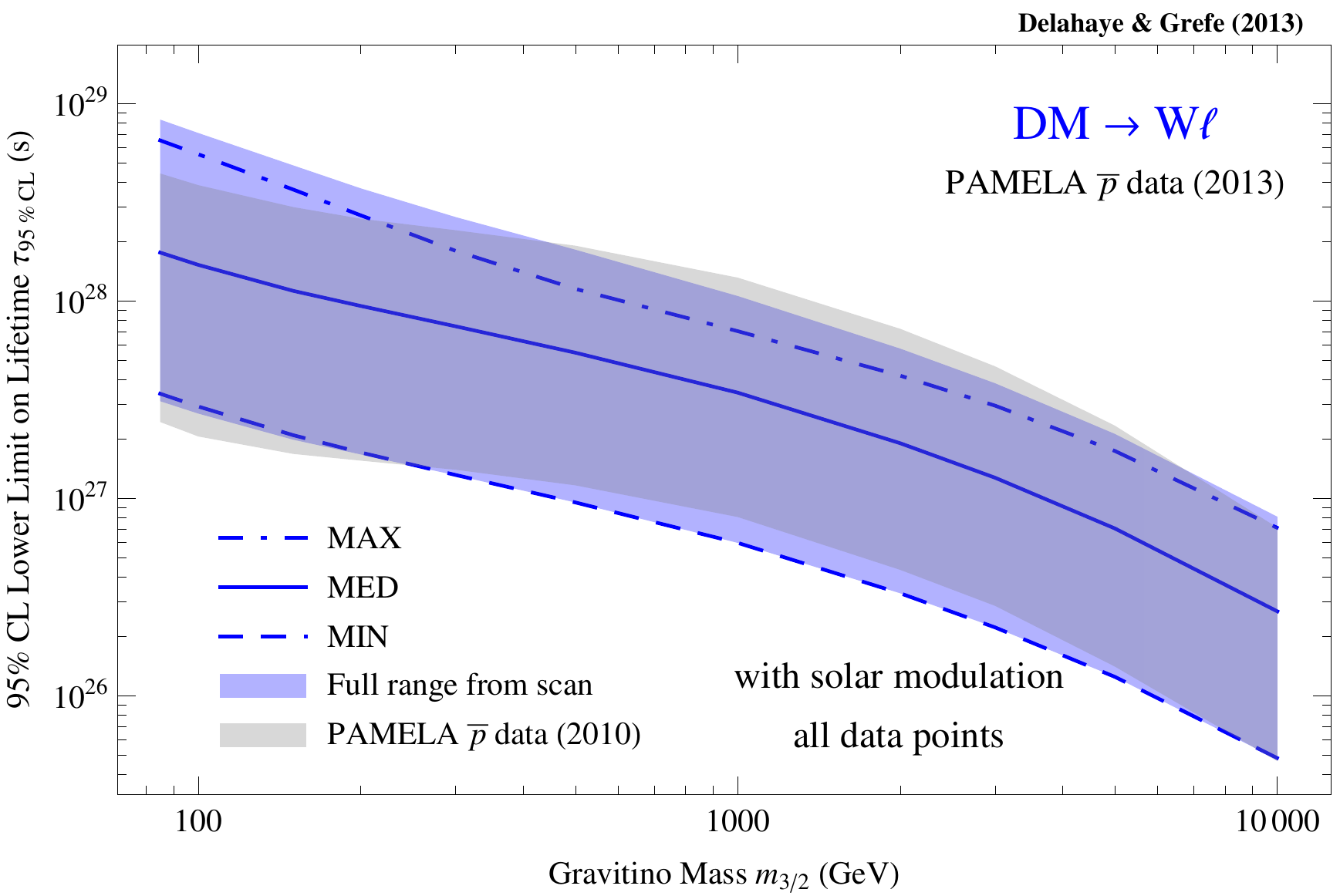}\\
  \includegraphics[width=0.49\linewidth]{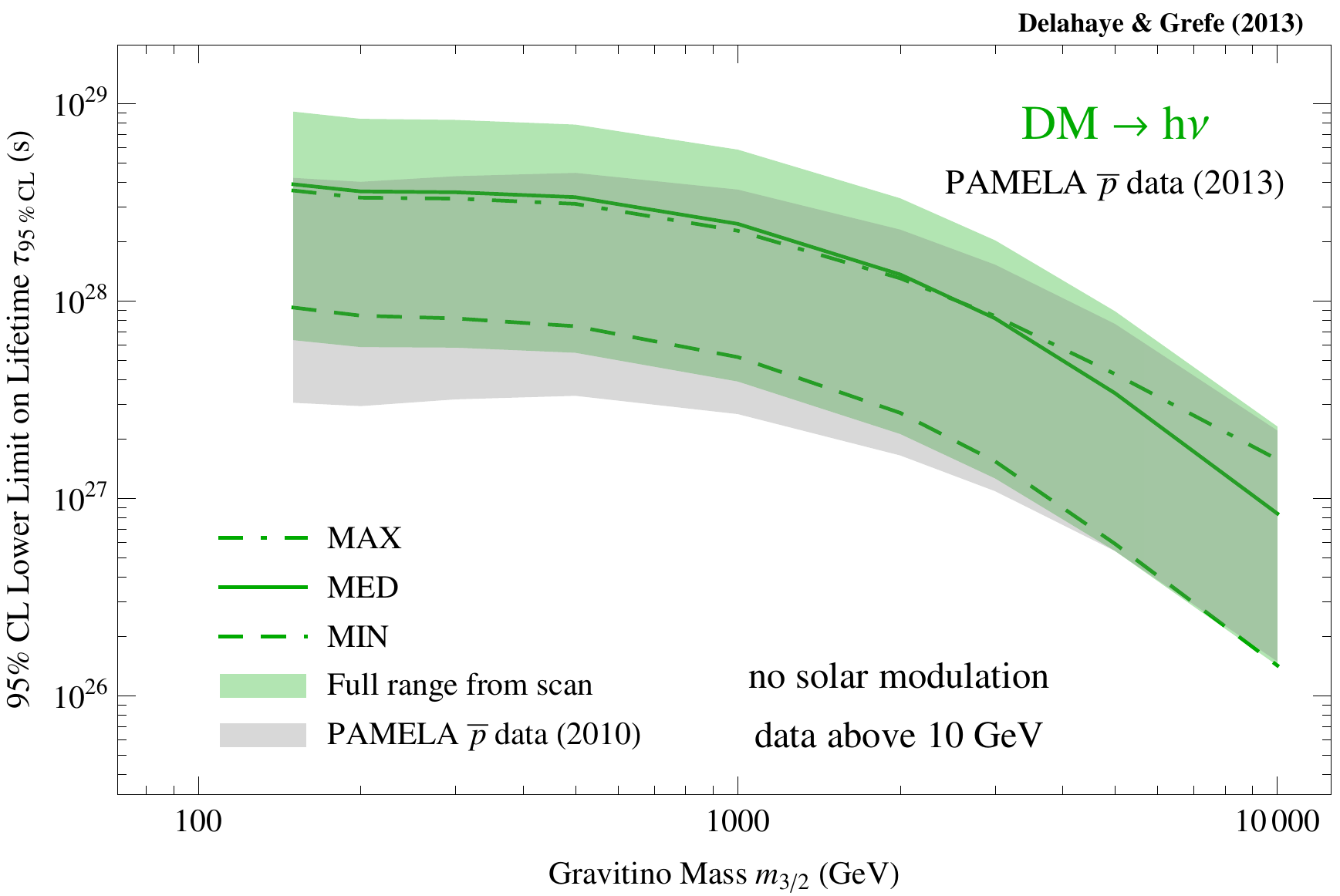}
  \includegraphics[width=0.49\linewidth]{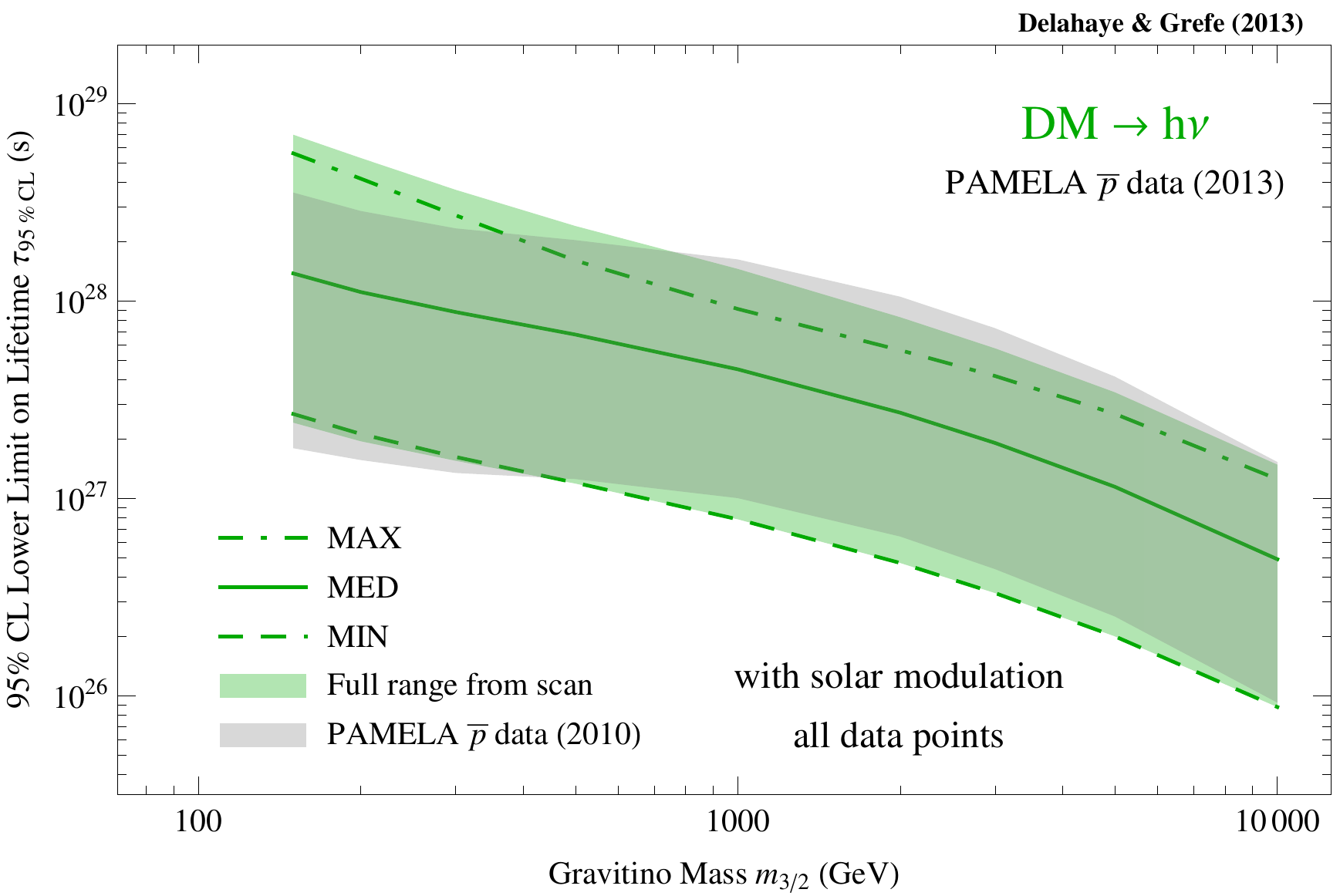}
  \caption{Constraints on the DM lifetime in individual two-body decay channels derived from antiproton observations by the PAMELA experiment~\cite{Adriani:2012paa}. From top to bottom the panels show the constraints for the decay channels $Z\nu$, $W\ell$ and $h\nu$. The left column presents the results without taking into account the effects of solar modulation while the right column includes these effects. In each case DM lifetimes below the dashed/solid/dash-dotted line are excluded at 95\,\% CL according to the method described in the text for MIN/MED/MAX propagation parameters. In addition, the coloured band shows the whole range of propagation uncertainties derived from a scan over allowed propagation parameters as described in the text. For comparison, we also present the range of limits derived from a previous PAMELA data release~\cite{Adriani:2010rc}.}
  \label{fig:channellifetime}
\end{figure}
We follow this procedure individually for each of the three gravitino decay channels that produce antiprotons in the final state (\textit{i.e.} $Z\nu$, $W\ell$ and $h\nu$). Fig.~\ref{fig:channellifetime} shows the lower limits on the DM lifetime obtained for each channel as a function of the DM mass. The plots in the left column present the limits obtained neglecting the effects of solar modulation and only using data points above 10\,GeV, while the plots in the right column present the limits obtained including the effects of solar modulation and using all data points. In each case, the coloured band corresponds to the full range of uncertainty on the lifetime limits as obtained from the scan over the propagation parameters, while the coloured lines indicate the cases for the MIN/MED/MAX propagation sets. The limits including the effects of solar modulation are also summarized in Table~\ref{tab:limits} at the end of the paper. In order to study how the limits change with increased statistics and improved analysis techniques, we show for comparison as a grey band the range of limits derived from 2.5 years of data collected with the PAMELA experiment from July 2006 to December 2008 and published in 2010~\cite{Adriani:2010rc}.

\begin{figure}[t]
  \includegraphics[width=0.49\linewidth]{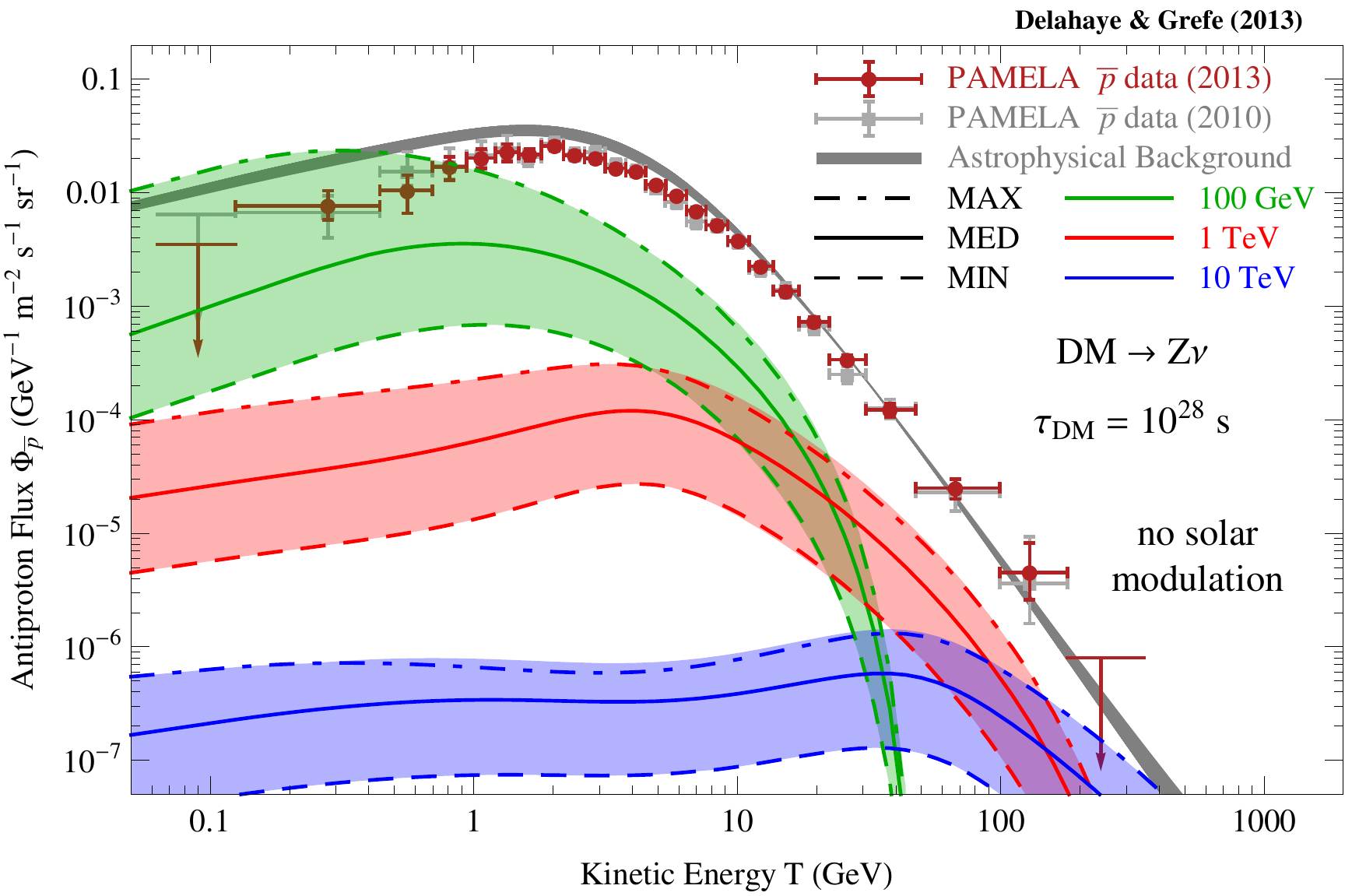}
  \hfill
  \includegraphics[width=0.49\linewidth]{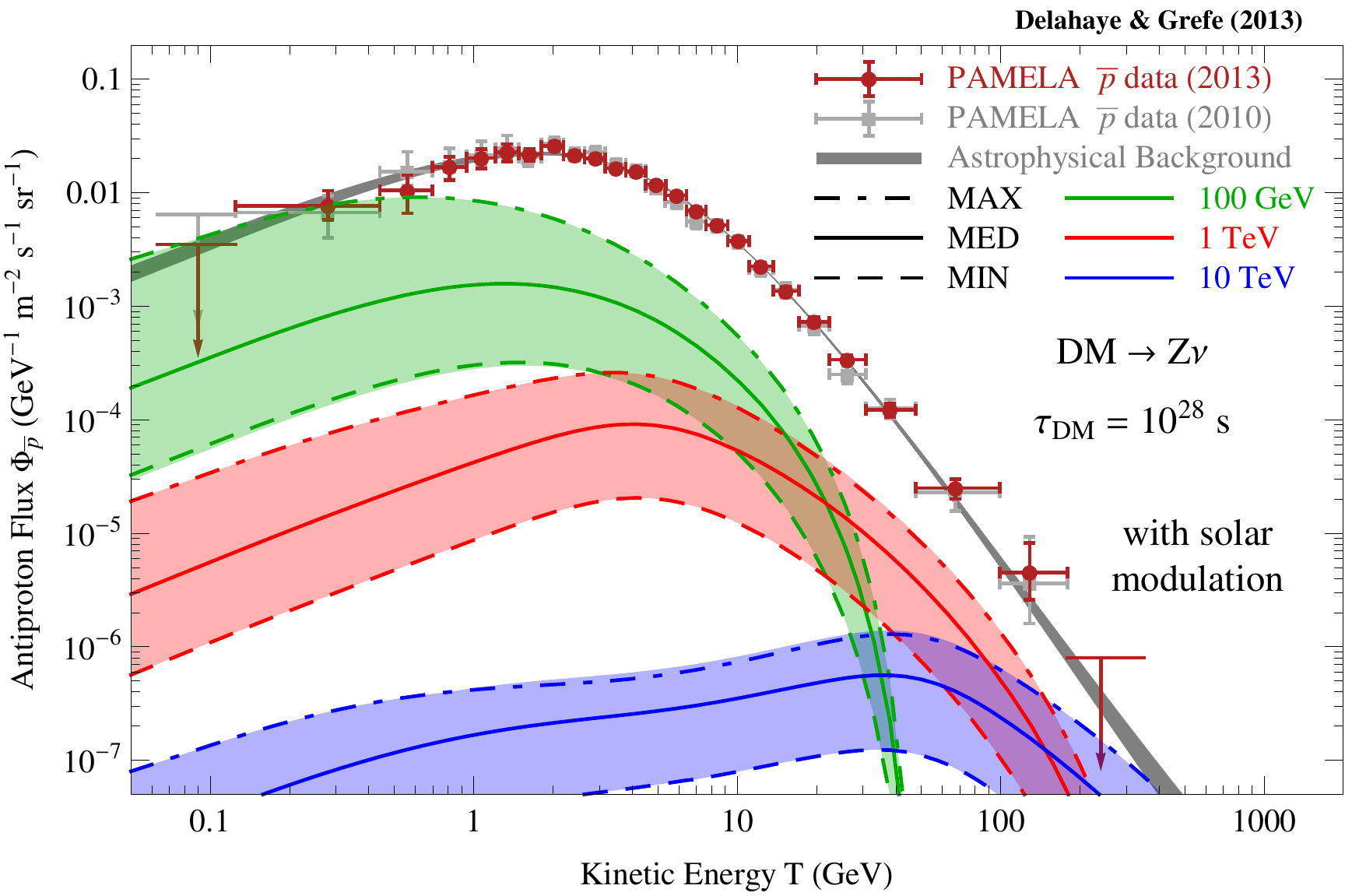}
  \caption{Antiproton flux expected from the DM decay into $Z\nu$ compared to the expected flux of antiprotons from secondary production and data from PAMELA~\cite{Adriani:2012paa,Adriani:2010rc}. The DM signal is calculated for a lifetime of $10^{28}$\,s and for three different masses: 100\,GeV, 1\,TeV and 10\,TeV. It is shown for MIN/MED/MAX propagation parameters and the coloured bands also show the allowed range of spectra from the scan over propagation parameters. The grey band shows the propagation uncertainty for secondary antiprotons. \textit{Left:} No correction for solar modulation. \textit{Right:} Correcting for solar modulation.}
  \label{fig:Znuantiproton}
\end{figure}

Comparing the constraints on the DM lifetime with and without solar modulation, we observe that those neglecting solar modulation and using less data points lead to somewhat stronger limits. This result is counter-intuitive as one would expect that taking into account the low-energy bins with very small error bars would lead to stronger limits, at least for lower DM masses. Our conclusion is that since the background flux of astrophysical antiprotons does not give a satisfactory fit to the data without taking into account solar modulation (in fact, as can be seen from Fig.~\ref{fig:Znuantiproton}, at energies around 10\,GeV the background model alone overshoots the data in that case), these limits are not completely reliable. Only for the largest DM masses, where the DM flux does not significantly contribute to the energy bins affected by solar modulation, both methods give almost identical results. Finally, let us stress that the limits without solar modulation do not suffer from hypotheses made on the modelling of solar modulation, and we believe it is important to also display them for the sake of completeness. 

From Fig.~\ref{fig:channellifetime} one can also see why it is important to perform a scan over all propagation parameter sets that are compatible with the B/C constraints rather than just using MIN/MED/MAX parameters. The differences between the lowest or highest lifetime limits derived from one technique and the other can amount up to 60\,\% in the case including solar modulation. For the case without solar modulation the difference can even be up to 150\,\%. This clearly shows the importance of using CPU-friendly methods, like the semi-analytical cosmic-ray propagation code used in this work, that allow for scans over large parameter spaces. This will become even more true once AMS-02 publishes new high-precision data for a large set of cosmic-ray species.

When we compare the lifetime constraints derived from the most recent PAMELA data with those derived from the 2010 data release, we find that in the case including the effects of solar modulation there is no overall improvement of the constraints. For DM masses around 100\,GeV the limits became slightly stronger, while they became slightly weaker for DM masses around 1\,TeV and practically did not change for a mass of 10\,TeV. The main differences between the 2013 and 2010 data sets are the improved statistics and therefore reduced error bars, and the presence of less outlier data points with respect to the expected spectral shape of secondary antiprotons. In fact, the $\chi^2$ statistics for the background of secondaries improved from $\chi^2\sim15$--17 to $\chi^2\sim8$--11 (for 22 degrees of freedom in both cases). At low DM masses the limits became stronger mainly due to the reduced error bars, but also due to a reduced central value of the 0.56\,GeV data bin. At intermediate masses, the 2010 data lead to stronger limits mainly driven by two outliers with low central values at 7.0\,GeV and 26.2\,GeV (see also Fig.~\ref{fig:Znuantiproton}).

In the case without solar modulation the limits became stronger by up to a factor of 2 for low DM masses. The main reason for this effect are the reduced error bars. For both data sets, the calculated secondary flux overshoots the data points around 10\,GeV. Due to the smaller error bars of the new data, however, the $\chi^2$ statistics for the background of secondaries worsened from $\chi^2\sim17$--25 to $\chi^2\sim22$--38 (for 8 degrees of freedom in both cases). Due to this effect the constraints strengthened in particular for lower DM masses, where the DM signal strongly contributes to the energy bins affected by solar modulation. However, since the $\chi^2$ statistics of the secondaries is unacceptably bad, we conclude that this improvement of the limits is not reliable.

Comparing our lifetime limits to similar results in the literature, we find a few differences. In~\cite{Garny:2012vt} lifetime limits at 95\,\% CL on the same set of decay channels were derived from PAMELA $\bar{p}/p$ data~\cite{Adriani:2010rc} for MIN/MED/MAX propagation sets. Their limits are similar to our results but exhibit a slightly different dependence on the DM mass. For the largest masses our limits are roughly a factor of 2 stronger. Possible sources for the differences are most probably the use of $\bar{p}/p$ data instead of $\bar{p}$ data, differences in the $\chi^2$ method employed to derive the limits, maybe differences in the model for secondary antiprotons, or the use of a fixed value of 550\,MV for the Fisk potential instead of determining the value from the best fit to the data. 

As an example of the propagated antiproton fluxes, we show in Fig.~\ref{fig:Znuantiproton} the expected fluxes in the DM decay channel $Z\nu$ for a lifetime of $10^{28}\,$s and three different DM masses: 100\,GeV, 1\,TeV and 10\,TeV. We compare them to the expected antiproton flux from secondary production (labelled as astrophysical background) and data of the PAMELA experiment~\cite{Adriani:2012paa,Adriani:2010rc}. For the calculated fluxes we present both cases, without and with correcting for solar modulation. It is well visible that the propagation uncertainties on the flux of secondary antiprotons is much lower than the uncertainty on the antiproton flux from DM decay which amounts to roughly one order of magnitude at large energies and even more than that at lower energies.

Clearly it is expected that higher statistics antiproton data from PAMELA could improve the limits on the DM lifetime found in this work. And of course precision data from the AMS-02 experiment are highly anticipated and could significantly improve the current limits or even exhibit a DM signal~\cite{Cirelli:2013hv}. The possible improvement of the limits presented in this work will heavily depend on how high in energy the measurement will go. As the comparison between the PAMELA 2010 and 2013 data shows, there is not much improvement to expect in only reducing the error bars if the energy resolution and range are not improved.

Yet, already the very high quality of the most recent PAMELA data, and especially the small size of their error bars points out the limitations of the indirect detection method. Indeed, this work being dedicated to the study of a DM signal, we did not go into the details of the theoretical uncertainties affecting the estimation of the secondary background; however, it is important to stress that the background suffers uncertainties from three different sources: The first one is due to the lack of knowledge of the correct propagation parameters, but it has been sized here making use of the boron-to-carbon ratio constraints. The second one comes from uncertainties in the knowledge of the flux of primary cosmic rays (mainly protons and $\alpha$-particles). Any variation in these fluxes translates almost linearly to the antiproton flux estimate. Though these are the most abundant cosmic-ray species, recent experimental data are not always in agreement with each other. Some spectral features have been observed by the PAMELA~\cite{Adriani:2011cu} and CREAM~\cite{Yoon:2011aa} experiments and various explanations for them have been proposed (see for instance~\cite{Bernard:2012pia,Tomassetti:2012ga,Blasi:2011fi}). However, these features do not appear in the very recent and so far unpublished AMS-02 data presented at the ICRC 2013 conference~\cite{AMS:2013}. The third source of uncertainty is our lack of knowledge of the antiproton production cross-section in cosmic-ray spallation processes in the interstellar medium. These uncertainties affect the dominant proton--proton process as well as the scaling that is used for taking into account processes involving heavier cosmic rays and heavier element components of the interstellar medium (\textit{i.e.} processes like $p+\text{He}$, $\alpha+\text{H}$ and $\alpha+\text{He}$). According to~\cite{Salati:2010rc}, these uncertainties can amount up to 25\,\%. The combined uncertainty from these three sources is larger than the current experimental errors and thus limits our results.

A systematic study of data on all possible secondary-to-primary ratios (including isotopes) as well as on proton and $\alpha$-fluxes, measured by the same experiment to avoid diverging systematics or variations in the effect of solar modulation, will probably allow to reduce the first two sources of uncertainty. In fact, the upcoming AMS-02 data on various cosmic-ray species are expected to allow for such an analysis. As shown by~\cite{Putze:2010zn}, a sizeable reduction of the uncertainties can only be achieved if the data extend to a few TeV per nucleon. Indeed, the primary antiproton flux depends almost linearly on the parameter $L$, the size of the diffusion zone. Reducing the uncertainty on $L$ would hence almost linearly translate into our antiproton flux estimates from DM decay. AMS-02 data on the ratio B/C will constrain $\delta$ and the ratio $K_0/L$ but not the latter two parameters independently. Other secondary-to-primary ratios like $^{10}$Be/$^{9}$Be, $^{26\!}$Al/$^{27\!}$Al, or $^{36}$Cl/Cl in principle can be used as well but the data are not very constraining yet. From this point of view AMS-02 carries major hopes for future improvements on the understanding of cosmic-ray propagation~\cite{Pato:2010ih}. These species could give interesting results if the data are good enough to break the degeneracy between $K_0$ and $L$ but this possibility depends on the actual value of $L$ and is impossible to guess; if $L$ is too small, it will be very hard to achieve considerable improvement.

The study of secondary species alone will not be enough in the future. Indeed, as stressed by~\cite{Donato:2010vm}, the effect of a possible change of slope in the spectra of cosmic-ray primaries like proton and helium can affect the antiproton flux above 100\,GeV where no data are available yet but might be available soon.

Little progress is to be expected in the nuclear physics part though, as there is no experimental effort going on right now in this direction. Estimating the impact due to uncertainties in the cross-sections is very difficult as these quantities affect the results in various ways: by affecting boron production, by affecting the propagation parameters, by affecting antiproton production and also pair-annihilation or energy losses during propagation. The nuclear physics community is not very much aware of this issue and currently no major efforts are undertaken to improve these results.\footnote{Recently David Maurin organized a meeting about this issue, but the discussion with experimentalists is still going on and it is not clear if there will be significant progress in this direction any time soon: \url{https://indico.in2p3.fr/conferenceDisplay.py?confId=7012.}}

Without a deep study for the reduction of these uncertainties, no significant progress on DM indirect detection should be expected as the comparison between the limits obtained from the 2010 and 2013 PAMELA antiproton data sets clearly shows. The limits given in this work should hence not be considered as being absolutely strict and variations are to be expected with better statistics, including the possibility of getting weaker. However, improvement can be expected if the energy range of future data extends to higher energies and if the energy resolution gets better.

\section{Constraints on the Gravitino Lifetime and RPV}
\label{lifetime}

In this section we apply the constraints found in the previous section to the particular case of gravitino DM. In addition, we convert these limits on the gravitino lifetime into limits on the strength of $R$-parity violation.

\subsection{Gravitino Lifetime}

The antiproton constraints on the gravitino lifetime are derived in a way analogous to the constraints for the individual decay channels. The local flux of antiprotons from gravitino decay is given as a linear combination of the fluxes in the individual channels according to the branching ratios discussed in Section~\ref{gravitinoRPV}:
\begin{equation}
 \Phi_{\bar{p}}=\BR(Z\nu)\,\Phi_{\bar{p}}^{Z\nu}+\BR(W\ell)\,\Phi_{\bar{p}}^{W\ell}+\BR(h\nu)\,\Phi_{\bar{p}}^{h\nu}.
\end{equation}
The two-body decay into photon and neutrino does not give any contribution to the antiproton flux. Therefore, for gravitino masses below the $W$ mass there are no antiprotons from gravitino decay as long as three-body decays and corrections from electroweak bremsstrahlung are neglected.

\begin{figure}[t!]
\centering
  \includegraphics[width=0.49\linewidth]{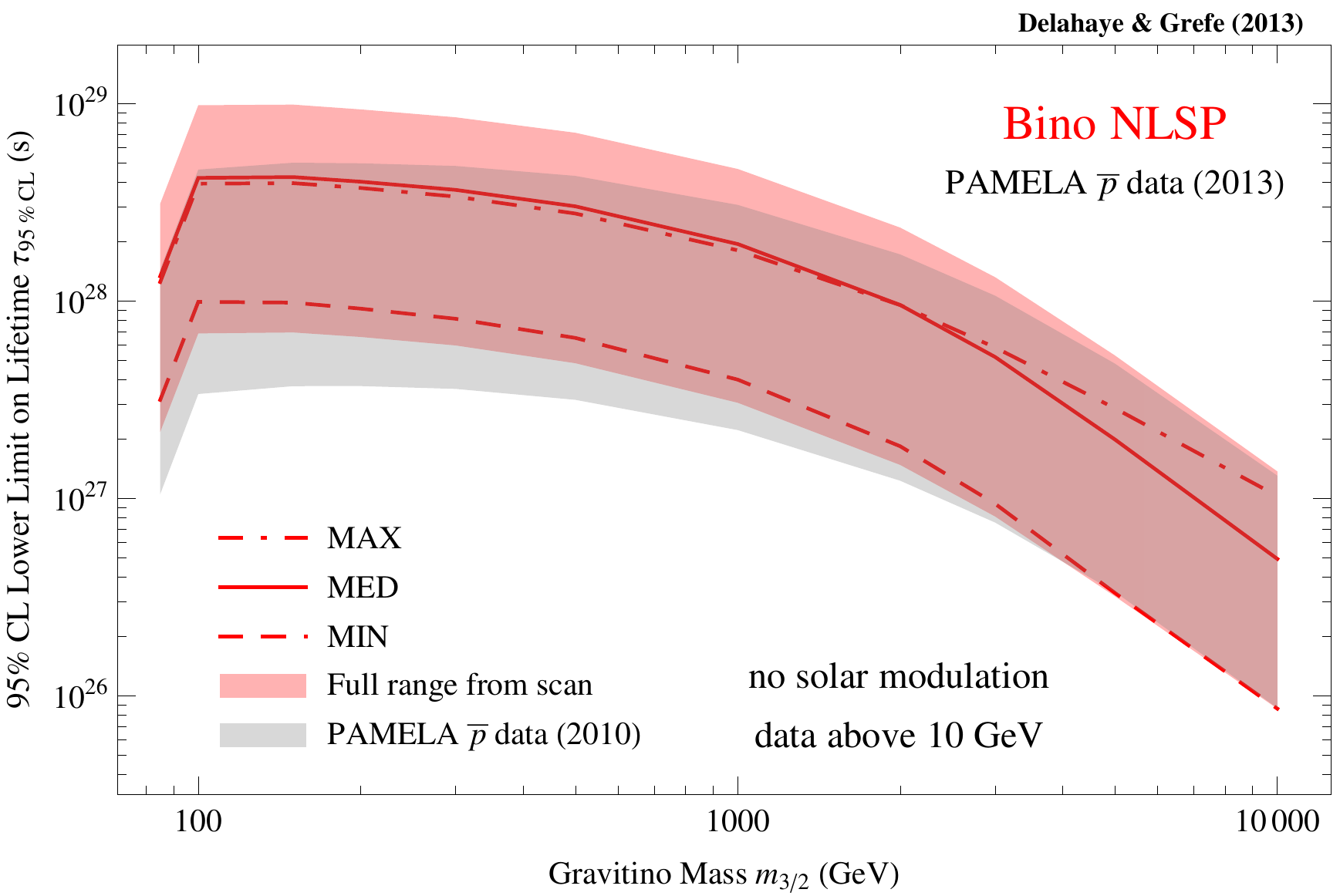}
  \includegraphics[width=0.49\linewidth]{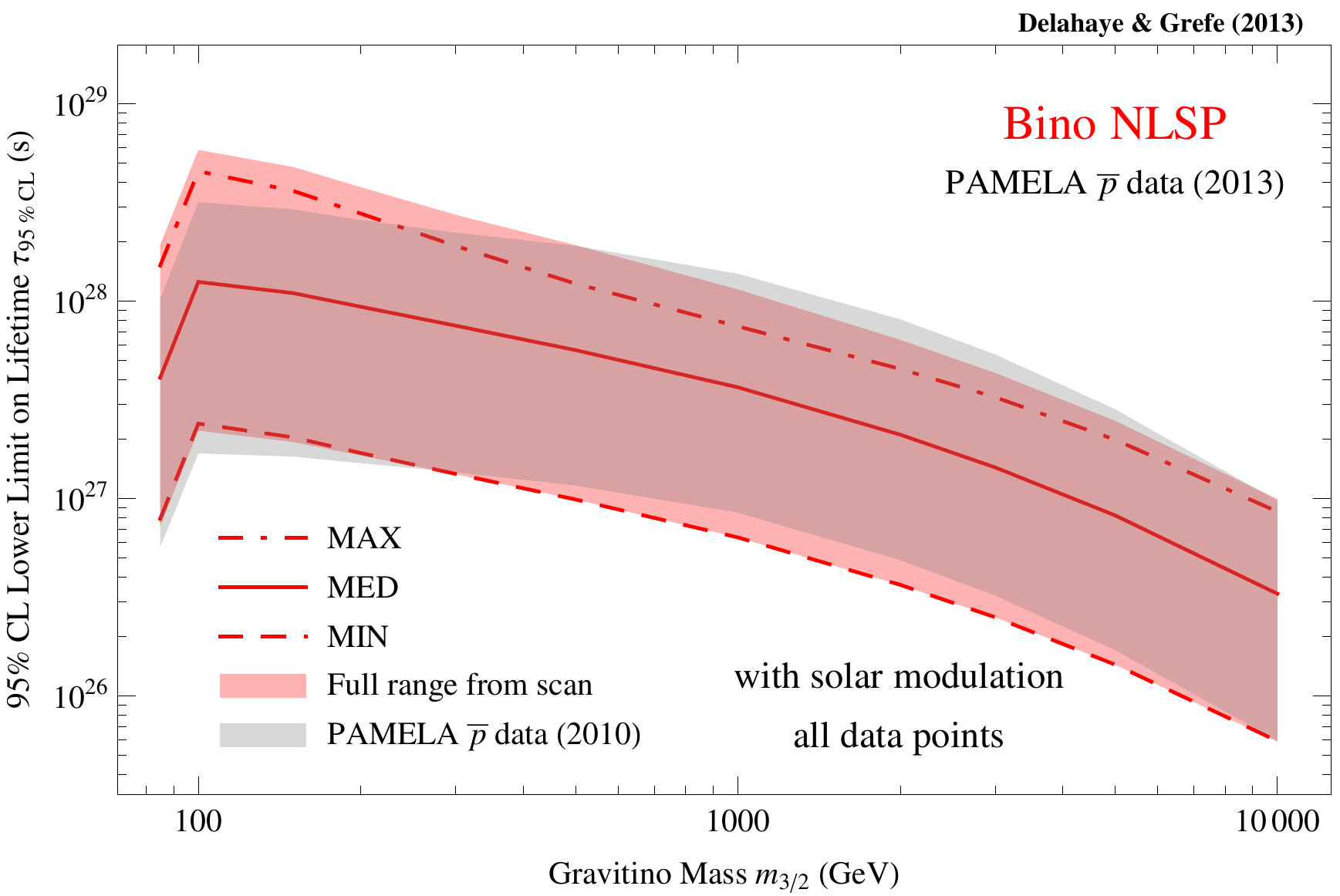}\\
  \includegraphics[width=0.49\linewidth]{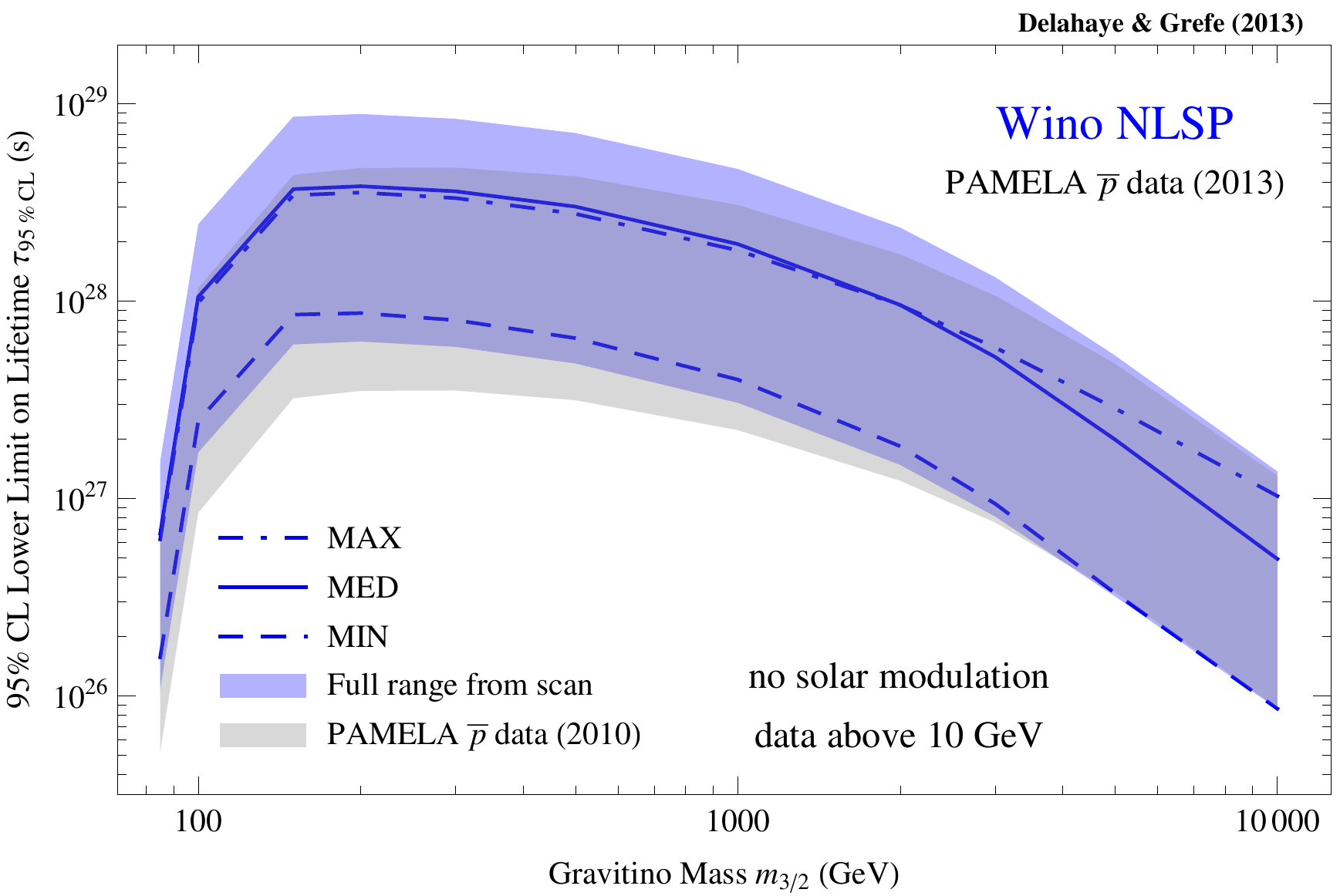}
  \includegraphics[width=0.49\linewidth]{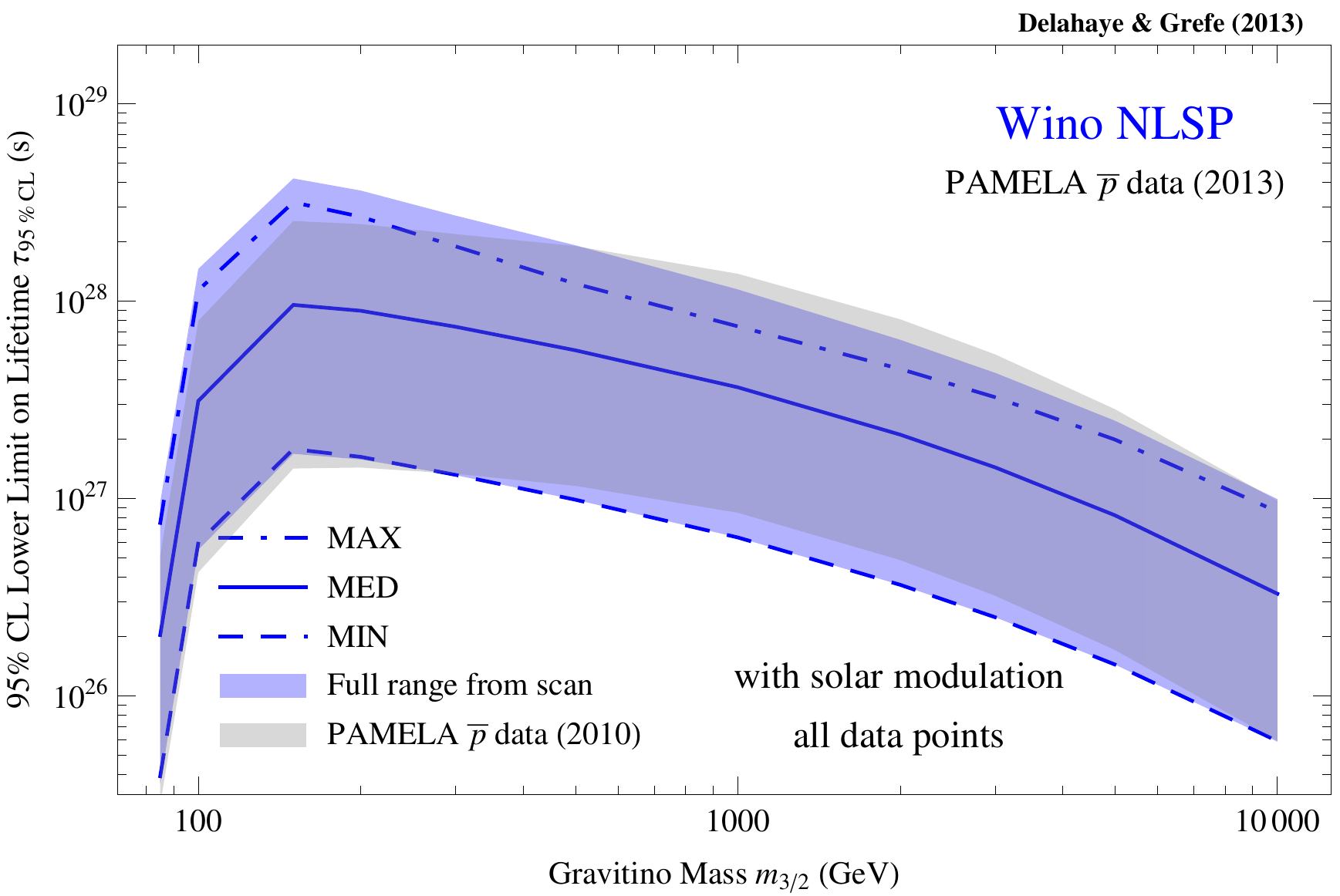}\\
  \includegraphics[width=0.49\linewidth]{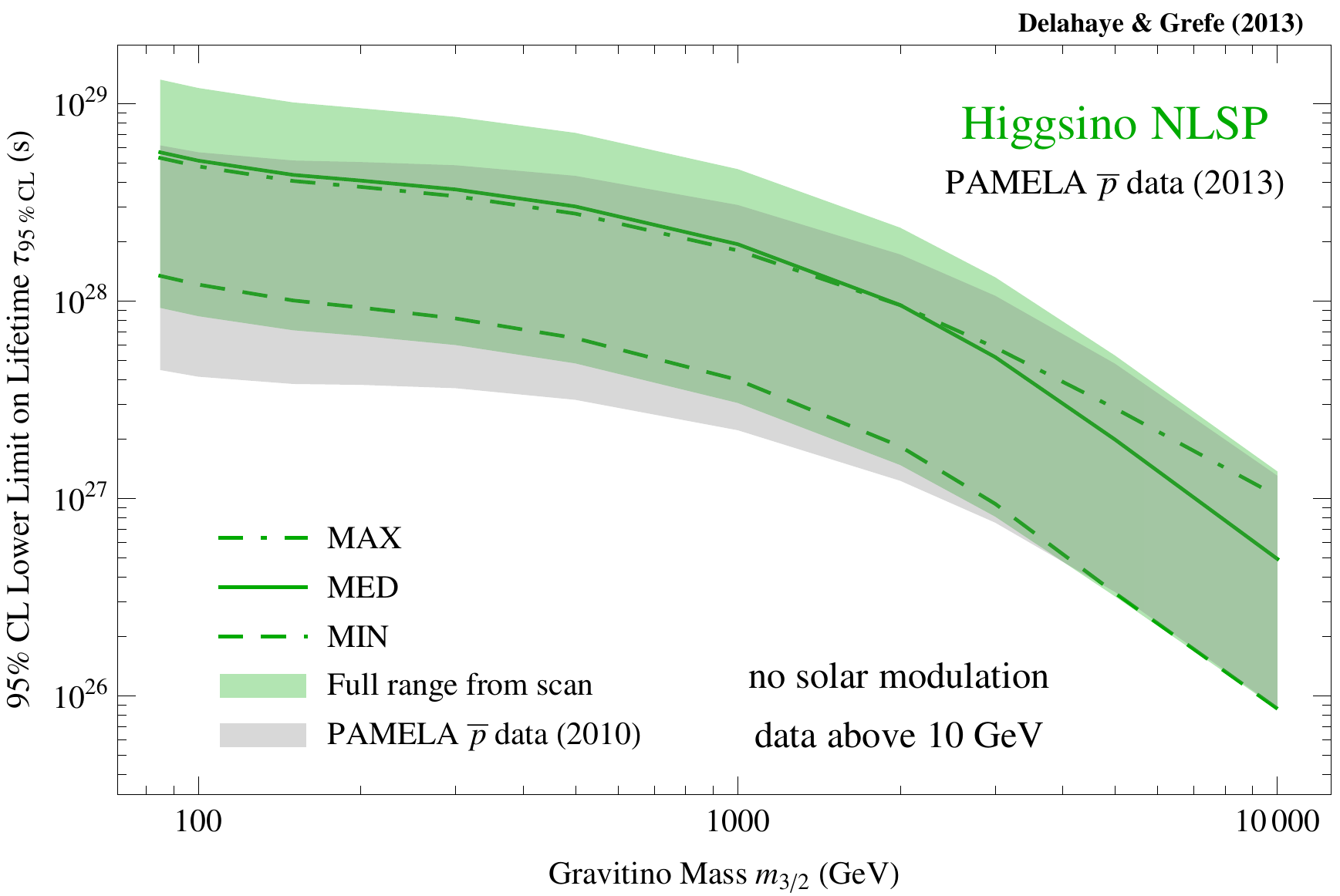}
  \includegraphics[width=0.49\linewidth]{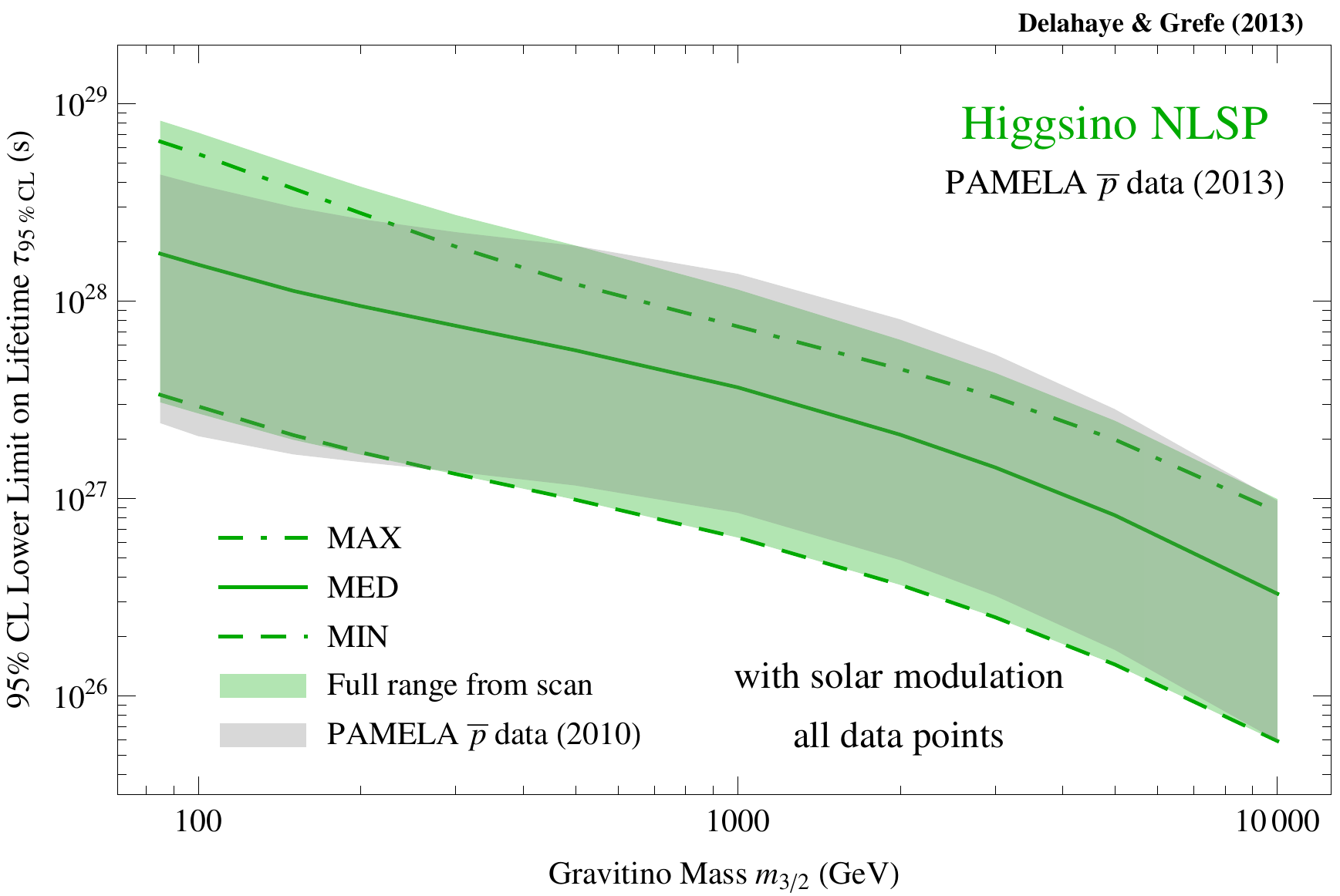}
  \caption{Constraints on the gravitino lifetime derived from antiproton observations by the PAMELA experiment~\cite{Adriani:2012paa}. From top to bottom the panels show the constraints for the cases of Bino NLSP, Wino NLSP and Higgsino NLSP. The left column presents the results without taking into account the effects of solar modulation while the right column includes these effects. In each case gravitino lifetimes below the dashed/solid/dash-dotted line are excluded at 95\,\% CL for MIN/MED/MAX propagation parameters. In addition, the coloured band shows the whole range of propagation uncertainties derived from a scan over allowed propagation parameters. For comparison, we also present the range of limits derived from a previous PAMELA data release~\cite{Adriani:2010rc}. For the cases of Bino and Wino NLSP the limits are reduced for low gravitino masses since in these cases the gravitino has a large branching ratio for the decay into $\gamma\nu$ in the mass range around 100\,GeV.}
  \label{fig:gravitinolifetime}
\end{figure}
Following the same procedure as in the previous section, we calculate 95\,\% CL lower limits on the gravitino lifetime for the three sets of supersymmetry parameters considered in Section~\ref{gravitinoRPV}. The results are presented in Fig.~\ref{fig:gravitinolifetime} and the limits including the effects of solar modulation are also summarized in Table~\ref{tab:gravitinolimits} at the end of the paper. As expected from the discussion of the gravitino branching ratios, it turns out that the lifetime limits for the different supersymmetry parameter sets are practically equivalent for gravitino masses above a few hundred GeV. For masses close to the electroweak scale, however, the difference is quite significant. This is mainly caused by the large differences in the branching ratio for the photon + neutrino channel. Thus, for instance, a gravitino around 100\,GeV produces much less antiprotons in the case of Wino NLSP than in the case of Higgsino NLSP, leading to weaker constraints in the first case.

One can see from Fig.~\ref{fig:gravitinolifetime} that as in the case of the individual decay channels discussed in the previous section, not taking into account the low-energy bins in order not to make any assumptions about solar modulation slightly changes the obtained constraints. Likewise, also the effect of scanning over the propagation parameter space compatible with the B/C data instead of solely using the MIN/MED/MAX parameter sets is practically the same as for the individual decay channels. The same holds for the differences in the limits between the 2010 and 2013 releases of the PAMELA antiproton data.

Lower limits on the gravitino lifetime derived from PAMELA antiproton data and fully taking into account the background of astrophysical secondary antiprotons as well as all propagation uncertainties have not been discussed in the literature so far. In~\cite{Buchmuller:2009xv} pre-PAMELA antiproton data were used to estimate the maximal gamma-ray flux from gravitino decays compatible with antiproton constraints. In~\cite{Grefe:2011dp} the PAMELA antiproton data were used to estimate constraints on the gravitino lifetime using MED propagation parameters and neglecting the contribution of the astrophysical background. Finally, in~\cite{Buchmuller:2012rc} the antiproton-to-proton ratio measured by PAMELA was used to constrain the lifetime of a 260\,GeV gravitino studied in the context of the Fermi 130\,GeV line. In this case the propagation uncertainties for the gravitino signal were taken into account using the MIN/MED/MAX parameter sets, but the propagation uncertainty on the astrophysical background was neglected. Therefore, our results clearly present an important improvement over the existing studies for the case of gravitino DM in models with bilinear $R$-parity violation.

Recently, decaying gravitino DM was studied in the context of the new AMS-02 result on the rising positron fraction~\cite{Aguilar:2013qda}. It was found that a gravitino heavier that 500\,GeV with a lifetime around $10^{26}\,$s could give a reasonable fit to the data~\cite{Ibe:2013nka}. From our results in Fig.~\ref{fig:gravitinolifetime}, however, it is clear that this possibility is in strong tension with antiproton observations, even for the most optimistic choices of cosmic-ray propagation parameters. We thus conclude that the rise in the positron fraction cannot be explained by gravitino DM in models with bilinear $R$-parity violation. As the authors of~\cite{Ibe:2013nka} point out, however, decaying gravitino DM with trilinear $R$-parity violation that only involves an operator of the type $LL\bar{E}$ in the superpotential could be compatible with the AMS-02 result and at the same time avoid all antiproton constraints.

\begin{figure}[t]
  \includegraphics[width=0.49\linewidth]{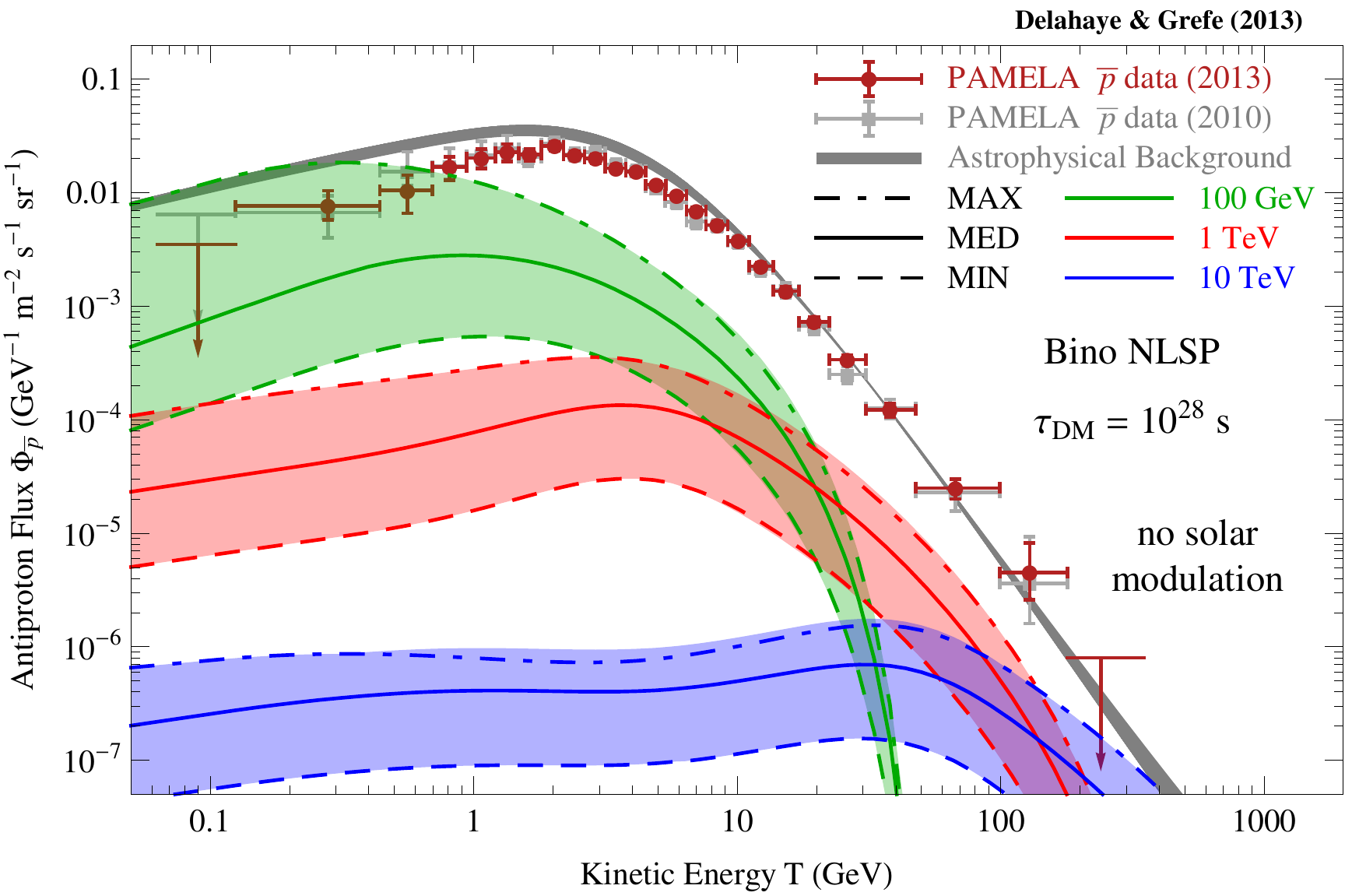}
  \hfill
  \includegraphics[width=0.49\linewidth]{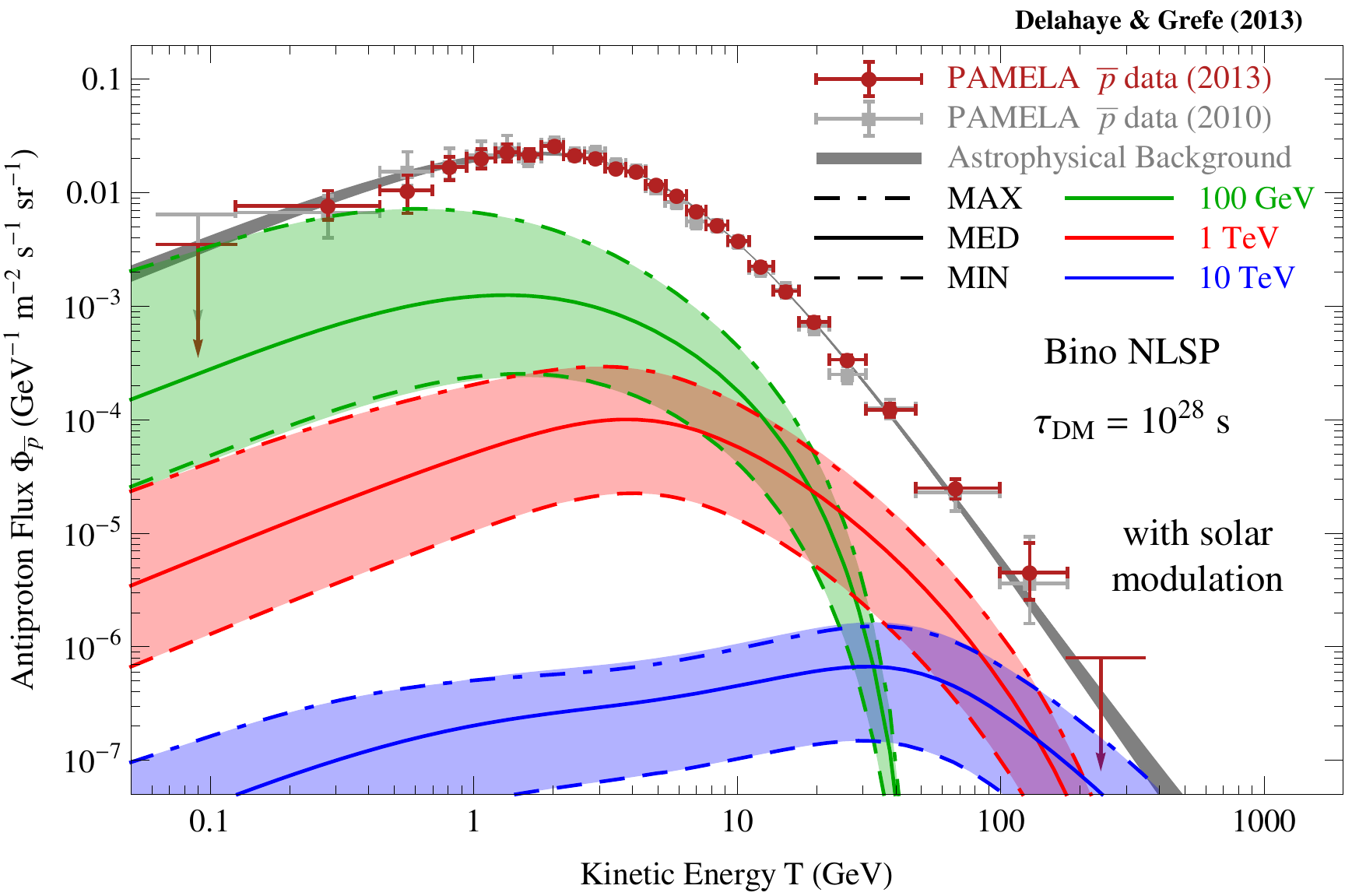}
  \caption{Antiproton flux expected from gravitino decay compared to the expected flux of antiprotons from secondary production and data from PAMELA~\cite{Adriani:2012paa,Adriani:2010rc}. The gravitino signal is calculated for the case of Bino NLSP, a lifetime of $10^{28}$\,s and for three different masses: 100\,GeV, 1\,TeV and 10\,TeV. It is shown for MIN/MED/MAX propagation parameters and the coloured bands also show the allowed range of spectra from the scan over propagation parameters. The grey band shows the propagation uncertainty for secondary antiprotons. \textit{Left:} No correction for solar modulation. \textit{Right:} Correcting for solar modulation.}
  \label{fig:gravitinoantiproton}
\end{figure}
In Fig.~\ref{fig:gravitinoantiproton} we show the antiproton flux expected from gravitino decay for Bino NLSP parameters compared to the PAMELA data~\cite{Adriani:2012paa,Adriani:2010rc} and the expected flux of secondary astrophysical antiprotons. The lifetime of $\tau_{3/2}=10^{28}\,$s is just chosen as an example and is already excluded for several gravitino masses and cosmic-ray propagation models. The plot clearly shows that depending on the gravitino mass the most-constraining energy bins are not the same. There is not much progress to expect from AMS-02 for energies in the 1--10\,GeV range as the error bars of the PAMELA data are rather small already. However, extending and improving the measurements at low energies is expected to strengthen the limits for gravitino masses lower than a couple of hundreds of GeV, in particular if new data lead to a better understanding of the effects of solar modulation. Increasing the data range to higher energies and improving on statistics for these energies would in particular allow to improve constraints for gravitino masses above the TeV scale. 

It is interesting to compare our limits on the gravitino lifetime to constraints coming from other cosmic-ray channels. Very strict lifetime limits on the order of a few times $10^{29}\,$s can be derived from searches for gamma-ray lines with Fermi-LAT~\cite{Fermi-LAT:2013uma}. However, these limits are mainly important for gravitino masses below the electroweak scale where the decay channel $\gamma\nu$ dominates~\cite{Grefe:2011dp}. In addition, expected gamma-ray signals from gravitino decays in galaxy clusters have been used to derive lifetime limits on the order of $10^{26}$\,s~\cite{Huang:2011xr}. Moreover, gravitino decays in the Galactic halo as well as extragalactic gravitino decays are expected to contribute to the isotropic diffuse gamma-ray flux. Lifetime limits of a few times $10^{26}$\,s over a wide range of gravitino masses have been estimated from Fermi-LAT data even without taking into account astrophysical contributions~\cite{Huang:2011xr,Grefe:2011dp}. Finally, also lifetime limits from neutrino telescopes are expected to play some role, at least for gravitinos of multi-TeV masses~\cite{Grefe:2011dp}.

Currently, antiproton observations and gamma-ray line searches provide the best-understood environment for setting constraints on the gravitino lifetime and to realistically estimate the related uncertainties. As stated above, however, other cosmic-ray channels clearly have the potential to provide competitive constraints. We think more sophisticated analyses of the limits on the gravitino lifetime from several of these other cosmic-ray channels are needed in order to fully explore their constraining potential.

\subsection{Constraints on the Strength of \texorpdfstring{\boldmath$R$}{R}-Parity Violation}
\label{RPVconstraints}

The constraints on the gravitino lifetime can be converted in a straightforward way into constraints on the strength of $R$-parity violation. Since the gravitino lifetime is simply given by the inverse of its total decay width, $\tau_{3/2}=1/\Gamma_{3/2}$, and the decay widths of all the individual channels are proportional to $\xi^2\,m_{3/2}^3$, the constraints will scale like $\xi\propto(\tau_{3/2}\,m_{3/2}^3)^{-1/2}$ as a function of the gravitino mass. On top of that, of course, also the dependence on the supersymmetric mass parameters enters (\textit{cf.} Eq.~(\ref{gammanu})--(\ref{hnu})). See also the discussion about the behaviour of $\Gamma_{3/2}$ as a function of the gravitino mass in Section~\ref{gravitinoRPV}. For the relation between the lifetime and the decay width remember that in natural units $1\,\text{GeV}^{-1}\simeq 6.58\times 10^{-25}\,$s.
\medskip

In addition, we compare our constraints on the amount of $R$-parity violation derived from the antiproton limits to a few generic constraints for this type of model: First of all, bilinear $R$-parity breaking contributes to the generation of neutrino masses. Although we allow for other mechanisms of neutrino mass generation like the see-saw mechanism and do not tie our $R$-parity violating parameter to neutrino parameters, the contribution from $R$-parity violation must not overshoot the measured limits on the neutrino masses. In particular, the mixing of the neutrinos with the neutralinos leads to one nonvanishing neutrino mass given by~\cite{Buchmuller:2007ui}
\begin{equation}
  m_{\nu_i}=\frac{g_Z^2v_i^2}{2}\sum_{\alpha=1}^4\frac{\abs{C_{\tilde{Z}\alpha}}^2}{m_{\chi_\alpha^0}}=\xi_i^2m_ZU_{\tilde{Z}\tilde{Z}}\simeq \xi_i^2m_Z^2\left( \frac{\sin^2\theta_W}{M_1}+\frac{\cos^2\theta_W}{M_2}\right) ,
\end{equation}
where we used Eq.~(\ref{UZZ}) for the latter relation. Cosmological bounds on the sum of neutrino masses, $\sum m_{\nu_i}<0.23\,$eV, are derived from most recent CMB and BAO data~\cite{Ade:2013lta}. We can then set an upper bound on the size of $R$-parity violation on the order of
\begin{equation}
  \xi^2\equiv\sum_i\xi_i^2\lesssim\frac{0.23\,\text{eV}}{m_Z^2\left( \frac{\sin^2\theta_W}{M_1}+\frac{\cos^2\theta_W}{M_2}\right)}\lesssim4.3\times10^{-12}\,\frac{M_1}{100\,\text{GeV}}\,,
\end{equation}
where the latter relation holds in the case of a GUT relation between the gaugino masses. For other gaugino mass hierarchies the prefactor can change by an $\mathcal{O}(1)$ factor and the limit scales with the mass of the lightest gaugino.\smallskip

A constraint of similar strength comes from a completely different source. If $R$-parity violation is too strong, a primordial baryon asymmetry produced in the very early Universe could be washed out before the time of the electroweak phase transition. This condition leads to the following constraint~\cite{Buchmuller:2007ui,Endo:2009cv}:\footnote{In the derivation of this constraint squark masses in the range 200\,GeV--1.2\,TeV and slepton masses in the range 100\,GeV--400\,GeV are assumed~\cite{Endo:2009cv}. For the discussion in this section we neglect the dependence on the supersymmetric mass spectrum although we consider gravitino LSP masses up to 10\,TeV.}
\begin{equation}
  \sqrt{\sum_i\abs{\frac{\mu_i}{\mu}}^2}\lesssim(1\text{--}2)\times10^{-6}\left(\frac{\tan\beta}{10}\right)^{-1}.
\end{equation}
From this limit we can derive a constraint on the sneutrino VEV $\xi_i$ if we assume typical values for the soft bilinear $R$-parity breaking terms, \textit{i.e.}
\begin{equation}
  B_i\sim\mu_i\,\tilde{m}\sim\frac{\mu_i}{\mu}\,\tilde{m}^2\qquad\text{and}\qquad m_{H_d\ell_i}\sim\frac{\mu_i}{\mu}\,\tilde{m}^2,
\end{equation}
where $\tilde{m}$ is the supersymmetry breaking scale~\cite{Buchmuller:2007ui}. Using these assumptions and Eq.~(\ref{sneutrinoVEV}) we can estimate
\begin{equation}
  \xi_i=\frac{v_i}{v}\sim\frac{\frac{\mu_i}{\mu}\,\tilde{m}^2\,(\sin\beta-\cos\beta)}{m_{\tilde{\ell}_{ij}}^2+\frac{1}{2}\,m_Z^2\cos2\beta}\sim\frac{\mu_i}{\mu}\,(\sin\beta-\cos\beta)\,,
\end{equation}
since the denominator is of the order of the supersymmetry breaking scale squared. The constraint then reads
\begin{equation}
  \xi\lesssim(1\text{--}2)\times10^{-5}\left(\frac{\sin\beta-\cos\beta}{\tan\beta}\right)\simeq(1\text{--}2)\times10^{-6}\left(\frac{\tan\beta}{10}\right)^{-1},
\end{equation}
where the latter relation only holds for values $\tan\beta\gg1$.\smallskip

As mentioned already in Section~\ref{cosmo}, a lower limit on the size of $R$-parity violation can be obtained from the requirement that the NLSP decays through $R$-parity breaking interactions before the time of BBN. The detailed constraints depend on the particle nature of the NLSP and the particular decay channel, but in general NLSP lifetimes are required to be smaller than $\mathcal{O}(100)$--$\mathcal{O}(1000)$\,s~\cite{Buchmuller:2007ui}.

Let us consider a neutralino as one example for the NLSP. There are three $R$-parity breaking two-body decay channels for electroweak scale neutralinos: $Z\nu$, $W\ell$ and $h\nu$. Assuming hierarchies of the gauginos as presented in Table~\ref{parameters} and neglecting phase space factors at kinematic thresholds, the lifetimes of the different neutralino NLSPs can be approximated as~\cite{Ishiwata:2008cu,Bobrovskyi:2012dc}
\begin{align}
  \tau_{\text{Bino NLSP}}&\sim\frac{2\sqrt{2}\,\pi\,\xi^{-2}\,M_1^{-1}}{G_F\sin^2\theta_W\,m_Z^2}\approx2.8\,\text{s}\left(\frac{\xi}{10^{-12}}\right)^{-2}\!\!\left(\frac{M_1}{100\,\text{GeV}}\right)^{-1}\!\!,\\
  \tau_{\text{Wino NLSP}}&\sim\frac{2\sqrt{2}\,\pi\,\xi^{-2}\,M_2^{-1}}{G_F\cos^2\theta_W\,m_Z^2}\approx0.83\,\text{s}\left(\frac{\xi}{10^{-12}}\right)^{-2}\!\!\left(\frac{M_2}{100\,\text{GeV}}\right)^{-1}\!\!,\\
  \tau_{\text{Higgsino NLSP}}&\sim\frac{\sqrt{2}\,\pi\,\xi^{-2}\,\mu^{-1}M_2^2}{2\,G_F\cos^4\theta_W\,m_Z^4\sin^2\!\beta}\approx33\,\text{s}\left(\frac{\xi}{10^{-12}}\right)^{-2}\!\!\left(\frac{\mu}{100\,\text{GeV}}\right)^{-1}\!\left(\!\frac{M_{2}}{1\,\text{TeV}}\!\right)^2\!\!.\!\!
\end{align}
While the Bino and Wino decay widths increase with their masses, the Higgsino decay width is suppressed compared to these cases by roughly a factor $(m_Z/M_2)^2$. This fact leads to a different behaviour when the mass spectrum is shifted to higher energies.

We also consider the case of a stau as an example of a non-neutralino NLSP. If the lightest stau is mainly right-handed its dominant decay channels are two-body decays into $\tau\,\nu$ and $\mu\,\nu$~\cite{Bobrovskyi:2010ps}. Its lifetime can then be estimated as~\cite{Ishiwata:2008cu}
\begin{equation}
  \tau_{\tilde{\tau}_R}\sim\frac{\pi\,\xi^{-2}\,m_{\tilde{\tau}_R}^{-1}\,m_{\tilde{\chi}^0}^2}{4\,G_F\sin^4\theta_W\,m_Z^4\,v^2}\approx8.6\,\text{s}\left(\frac{\xi}{10^{-12}}\right)^{-2}\!\!\left(\frac{m_{\tilde{\tau}_R}}{100\,\text{GeV}}\right)^{-1}\!\left(\frac{m_{\tilde{\chi}^0}}{150\,\text{GeV}}\right)^2,
\end{equation}
where $m_{\tilde{\chi}^0}$ is the mass of the lightest neutralino and $v=174\,$GeV is the Higgs VEV. If we choose the mass of the lightest stau to be slightly below the mass of the lightest neutralino, but scaling in the same way with the gravitino mass, the stau decay width scales in the same way as the Higgsino decay width.

\begin{figure}[t]
\centering
  \includegraphics[width=0.8\linewidth]{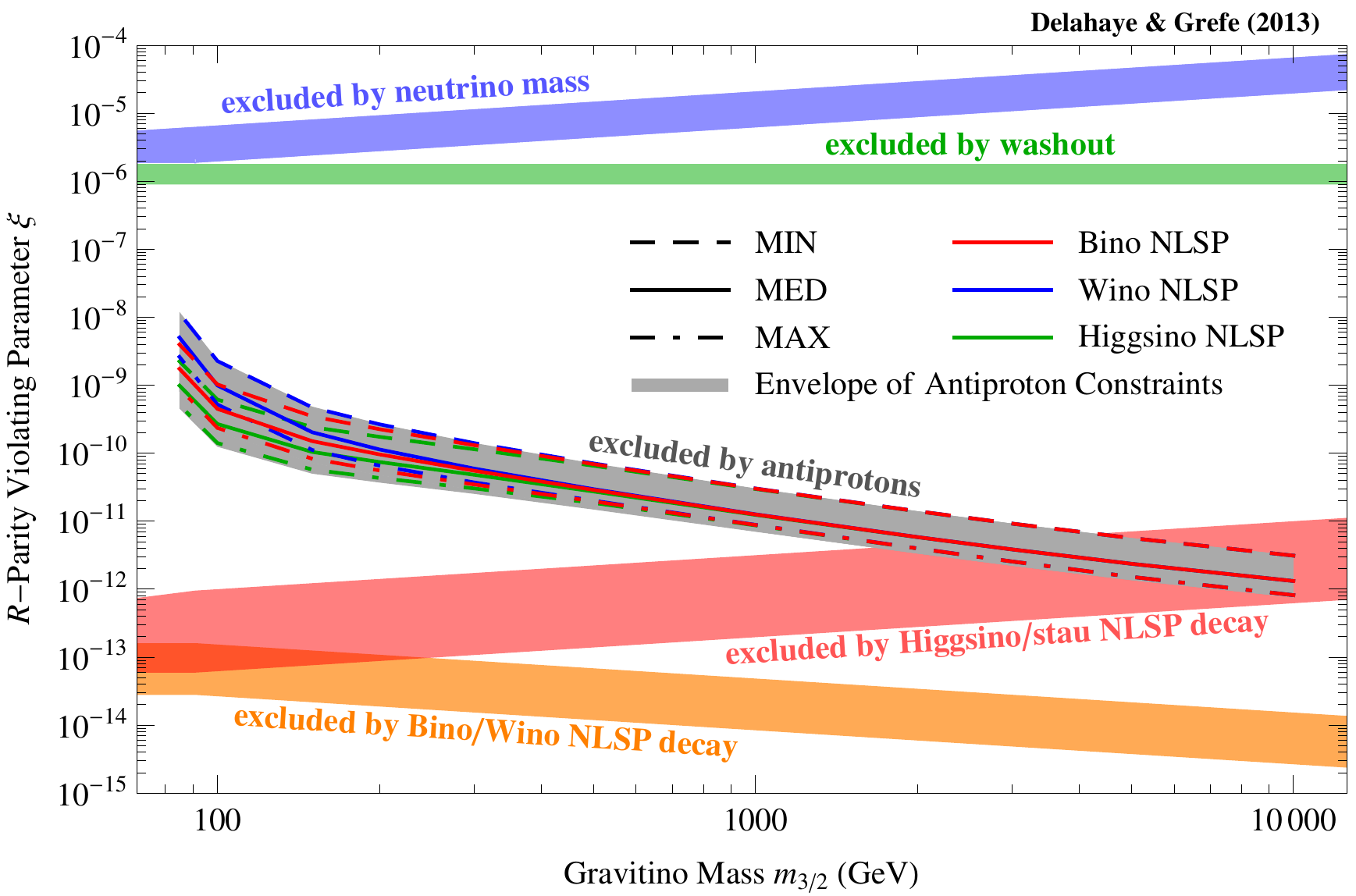}
  \caption{Constraints on the amount of $R$-parity violation as a function of the gravitino mass. The constraints derived from antiproton data are shown in the central part of the plot. The area above the lines is excluded. For comparison we also show limits from other sources: The area above the blue band is excluded by contributions to the neutrino mass, the area above the green band is excluded by the washout limit, the area below the red band is excluded by BBN bounds in the case of a Higgsino or stau NLSP, and the area below the orange band is excluded by BBN bounds in the case of a Bino or Wino NLSP. See text for details.}
  \label{fig:xibound}
\end{figure}

The limits mentioned above are shown together with the antiproton constraints derived in this work in Fig.~\ref{fig:xibound}. The antiproton constraints are shown in the central part of the plot. We present the individual limits for the different gravitino benchmark models for MIN/MED/MAX propagation parameters, but also the envelope of the constraints including the uncertainties from cosmic-ray propagation and from the dependence on the supersymmetry parameters. In this case the parameter space above the lines is excluded. The limits from other sources are the following: The area above the blue band is excluded by contributions to the neutrino mass, the area above the green band is excluded by the washout limit, the area below the red band is excluded by BBN bounds in the case of a Higgsino or stau NLSP, and the area below the orange band is excluded by BBN bounds in the case of a Bino or Wino NLSP. The widths of the bands present an estimate of the uncertainties on the positions of the limits.

It is obvious that the antiproton constraints are much stricter than the neutrino mass and washout limits. However, they are not sufficient to rule out the parameter space completely. Depending on the nature of the NLSP there can be plenty of viable parameter space at lower values of the $R$-parity breaking parameter. In particular if one takes into account the cosmological motivation for a gravitino lighter than a few hundred GeV, there is a big gap between the BBN constraints and the limits derived from antiproton data. Unfortunately, since the gravitino lifetime scales with $\xi^{-2}$, it is highly unlikely that indirect searches will be able to completely probe the remaining parameter space in the foreseeable future. However, forthcoming AMS-02 results, extending the range of antiproton measurements both to lower and higher energies and having significantly smaller error bars, could improve these limits appreciably. 

Collider searches for signals from the decay of the metastable NLSP have a sensitivity to the strength of $R$-parity violation that can be complementary to that of indirect DM searches. In particular, colliders can have a better sensitivity to $\xi$ in the case of small supersymmetry mass parameters. In the case of a neutralino NLSP, values of $\xi$ of the order of $10^{-9}$--$10^{-10}$ have been found to be in the reach of the LHC with roughly 100\,fb$^{-1}$ of data for gravitino masses of the order of 100\,GeV~\cite{Bobrovskyi:2011vx}. For a Higgsino NLSP, the sensitivity is a bit worse and values of $\xi$ of the order of $10^{-8}$--$10^{-9}$ are expected to be testable with roughly 100\,fb$^{-1}$ of LHC data for gravitino masses of the order of 100\,GeV~\cite{Bobrovskyi:2012dc}.

\section{Conclusions}

In this work we studied gravitino dark matter in a model with bilinear $R$-parity violation. This type of models is well motivated from cosmology as it can reconcile the generation of the baryon asymmetry in the Universe via thermal leptogenesis with the constraints on late-decaying particles from big bang nucleosynthesis: In fact, the reheating temperature needed for thermal leptogenesis can lead to a gravitino relic abundance from production in thermal scatterings that matches the observed dark matter density in the Universe if the gravitino mass is in the range of roughly ten to several hundreds of GeV. In addition, a small violation of $R$-parity can trigger the decay of the next-to-lightest supersymmetric particle before the time of big bang nucleosynthesis, thus not spoiling the successful predictions of standard cosmology. At the same time, the double suppression of the gravitino decay width by the Planck scale and the small amount of $R$-parity violation leads to a lifetime exceeding the age of the Universe. Therefore the unstable gravitino is a viable candidate for the dark matter in the Universe.

After a short review of the gravitino two-body decay widths and the resulting branching ratios of the different decay channels, we presented the results of a new simulation of the proton and antiproton spectra generated in these decays. Making use of these spectra we computed the cosmic-ray antiproton fluxes at the Earth coming from dark matter decays in the Galactic halo. We individually considered the different two-body decay channels of the gravitino that contain antiprotons in the final state, \textit{i.e.} $Z\nu$, $W\ell$ and $h\nu$, and also three benchmark models for the case of gravitino dark matter, \textit{i.e.} models with supersymmetry mass parameters chosen such that the next-to-lightest supersymmetric particle is either Bino-like, Wino-like or Higgsino-like. We explored a gravitino mass range from 85\,GeV to 10\,TeV, thus going somewhat beyond the mass range motivated by cosmology. 

We compared these antiproton fluxes to the data of the PAMELA experiment and, taking into account also the astrophysical antiproton background, derived constraints on the dark matter lifetime in each individual channel. To achieve this we employed a $\chi^2$ method and derived 95\,\% CL lower limits on the lifetime. In order to properly estimate the uncertainties on these constraints coming from our poor knowledge of the Galactic cosmic-ray propagation parameters, we performed a scan over the parameter sets allowed by other cosmic-ray nuclei data. We found lower limits ranging from roughly $8\times10^{28}$\,s to roughly $5\times10^{25}$\,s. The propagation model uncertainties amount to roughly one order of magnitude.

Then we applied the same method to the three benchmark models for gravitino dark matter and translated the resulting lifetime limits into constraints on the amount of $R$-parity violation. We found upper limits in the range of roughly $10^{-8}$ to roughly $8\times10^{-13}$. These limits are much stricter than those coming from contributions to the neutrino mass or from the washout of the baryon asymmetry in the early Universe. In combination with lower limits coming from big bang nucleosynthesis constraints on the lifetime of the next-to-lightest supersymmetric particle, this allows to narrow the available parameter space for gravitino dark matter with bilinear $R$-parity violation.

Updated data from PAMELA with even higher statistics might allow to strengthen the constraints found in this work. However, as we have seen by comparing constraints derived from the 2010 and 2013 PAMELA antiproton data sets, variations of the central values of the data points can also lead to a weakening of constraints. Beyond that, new data on cosmic-ray spectra from the AMS-02 experiment with better energy resolution and reaching higher energies are expected to significantly improve the current limits or hopefully even to exhibit a signal of dark matter decay or annihilation. In addition, AMS-02 data on secondary-to-primary ratios and on the spectra of radioactive nuclei are expected to improve the constraints on the propagation parameters for charged cosmic rays, thereby reducing the uncertainty on the determination of local antiproton spectra from dark matter decays. Moreover, improved measurements of the spectra of primary cosmic rays will allow to improve the predictions of the secondary antiproton flux. Then again, the reduction of experimental error bars and propagation uncertainties, as well as improved knowledge of the primary comic-ray spectra demand for a better understanding of nuclear physics properties that play an important role for the production of secondary particles in cosmic-ray spallation processes. It has to be stressed though, that the progress to be expected from future AMS-02 results will not show up immediately as only a careful analysis of all the different channels at once will allow for significant steps forward in the understanding of cosmic-ray physics.

But already current antiproton limits are in strong tension with interpretations of the rising positron fraction measured by PAMELA and AMS-02 in terms of decaying gravitino dark matter with bilinear $R$-parity violation. We conclude that this rise must be at least partially due to astrophysical sources if the unstable gravitino makes up the dark matter in the Universe.

Very strong constraints on the gravitino lifetime also come from other cosmic-ray observations. In particular searches for gamma-ray lines dominate the limits for gravitino masses below the electroweak scale. But also contributions of gravitino decays to the isotropic diffuse gamma-ray flux can be used to set limits on the gravitino lifetime over a wide range of gravitino masses. These limits are generally estimated to be similar in strength to constraints derived from antiprotons.

Of course also other cosmic-ray species like radio photons, neutrinos and antideuterons could contribute in strengthening the constraints presented in this work and we have plans to discuss prospects and updated limits for these channels in future works.

\section*{Acknowledgements}

We would like to thank Laura Covi for useful comments on the manuscript. We are also grateful to Savvas Nesseris for fruitful discussions about chi-squared analyses and to Andrey Mayorov for drawing our attention to a recent update of the PAMELA antiproton data.

The work of TD is supported in part by the ANR project DMAstroLHC, ANR-12-BS05-0006-01. The work of MG is supported by the Marie Curie ITN "UNILHC" under grant number PITN-GA-2009-237920 and by the Comunidad de Madrid under the grant HEPHACOS S2009/ESP-1473.

{\appendix

\section[Bilinear R-Parity Violation]{Bilinear \boldmath$R$-Parity Violation}
\label{App:RPV}

In this work we consider a supergravity theory with $R$-parity violation introduced through the bilinear term $W_{\slashed{R}}=\mu_iH_u\tilde{\ell}_i$ in the superpotential and the bilinear terms $\mathscr{L}_{\slashed{R}}^{\text{soft}}=-B_iH_u\tilde{\ell}_i-m_{H_d\ell_i}^2H_d^*\tilde{\ell}_i+\text{h.c.}$ in the soft supersymmetry-breaking Lagrangian~\cite{Buchmuller:2007ui}. The choice of only lepton number violating operators guarantees the stability of the proton.

These $R$-parity violating terms in general lead to nonvanishing vacuum expectation values (VEVs) of the sneutrinos. After a field redefinition of the down-type Higgs and lepton doublets, the bilinear $R$-parity breaking term in the superpotential vanishes and the sneutrino VEVs $v_i$ are given in terms of the $R$-parity breaking soft terms as~\cite{Grefe:2011dp}
\begin{equation}
 \xi_i=\frac{v_i}{v}\simeq\frac{B_i\sin\beta-m_{H_d\ell_i}^{*2}\cos\beta}{m_{\tilde{\ell}_{ij}}^2+\frac{1}{2}\,m_Z^2\cos2\beta}\,.
 \label{sneutrinoVEV}
\end{equation} 
In this expression we introduced the ratio of the sneutrino VEVs and the Higgs VEV $\xi_i=v_i/v$ for a dimensionless parametrization of $R$-parity violation. 

In models of this type the standard $4\times4$ neutralino mixing matrix of the MSSM is extended to a $7\times7$ matrix that also includes mixings with the three flavours of the light neutrinos. This neutralino--neutrino mixing matrix can be written as~\cite{Grefe:2011dp}
\begin{equation}
  M_N=
  \begin{pmatrix}
   M_1c_W^2+M_2\,s_W^2 & \left( M_2-M_1\right) s_W\,c_W & 0 & 0 & 0 \\
   \left( M_2-M_1\right) s_Wc_W & M_1s_W^2+M_2\,c_W^2 & -m_Z\,s_\beta & m_Z\,c_\beta & m_Z\,\xi_j \\
   0 & -m_Z\,s_\beta & 0 & -\mu & 0 \\
   0 & m_Z\,c_\beta & -\mu & 0 & 0 \\
   0 & m_Z\,\xi_i & 0 & 0 & 0
  \end{pmatrix},
\end{equation}
where the basis is given by $\psi_i^0=(-i\tilde{\gamma},\,-i\tilde{Z},\,\tilde{H}_u^0,\,\tilde{H}_d^0,\,\nu_i)^T$.

In our case of small $R$-parity violation the mixing to the neutrinos is a small perturbation to the standard neutralino mass matrix. Therefore, the mixing parameters $N$ between the neutrinos and the neutralinos can be expressed in terms of a product of the sneutrino VEVs and the mixing parameters between the different states in the standard neutralino mass matrix. As described in~\cite{Grefe:2011dp} we can find analytical approximations for these parameters that show explicitly the dominant dependence on the relevant supersymmetry parameters:\footnote{For the first two parameters we find a different sign compared to our result in~\cite{Grefe:2011dp}. This is due to an erroneous sign in the gaugino--Higgsino mixing block of the neutralino mass matrix in the previous analysis.}
\begin{align}
  N_{\nu_i\tilde{\gamma}}&\simeq-\xi_i\,U_{\tilde{\gamma}\tilde{Z}}\,, &U_{\tilde{\gamma}\tilde{Z}}&\simeq -m_Z\sin\theta_W\cos\theta_W\,\frac{M_2-M_1}{M_1\,M_2}\,, \label{UgammaZ}\\
  N_{\nu_i\tilde{Z}}&\simeq-\xi_i\,U_{\tilde{Z}\tilde{Z}}\,, &U_{\tilde{Z}\tilde{Z}}&\simeq m_Z\left( \frac{\sin^2\theta_W}{M_1}+\frac{\cos^2\theta_W}{M_2}\right) \,, \label{UZZ}\\
  N_{\nu_i\tilde{H}_u^0}&\simeq-\xi_i\,U_{\tilde{H}_u^0\tilde{Z}}\,, &U_{\tilde{H}_u^0\tilde{Z}}&\simeq m_Z^2\cos\beta\,\frac{M_1\cos^2\theta_W+M_2\sin^2\theta_W}{M_1\,M_2\,\mu}\,, \label{UHuZ}\\
  N_{\nu_i\tilde{H}_d^0}&\simeq-\xi_i\,U_{\tilde{H}_d^0\tilde{Z}}\,, &U_{\tilde{H}_d^0\tilde{Z}}&\simeq -m_Z^2\sin\beta\,\frac{M_1\cos^2\theta_W+M_2\sin^2\theta_W}{M_1\,M_2\,\mu}\,. \label{UHdZ}
\end{align}\smallskip

Similar to the case of the neutralino mass matrix, the $2\times2$ chargino mixing matrix of the MSSM is extended to a $5\times5$ matrix that also includes mixings with the three flavours of charged leptons. This chargino--lepton mixing matrix can be written as~\cite{Grefe:2011dp}
\begin{equation}
 M_C=
 \begin{pmatrix}
  M_2 & \sqrt{2}\,m_W\,s_\beta & 0 \\
  \sqrt{2}\,m_W\,c_\beta & \mu & -m_{\ell_{ij}}\,\xi_i\,c_\beta \\
  \sqrt{2}\,m_W\,\xi_i & 0 & m_{\ell_{ij}}
 \end{pmatrix},
\end{equation}
where the basis vectors are $\psi^-=(-i\tilde{W}^-,\,\tilde{H}_d^-,\,\ell_i^-)^T$ and $\psi^+=(-i\tilde{W}^+,\,\tilde{H}_u^+,\,e_i^{c\,+})^T$. 

The mixing parameters $U$ between the left-handed charged leptons and the charginos can be expressed in terms of a product of the sneutrino VEVs and the mixing parameters between the different states in the standard chargino mass matrix. Also in this case we can find analytical approximations~\cite{Grefe:2011dp}:
\begin{align}
  U_{\ell_i\,\tilde{W}}&\simeq-\sqrt{2}\,\xi_i\,U_{\tilde{W}\tilde{W}}\,,& U_{\tilde{W}\tilde{W}}&\simeq \frac{m_W}{M_2}\,, \label{UWW}\\
  U_{\ell_i\,\tilde{H}_d^-}&\simeq-\sqrt{2}\,\xi_i\,U_{\tilde{H}_d^-\tilde{W}}\,,& U_{\tilde{H}_d^-\tilde{W}}&\simeq -\frac{\sqrt{2}\,m_W^2\sin\beta}{M_2\,\mu}\,. \label{UHW}
\end{align}
The mixing of the right-handed charged leptons to the charginos is suppressed and can be neglected~\cite{Grefe:2011dp}.\smallskip

In addition to the extended mixing in the fermionic sector, $R$-parity violation also introduces new mixings in the scalar sector between the Higgs bosons and the slepton doublets. In particular the mixing between the sleptons and the lightest Higgs boson contributes to the gravitino two-body decay into Higgs + neutrino.

\section{Gravitino Decay Widths}
\label{App:decaywidths}
The decay widths for the gravitino two-body decay channels have been studied in various works in the literature~\cite{Takayama:2000uz,Ibarra:2007wg,Ishiwata:2008cu,Covi:2008jy}. The most recent and most complete results were presented in~\cite{Buchmuller:2012rc} and~\cite{Grefe:2011dp}. In this work we stick to the results found in~\cite{Grefe:2011dp}:\footnote{The authors of~\cite{Buchmuller:2012rc} use a different notation for the analytical expressions and a different parametrization of bilinear $R$-parity violation. After matching the notation there are a few slight differences remaining between the results of~\cite{Buchmuller:2012rc} and~\cite{Grefe:2011dp}, most of which, however, appear to vanish in an even more recent recalculation~\cite{Hajer:2013jla} following the same approach as~\cite{Buchmuller:2012rc}. Recently, the authors of~\cite{Buchmuller:2012rc} have revised their calculation and now agree with our results~\cite{Garny:2013}. In any case the numerical results for the branching ratios agree very well throughout all these works.}
\begin{align}
  \Gamma_{\psi_{3/2}\rightarrow\gamma\nu_i} &\simeq\frac{\xi_i^2\,m_{3/2}^3}{32\,\pi\,\MP^2}\abs{U_{\tilde{\gamma}\tilde{Z}}}^2, \label{gammanuA}\\
  \Gamma_{\psi_{3/2}\rightarrow Z\nu_i} &\simeq\frac{\xi_i^2\,m_{3/2}^3\,\beta_Z^2}{32\,\pi\,\MP^2}\left\lbrace \,U_{\tilde{Z}\tilde{Z}}^2f_Z+\frac{1}{6}\abs{1+s_\beta\,U_{\tilde{H}_u^0\tilde{Z}}-c_\beta\,U_{\tilde{H}_d^0\tilde{Z}}}^2h_Z\right. \nonumber\\
    &\left. \quad\qquad-\frac{8}{3}\frac{m_Z}{m_{3/2}}\,U_{\tilde{Z}\tilde{Z}}\left( 1+s_\beta \RE U_{\tilde{H}_u^0\tilde{Z}}-c_\beta \RE U_{\tilde{H}_d^0\tilde{Z}}\right) j_Z \right\rbrace, \label{gravitinowidths}\\
  \Gamma_{\psi_{3/2}\rightarrow W\ell_i} &\simeq\frac{\xi_i^2\,m_{3/2}^3\,\beta_W^2}{16\,\pi\,\MP^2}\left\lbrace \,U_{\tilde{W}\tilde{W}}^2f_W+\frac{1}{6}\abs{1-\sqrt{2}\,c_\beta\,U_{\tilde{H}_d^-\tilde{W}}}^2h_W\right. \nonumber\\
    &\left. \quad\qquad-\frac{8}{3}\frac{m_W}{m_{3/2}}\,U_{\tilde{W}\tilde{W}}\left( 1-\sqrt{2}\,c_\beta \RE U_{\tilde{H}_d^-\tilde{W}}\right) j_W \right\rbrace , \\
  \Gamma_{\psi_{3/2}\rightarrow h\nu_i} &\simeq\frac{\xi_i^2\,m_{3/2}^3\,\beta_h^4}{192\,\pi\,\MP^2}\abs{\frac{m_{\tilde{\nu}_i}^2+\frac{1}{2}\,m_Z^2\cos2\beta}{m_h^2-m_{\tilde{\nu}_i}^2}+2\,s_\beta\,U_{\tilde{H}_u^0\tilde{Z}}+2\,c_\beta\,U_{\tilde{H}_d^0\tilde{Z}}}^2, \label{hnuA}
\end{align}
where we used $s_\beta\equiv\sin{\beta}$ and $c_\beta\equiv\cos{\beta}$ for a compact notation. Plugging the approximations for the mixing parameters~(\ref{UgammaZ})--(\ref{UHdZ}) and (\ref{UWW}), (\ref{UHW}) into the expressions for the gravitino decay widths~(\ref{gammanuA})--(\ref{hnuA}), we find the result quoted in Eq.~(\ref{gammanu})--(\ref{hnu}) in the text.

\section{Generation of Decay Spectra in PYTHIA}
\label{App:spectra}

As discussed in Section~\ref{spectra} we generated the decay spectra with \textsc{Pythia}~6.4~\cite{Sjostrand:2006za}. Here we want to describe briefly the routines we used to produce the spectra. In principle one could start the event record by calling the routine \verb PY2ENT \ with the centre of mass energy given by the DM mass and the corresponding set of particles produced in the two-body decay: $Z\nu$, $We$, $W\mu$, $W\tau$, or $h\nu$. Unfortunately, this method can lead to errors and an abort of the program if the DM mass is close to the nominal mass of the massive boson and the bosons are treated with a Breit--Wigner shaped width. Therefore, we used a slightly different approach.

We started by redefining the properties of the fourth-generation neutrino $\nu_\tau'$ (particle code 18): We set its mass to the mass of the DM particle and set its only decay channel to be $Z\nu$, $We$, $W\mu$, $W\tau$, or $h\nu$, respectively. Since the pair of two particles is generated in a decay, it is guaranteed that the daughter particle masses do not exceed the DM mass -- even if the boson widths are treated correctly.

Another issue is to allow final state radiation (FSR) from the decay products. Since the initial particles in the event record are defined manually, the \textsc{Pythia} switch for FSR does not apply to them.\footnote{In any case \textsc{Pythia} only supports QED and QCD FSR. Massive gauge bosons like $W^\pm$ do not radiate in \textsc{Pythia}.} Thus we have to force FSR for the initial particles by calling the \verb PYSHOW \ routine.

In the following we present a brief example of the routines used to start the event record for the case of the $W\tau$ decay channel (particle codes 24 and 15, respectively): 
\begin{verbatim}
      CALL PYINIT('None','','','')

      PMAS(24,3)=PMAS(24,1)

      DO 100 NEV=1,NEVENTS

      MDCY(PYCOMP(15),1)=0
      MDCY(PYCOMP(24),1)=0

      CALL PY1ENT(0,18,mDM,0,0)
      CALL PYSHOW(2,3,mDM)
      CALL PYEXEC

      MDCY(PYCOMP(15),1)=1
      MDCY(PYCOMP(24),1)=1
      CALL PYEXEC
\end{verbatim}
We initialise \textsc{Pythia} without initial beams since we want to start the process with a single particle that subsequently decays. Then we increase the allowed deviation of the $W$ mass from its nominal value in the Breit--Wigner shape to avoid artefacts from truncation in the spectra. After that we loop over a predefined number of events -- in our case $5\times10^7$. At the beginning of each event we forbid the decay of the initial decay products, then we call the redefined $\nu_\tau'$ with the corresponding DM mass as the initial particle in the event record, thereafter we call the showering routine for its decay products, and finally call \verb PYEXEC \ to execute the event generation. In a second step we allow again the decay of the initial decay products and execute the event generation again. In this way we avoid duplicates of the subsequent decay products as only the $W\tau$ pair already processed by the \verb PYSHOW \ routine is allowed to decay.

Before the initialisation of the event generation we had to make sure that the decays of all unstable particles are switched on since we are only interested in stable final state particles.\footnote{Several long-lived particles like the muon by default are not allowed to decay in \textsc{Pythia} as they can be considered stable in the context of collider studies.} In particular, this also means that we had to define neutron decay as it is not implemented in \textsc{Pythia}~6.4.

\section{Cosmic-Ray Propagation}
\label{App:CR}

The propagation of Galactic cosmic rays in the interstellar medium is a long-standing research subject (see for instance~\cite{1969SSRv....9..651P}). The current way of modelling propagation is through the use of a diffusion equation that is a differential equation for the stationary cosmic-ray density $\psi$~\cite{1976RvMP...48..161G,Strong:2007nh,Maurin:2001sj,1992ApJ...390...96W}:
\begin{equation}
 \begin{split}
\vec{\nabla} \cdot(\vec{V}_c\, \psi - K_0\, \beta\, p^\delta\, \vec{\nabla} \psi ) + 2\,h\,\delta(z)\, \partial_E\left(b_{\text{loss}}\,\psi -D_{EE}\, \partial_E\psi\right) \qquad\qquad\qquad\quad \\
\qquad\qquad\quad = Q^{\text{prim}} + 2\,h\,\delta(z) \left(Q^{\text{sec}} + Q^{\text{ter}}\right) - 2\,h\,\delta(z)\,\Gamma^{\text{ann}}\,\psi\,,
 \end{split}
\end{equation}
where $\vec{V}_c$ is the convective wind from stars in the Galactic plane and $K_0\, \beta\, p^\delta$ is the diffusion term which is a power-law in momentum. The coefficient $D_{EE}=\frac{2}{9}\,V_a^2\,\frac{E^2\,\beta^4}{K_0\,E^\delta}$ represents diffusion in energy space, \textit{i.e.} re-acceleration by turbulences of the Galactic magnetic field characterised by their Alfv\'en speed $V_a$, while the term $b_{\text{loss}}$ stands for the energy losses which are mainly due to elastic interaction with the interstellar gas. The half-thickness $h$ of the Galactic disk is taken to the standard value of 100\,pc. The various source terms are 
\begin{equation}
 Q^{\text{prim}}(T,r)=\frac{\rho_{\text{halo}}(r)}{m_{\text{DM}}\,\tau_{\text{DM}}}\,\frac{dN}{dT}\,,
\end{equation}
corresponding to DM decay in this case, $Q^{\text{sec}}$, coming from the interaction of cosmic-ray protons and $\alpha$ particles with the interstellar medium (\textit{i.e.} hydrogen and helium), and $Q^{\text{ter}}$, coming from the inelastic interaction of cosmic-ray antiprotons with the gas. Another contribution to the source term is also the destruction of protons and antiprotons happening at a rate $\Gamma^{\text{ann}}$. Finally, one also needs to take into account the boundary conditions: outside the diffusive halo the flux is assumed to vanish, \textit{i.e.} $\psi(z=\pm L)=0$, where $L$ stands for the half-thickness of the diffusion zone. A detailed solution of this equation with the use of semi-analytical methods can be found in~\cite{Bringmann:2006im}. The cosmic-ray flux $\Phi$ is directly related to the density through $\Phi=\beta\,\psi/(4\pi)$.

The precise values of the parameters $K_0$, $\delta$, $\vec{V}_c$, $V_a$ and $L$ are not fixed by the theory. The way to proceed is to make use of the ratio of secondary to primary cosmic rays (such as boron to carbon for instance) and proceed to a $\chi^2$ analysis to find the parameters that give a decent fit to the data. In this paper we made use of all the parameters found by~\cite{Maurin:2001sj} to give a $\chi^2$ statistics lower than 40 (for 23 degrees of freedom). This corresponds to a grid of parameters made of about 1600 sets. The scan we performed has been done over all these parameter sets. The authors of~\cite{Donato:2003xg} selected three parameters sets they have labelled MIN/MED/MAX which are given in Table~\ref{tab:propa} as examples of possible values of these parameters.
\begin{table}[t]
 \centering
 \begin{tabular}{lccccc}
 \toprule
 & $L$ (kpc) & $K_0$ (kpc$^2$/Myr) & $\delta$ & $\Vert\vec{V}_c\Vert$ (km/s) & $V_a$ (km/s) \\ 
 \midrule
 MIN & 1 & 0.0016 & 0.85 & 13.5 & 22.4 \\
 MED & 4 & 0.0112 & 0.70 & 12 & 52.9 \\
 MAX & 15 & 0.0765 & 0.46 & 5 & 117.6 \\
 \bottomrule
 \end{tabular}
 \caption{Three sets of propagation parameters found in~\cite{Donato:2003xg} to give a good fit to the boron-to-carbon data and to size approximatively the full span of antiproton fluxes.}
 \label{tab:propa}
\end{table}

Each component (primaries, secondaries and tertiaries) has its own phenomenology and the point of this paper is not to detail them. For more details the interested reader might want to have a look at~\cite{Donato:2001ms,Donato:2003xg,Bringmann:2006im,Bottino:2005xy,Brun:2007tn} which were realized within the same semi-analytical framework. Alternative approaches, making use of a fully numerical solution of the diffusion equation, can also be found in the literature~\cite{Strong:1998pw,Evoli:2008dv}.

}

\begin{table}[p]
 \centering
 \begin{small}
 \setlength{\tabcolsep}{2mm}
 \renewcommand{\arraystretch}{0.95}
 \begin{tabular}{@{}c@{\hskip 4mm}ccccc}
  \toprule
  & \multicolumn{5}{c}{$\text{DM}\rightarrow Z\nu$ with solar modulation} \\
   \cmidrule(lr){2-6}
  $m_{\text{DM}}$ & Low & MIN & MED & MAX & High \\
  \midrule
 100\,GeV & $2.8\times 10^{27}$\,s & $3.0\times 10^{27}$\,s & $1.6\times 10^{28}$\,s & $5.7\times 10^{28}$\,s & $7.3\times 10^{28}$\,s \\
 150\,GeV & $2.0\times 10^{27}$\,s & $2.1\times 10^{27}$\,s & $1.1\times 10^{28}$\,s & $3.8\times 10^{28}$\,s & $5.0\times 10^{28}$\,s \\
 200\,GeV & $1.6\times 10^{27}$\,s & $1.7\times 10^{27}$\,s & $9.3\times 10^{27}$\,s & $2.8\times 10^{28}$\,s & $3.8\times 10^{28}$\,s \\
 300\,GeV & $1.3\times 10^{27}$\,s & $1.3\times 10^{27}$\,s & $7.3\times 10^{27}$\,s & $1.9\times 10^{28}$\,s & $2.7\times 10^{28}$\,s \\
 500\,GeV & $9.4\times 10^{26}$\,s & $9.4\times 10^{26}$\,s & $5.4\times 10^{27}$\,s & $1.2\times 10^{28}$\,s & $1.8\times 10^{28}$\,s \\
 1\,TeV & $6.0\times 10^{26}$\,s & $6.0\times 10^{26}$\,s & $3.5\times 10^{27}$\,s & $7.1\times 10^{27}$\,s & $1.1\times 10^{28}$\,s \\
 2\,TeV & $3.4\times 10^{26}$\,s & $3.4\times 10^{26}$\,s & $2.0\times 10^{27}$\,s & $4.2\times 10^{27}$\,s & $5.9\times 10^{27}$\,s \\
 3\,TeV & $2.3\times 10^{26}$\,s & $2.3\times 10^{26}$\,s & $1.3\times 10^{27}$\,s & $3.0\times 10^{27}$\,s & $4.0\times 10^{27}$\,s \\
 5\,TeV & $1.3\times 10^{26}$\,s & $1.3\times 10^{26}$\,s & $7.5\times 10^{26}$\,s & $1.8\times 10^{27}$\,s & $2.3\times 10^{27}$\,s \\
 10\,TeV & $5.3\times 10^{25}$\,s & $5.3\times 10^{25}$\,s & $3.0\times 10^{26}$\,s & $7.8\times 10^{26}$\,s & $9.0\times 10^{26}$\,s \\
  \midrule
  & \multicolumn{5}{c}{$\text{DM}\rightarrow W\ell$ with solar modulation} \\
   \cmidrule(lr){2-6}
  $m_{\text{DM}}$ & Low & MIN & MED & MAX & High \\
  \midrule
 85\,GeV & $3.1\times 10^{27}$\,s & $3.4\times 10^{27}$\,s & $1.8\times 10^{28}$\,s & $6.5\times 10^{28}$\,s & $8.3\times 10^{28}$\,s \\
 100\,GeV & $2.7\times 10^{27}$\,s & $2.9\times 10^{27}$\,s & $1.5\times 10^{28}$\,s & $5.6\times 10^{28}$\,s & $7.1\times 10^{28}$\,s \\
 150\,GeV & $2.0\times 10^{27}$\,s & $2.1\times 10^{27}$\,s & $1.1\times 10^{28}$\,s & $3.7\times 10^{28}$\,s & $4.9\times 10^{28}$\,s \\
 200\,GeV & $1.7\times 10^{27}$\,s & $1.7\times 10^{27}$\,s & $9.5\times 10^{27}$\,s & $2.7\times 10^{28}$\,s & $3.7\times 10^{28}$\,s \\
 300\,GeV & $1.3\times 10^{27}$\,s & $1.3\times 10^{27}$\,s & $7.5\times 10^{27}$\,s & $1.8\times 10^{28}$\,s & $2.7\times 10^{28}$\,s \\
 500\,GeV & $9.6\times 10^{26}$\,s & $9.6\times 10^{26}$\,s & $5.5\times 10^{27}$\,s & $1.2\times 10^{28}$\,s & $1.8\times 10^{28}$\,s \\
 1\,TeV & $6.0\times 10^{26}$\,s & $6.0\times 10^{26}$\,s & $3.4\times 10^{27}$\,s & $7.0\times 10^{27}$\,s & $1.1\times 10^{28}$\,s \\
 2\,TeV & $3.3\times 10^{26}$\,s & $3.3\times 10^{26}$\,s & $1.9\times 10^{27}$\,s & $4.2\times 10^{27}$\,s & $5.8\times 10^{27}$\,s \\
 3\,TeV & $2.2\times 10^{26}$\,s & $2.2\times 10^{26}$\,s & $1.3\times 10^{27}$\,s & $3.0\times 10^{27}$\,s & $3.8\times 10^{27}$\,s \\
 5\,TeV & $1.2\times 10^{26}$\,s & $1.2\times 10^{26}$\,s & $7.1\times 10^{26}$\,s & $1.7\times 10^{27}$\,s & $2.1\times 10^{27}$\,s \\
 10\,TeV & $4.8\times 10^{25}$\,s & $4.8\times 10^{25}$\,s & $2.7\times 10^{26}$\,s & $7.1\times 10^{26}$\,s & $8.1\times 10^{26}$\,s \\
  \midrule
  & \multicolumn{5}{c}{$\text{DM}\rightarrow h\nu$ with solar modulation} \\
   \cmidrule(lr){2-6}
  $m_{\text{DM}}$ & Low & MIN & MED & MAX & High \\
  \midrule
 150\,GeV & $2.4\times 10^{27}$\,s & $2.7\times 10^{27}$\,s & $1.4\times 10^{28}$\,s & $5.6\times 10^{28}$\,s & $7.0\times 10^{28}$\,s \\
 200\,GeV & $2.0\times 10^{27}$\,s & $2.1\times 10^{27}$\,s & $1.1\times 10^{28}$\,s & $4.2\times 10^{28}$\,s & $5.3\times 10^{28}$\,s \\
 300\,GeV & $1.6\times 10^{27}$\,s & $1.6\times 10^{27}$\,s & $8.8\times 10^{27}$\,s & $2.7\times 10^{28}$\,s & $3.7\times 10^{28}$\,s \\
 500\,GeV & $1.2\times 10^{27}$\,s & $1.2\times 10^{27}$\,s & $6.8\times 10^{27}$\,s & $1.6\times 10^{28}$\,s & $2.4\times 10^{28}$\,s \\
 1\,TeV & $7.8\times 10^{26}$\,s & $7.8\times 10^{26}$\,s & $4.5\times 10^{27}$\,s & $9.1\times 10^{27}$\,s & $1.5\times 10^{28}$\,s \\
 2\,TeV & $4.7\times 10^{26}$\,s & $4.7\times 10^{26}$\,s & $2.7\times 10^{27}$\,s & $5.6\times 10^{27}$\,s & $8.3\times 10^{27}$\,s \\
 3\,TeV & $3.3\times 10^{26}$\,s & $3.3\times 10^{26}$\,s & $1.9\times 10^{27}$\,s & $4.2\times 10^{27}$\,s & $5.8\times 10^{27}$\,s \\
 5\,TeV & $2.0\times 10^{26}$\,s & $2.0\times 10^{26}$\,s & $1.1\times 10^{27}$\,s & $2.7\times 10^{27}$\,s & $3.5\times 10^{27}$\,s \\
 10\,TeV & $8.8\times 10^{25}$\,s & $8.8\times 10^{25}$\,s & $4.9\times 10^{26}$\,s & $1.3\times 10^{27}$\,s & $1.5\times 10^{27}$\,s \\
  \bottomrule
 \end{tabular}
 \end{small}
 \caption{Summary table of the 95\,\% CL lower limits on the DM lifetime in the decay channels $Z\nu$, $W\ell$ and $h\nu$ derived from PAMELA antiproton data~\cite{Adriani:2012paa}. \textit{Low} and \textit{High} indicate the lowest and highest limits found in the scan over propagation parameter sets, respectively.}
 \label{tab:limits}
\end{table}
\begin{table}[p]
 \centering
 \begin{small}
 \setlength{\tabcolsep}{2mm}
 \renewcommand{\arraystretch}{0.95}
 \begin{tabular}{@{}c@{\hskip 4mm}ccccc}
  \toprule
  & \multicolumn{5}{c}{Bino NLSP with solar modulation} \\
   \cmidrule(lr){2-6}
  $m_{3/2}$ & Low & MIN & MED & MAX & High \\
  \midrule
 85\,GeV & $7.2\times 10^{26}$\,s & $7.9\times 10^{26}$\,s & $4.1\times 10^{27}$\,s & $1.5\times 10^{28}$\,s & $1.9\times 10^{28}$\,s \\
 100\,GeV & $2.2\times 10^{27}$\,s & $2.4\times 10^{27}$\,s & $1.3\times 10^{28}$\,s & $4.6\times 10^{28}$\,s & $5.8\times 10^{28}$\,s \\
 150\,GeV & $1.9\times 10^{27}$\,s & $2.0\times 10^{27}$\,s & $1.1\times 10^{28}$\,s & $3.6\times 10^{28}$\,s & $4.8\times 10^{28}$\,s \\
 200\,GeV & $1.7\times 10^{27}$\,s & $1.7\times 10^{27}$\,s & $9.4\times 10^{27}$\,s & $2.8\times 10^{28}$\,s & $3.8\times 10^{28}$\,s \\
 300\,GeV & $1.3\times 10^{27}$\,s & $1.3\times 10^{27}$\,s & $7.5\times 10^{27}$\,s & $1.9\times 10^{28}$\,s & $2.7\times 10^{28}$\,s \\
 500\,GeV & $9.9\times 10^{26}$\,s & $9.9\times 10^{26}$\,s & $5.7\times 10^{27}$\,s & $1.2\times 10^{28}$\,s & $1.9\times 10^{28}$\,s \\
 1\,TeV & $6.4\times 10^{26}$\,s & $6.4\times 10^{26}$\,s & $3.7\times 10^{27}$\,s & $7.5\times 10^{27}$\,s & $1.1\times 10^{28}$\,s \\
 2\,TeV & $3.7\times 10^{26}$\,s & $3.7\times 10^{26}$\,s & $2.1\times 10^{27}$\,s & $4.5\times 10^{27}$\,s & $6.4\times 10^{27}$\,s \\
 3\,TeV & $2.5\times 10^{26}$\,s & $2.5\times 10^{26}$\,s & $1.4\times 10^{27}$\,s & $3.3\times 10^{27}$\,s & $4.3\times 10^{27}$\,s \\
 5\,TeV & $1.4\times 10^{26}$\,s & $1.4\times 10^{26}$\,s & $8.2\times 10^{26}$\,s & $2.0\times 10^{27}$\,s & $2.5\times 10^{27}$\,s \\
 10\,TeV & $5.9\times 10^{25}$\,s & $5.9\times 10^{25}$\,s & $3.3\times 10^{26}$\,s & $8.6\times 10^{26}$\,s & $9.9\times 10^{26}$\,s \\
  \midrule
  & \multicolumn{5}{c}{Wino NLSP with solar modulation} \\
   \cmidrule(lr){2-6}
  $m_{3/2}$ & Low & MIN & MED & MAX & High \\
  \midrule
 85\,GeV & $3.6\times 10^{25}$\,s & $3.9\times 10^{25}$\,s & $2.\times 10^{26}$\,s & $7.5\times 10^{26}$\,s & $9.6\times 10^{26}$\,s \\
 100\,GeV & $5.5\times 10^{26}$\,s & $6.0\times 10^{26}$\,s & $3.1\times 10^{27}$\,s & $1.1\times 10^{28}$\,s & $1.5\times 10^{28}$\,s \\
 150\,GeV & $1.7\times 10^{27}$\,s & $1.8\times 10^{27}$\,s & $9.6\times 10^{27}$\,s & $3.2\times 10^{28}$\,s & $4.2\times 10^{28}$\,s \\
 200\,GeV & $1.6\times 10^{27}$\,s & $1.6\times 10^{27}$\,s & $8.9\times 10^{27}$\,s & $2.7\times 10^{28}$\,s & $3.6\times 10^{28}$\,s \\
 300\,GeV & $1.3\times 10^{27}$\,s & $1.3\times 10^{27}$\,s & $7.4\times 10^{27}$\,s & $1.9\times 10^{28}$\,s & $2.7\times 10^{28}$\,s \\
 500\,GeV & $9.9\times 10^{26}$\,s & $9.9\times 10^{26}$\,s & $5.6\times 10^{27}$\,s & $1.2\times 10^{28}$\,s & $1.9\times 10^{28}$\,s \\
 1\,TeV & $6.3\times 10^{26}$\,s & $6.3\times 10^{26}$\,s & $3.7\times 10^{27}$\,s & $7.5\times 10^{27}$\,s & $1.1\times 10^{28}$\,s \\
 2\,TeV & $3.7\times 10^{26}$\,s & $3.7\times 10^{26}$\,s & $2.1\times 10^{27}$\,s & $4.5\times 10^{27}$\,s & $6.4\times 10^{27}$\,s \\
 3\,TeV & $2.5\times 10^{26}$\,s & $2.5\times 10^{26}$\,s & $1.4\times 10^{27}$\,s & $3.3\times 10^{27}$\,s & $4.3\times 10^{27}$\,s \\
 5\,TeV & $1.4\times 10^{26}$\,s & $1.4\times 10^{26}$\,s & $8.2\times 10^{26}$\,s & $2.0\times 10^{27}$\,s & $2.5\times 10^{27}$\,s \\
 10\,TeV & $5.9\times 10^{25}$\,s & $5.9\times 10^{25}$\,s & $3.3\times 10^{26}$\,s & $8.6\times 10^{26}$\,s & $9.9\times 10^{26}$\,s \\
  \midrule
  & \multicolumn{5}{c}{Higgsino NLSP with solar modulation} \\
   \cmidrule(lr){2-6}
  $m_{3/2}$ & Low & MIN & MED & MAX & High \\
  \midrule
 85\,GeV & $3.1\times 10^{27}$\,s & $3.4\times 10^{27}$\,s & $1.7\times 10^{28}$\,s & $6.4\times 10^{28}$\,s & $8.2\times 10^{28}$\,s \\
 100\,GeV & $2.7\times 10^{27}$\,s & $2.9\times 10^{27}$\,s & $1.5\times 10^{28}$\,s & $5.6\times 10^{28}$\,s & $7.1\times 10^{28}$\,s \\
 150\,GeV & $2.0\times 10^{27}$\,s & $2.1\times 10^{27}$\,s & $1.1\times 10^{28}$\,s & $3.7\times 10^{28}$\,s & $4.9\times 10^{28}$\,s \\
 200\,GeV & $1.7\times 10^{27}$\,s & $1.7\times 10^{27}$\,s & $9.5\times 10^{27}$\,s & $2.8\times 10^{28}$\,s & $3.8\times 10^{28}$\,s \\
 300\,GeV & $1.3\times 10^{27}$\,s & $1.3\times 10^{27}$\,s & $7.5\times 10^{27}$\,s & $1.9\times 10^{28}$\,s & $2.7\times 10^{28}$\,s \\
 500\,GeV & $9.9\times 10^{26}$\,s & $9.9\times 10^{26}$\,s & $5.6\times 10^{27}$\,s & $1.2\times 10^{28}$\,s & $1.9\times 10^{28}$\,s \\
 1\,TeV & $6.3\times 10^{26}$\,s & $6.3\times 10^{26}$\,s & $3.7\times 10^{27}$\,s & $7.4\times 10^{27}$\,s & $1.1\times 10^{28}$\,s \\
 2\,TeV & $3.7\times 10^{26}$\,s & $3.7\times 10^{26}$\,s & $2.1\times 10^{27}$\,s & $4.5\times 10^{27}$\,s & $6.4\times 10^{27}$\,s \\
 3\,TeV & $2.5\times 10^{26}$\,s & $2.5\times 10^{26}$\,s & $1.4\times 10^{27}$\,s & $3.3\times 10^{27}$\,s & $4.3\times 10^{27}$\,s \\
 5\,TeV & $1.4\times 10^{26}$\,s & $1.4\times 10^{26}$\,s & $8.2\times 10^{26}$\,s & $2.0\times 10^{27}$\,s & $2.5\times 10^{27}$\,s \\
 10\,TeV & $5.9\times 10^{25}$\,s & $5.9\times 10^{25}$\,s & $3.3\times 10^{26}$\,s & $8.6\times 10^{26}$\,s & $9.9\times 10^{26}$\,s \\
  \bottomrule
 \end{tabular}
 \end{small}
 \caption{Summary table of the 95\,\% CL lower limits on the gravitino lifetime for the cases with Bino, Wino or Higgsino NLSP derived from PAMELA antiproton data~\cite{Adriani:2012paa}. \textit{Low} and \textit{High} indicate the lowest and highest limits found in the scan over propagation parameter sets, respectively.}
 \label{tab:gravitinolimits}
\end{table}

\end{document}